\documentclass[
  DIV=12,
  paper=letter,
  fontsize=12,
  toc=bib,
  captions=tableheading
]{scrartcl}

\addtokomafont{caption}{\small} 
\setkomafont{captionlabel}{\sffamily\bfseries} 
\setcapindent{0em} 

\usepackage[margin=1in]{geometry} 
\usepackage{lmodern} 
\usepackage[semibold]{sourcesanspro} 
\usepackage[scale=0.9]{sourcecodepro} 
\usepackage[utf8]{inputenc} 

\usepackage[
  backend=biber,
  sorting=none,
  style=phys,
  articletitle=true,
  biblabel=brackets,
  chaptertitle=false,
  eprint=true,
  pageranges=false
]{biblatex}
\usepackage{graphicx} 
\usepackage{float} 
\usepackage{caption} 

\usepackage{booktabs} 

\usepackage{footnote} 
\makesavenoteenv{tabular}
\makesavenoteenv{table}

\usepackage{enumitem} 

\usepackage{color} 
\usepackage{url} 
\usepackage{hyperref} 
\hypersetup{
  colorlinks=true, 
  allcolors=[rgb]{0, 0.19, 0.33}, 
  linktoc=section,
  bookmarksnumbered=true
}

\usepackage{mathtools}
\usepackage{amssymb}

\allowdisplaybreaks

\makeatletter
\newsavebox\myboxA
\newsavebox\myboxB
\newlength\mylenA

\newcommand*\xoverline[2][0.75]{%
    \sbox{\myboxA}{$\m@th#2$}%
    \setbox\myboxB\null
    \ht\myboxB=\ht\myboxA%
    \dp\myboxB=\dp\myboxA%
    \wd\myboxB=#1\wd\myboxA
    \sbox\myboxB{$\m@th\overline{\copy\myboxB}$}
    \setlength\mylenA{\the\wd\myboxA}
    \addtolength\mylenA{-\the\wd\myboxB}%
    \ifdim\wd\myboxB<\wd\myboxA%
      \rlap{\hskip 0.5\mylenA\usebox\myboxB}{\usebox\myboxA}%
    \else
      \hskip -0.5\mylenA\rlap{\usebox\myboxA}{\hskip 0.5\mylenA\usebox\myboxB}%
    \fi
}
\makeatother

\begin{document}


\begin{titlepage}

  \renewcommand{\thefootnote}{\fnsymbol{footnote}}

  \begin{flushright}
    UMN--TH--3805/18
  \end{flushright}

  \begin{centering}

    \bigskip
    \bigskip

    {\Large \textbf{Partially composite supersymmetry}}

    \bigskip
    \bigskip

    Yusuf Buyukdag,\footnote{\href{mailto:buyuk007@umn.edu}{buyuk007@umn.edu}}
    Tony Gherghetta,\footnote{\href{mailto:tgher@umn.edu}{tgher@umn.edu}}
    and Andrew S. Miller\footnote{\href{mailto:mill5738@umn.edu}{mill5738@umn.edu}}

    \medskip

    \textit{School of Physics and Astronomy, University of Minnesota,\\Minneapolis, Minnesota 55455, USA}

  \end{centering}

  \bigskip
  \bigskip

  \begin{abstract}

    We consider a supersymmetric model that uses partial compositeness
    to explain the fermion mass hierarchy and predict the sfermion mass spectrum.
    The Higgs and third-generation matter superfields are elementary, 
    while the first two matter generations are composite. Linear mixing between elementary
    superfields and supersymmetric operators with large anomalous dimensions
    is responsible for simultaneously generating the fermion and sfermion mass hierarchies.
    After supersymmetry is broken by the strong dynamics, partial compositeness causes
    the first- and second-generation  sfermions to be split from the much lighter
    gauginos and third-generation sfermions. This occurs even though the tree-level 
    soft masses of the elementary fields are subject to large radiative corrections from the 
    composite sector, which we calculate in the gravitational dual theory using the 
    AdS/CFT correspondence. The sfermion mass scale is constrained
    by the observed 125 GeV Higgs boson, leading to stop masses and gauginos
    around 10--100 TeV and the first two generation sfermion masses around 100--1000 TeV.
    This gives rise to a splitlike supersymmetric model that explains the fermion mass
    hierarchy while simultaneously predicting an inverted sfermion mass spectrum
    consistent with LHC and flavor constraints. Finally, the lightest supersymmetric particle 
    is a gravitino in the keV to TeV range, which can play the role of dark matter.


  \end{abstract}

\end{titlepage}

\setcounter{footnote}{0}


\tableofcontents


\section{Introduction}

The Higgs boson discovery at the LHC~\cite{Chatrchyan:2012xdj,Aad:2012tfa,Aad:2015zhl} has led to new constraints on the parameter space of supersymmetric models.  In particular, the explanation of the 125 GeV Higgs boson requires stop masses (or $A$-terms) which are large, $\gtrsim 1$~TeV, causing a significant increase in the tuning of supersymmetric models. In addition, in order to ameliorate the supersymmetric flavor problem without any additional structure, the first- and second-generation sfermions are required to have masses in the 100--1000 TeV range. Satisfying these two requirements leads to a version of split supersymmetry~\cite{Wells:2004di, ArkaniHamed:2004fb} dubbed \textit{minisplit}~\cite{Arvanitaki:2012ps, ArkaniHamed:2012gw} which explains the 125 GeV Higgs boson while simultaneously maintaining the successful features of supersymmetric models such as gauge coupling unification and a dark matter candidate. Of course, this comes at the price of a meso-tuning, which may be a sign that we live in a multiverse~\cite{ArkaniHamed:2004fb}, or instead could possibly be explained by a relaxion mechanism~\cite{Batell:2015fma, Evans:2016htp}.

A knowledge of the sfermion mass spectrum has important implications for collider and flavor experiments seeking to discover supersymmetry. In split supersymmetric models, the supersymmetry-breaking scale occurs near the PeV scale, with sfermion masses in the range 10--1000 TeV. Even though this hierarchy of sfermion masses seems unrelated to the fermion mass hierarchy, it begs the question as to whether these two hierarchies could in fact be explained by the same mechanism. For example, a novel way to account for the fermion mass hierarchy is the idea of partial compositeness~\cite{Kaplan:1991dc}. New strong dynamics is responsible for operators with large anomalous dimensions that linearly mix with elementary fermions. Assuming the Higgs boson is elementary, a large Yukawa coupling then arises for mostly elementary fermion mass eigenstates, while mostly composite fermion mass eigenstates have a correspondingly smaller Yukawa coupling. The hierarchy of Yukawa couplings is therefore explained by a set of operators with large, order-one anomalous dimensions.

If one now further assumes that the strong dynamics is responsible for breaking supersymmetry, then an interesting correlation between fermion and sfermion masses results from partial compositeness. Supersymmetric operators that linearly mix with elementary fermions can now communicate supersymmetry breaking to the elementary sector. In this way, composite sfermions obtain large supersymmetry breaking masses, while elementary sfermions obtain hierarchically smaller soft masses. The fermion mass hierarchy is therefore inversely related to the sfermion mass hierarchy: light (elementary) stops correspond to heavy (elementary) top quarks, while heavy (composite) selectrons are related to the light (composite) electron. Together with the fact that gauginos and Higgsinos are predominantly elementary---and therefore lighter than the composite sfermions---a ``split'' supersymmetric spectrum arises where the fermion mass hierarchy is naturally explained.  It is the anomalous dimensions of the corresponding supersymmetric operators that simultaneously controls the fermion and sfermion masses. This contrasts with an alternative approach that radiatively generates fermion masses from a sfermion anarchy~\cite{Altmannshofer:2014qha}.

A four-dimensional (4D), holographic description of partial compositeness for the fermion mass hierarchy was previously considered in Refs.~\cite{Contino:2004vy, Gherghetta:2010cj}. In this paper, we generalize this description to the supersymmetric case, using results from Ref.~\cite{Cacciapaglia:2008bi}. Assuming that electroweak symmetry is broken in the elementary sector (with an elementary Higgs), we consider the linear mixing of elementary superfields with supersymmetric operators in the 4D dual gauge theory. This mixing is responsible not only for the fermion mass hierarchy but also for the transmission of the (dynamical) supersymmetry breaking in the composite sector to the elementary sector. This leads to relations that determine the fermion and sfermion mass hierarchy in terms of the anomalous dimensions of supersymmetric operators. These type of theories are similar to single-sector models of supersymmetry breaking which were originally proposed in Refs.~\cite{ArkaniHamed:1997fq, Luty:1998vr} and further studied in Refs.~\cite{Gherghetta:2000kr, Gabella:2007cp, Franco:2009wf, Craig:2009hf, Aharony:2010ch}. A shortened version of this work summarizing the main results can be found in Ref.~\cite{Buyukdag:2018ose}.

\begin{figure}[t]
  \centering
  \includegraphics{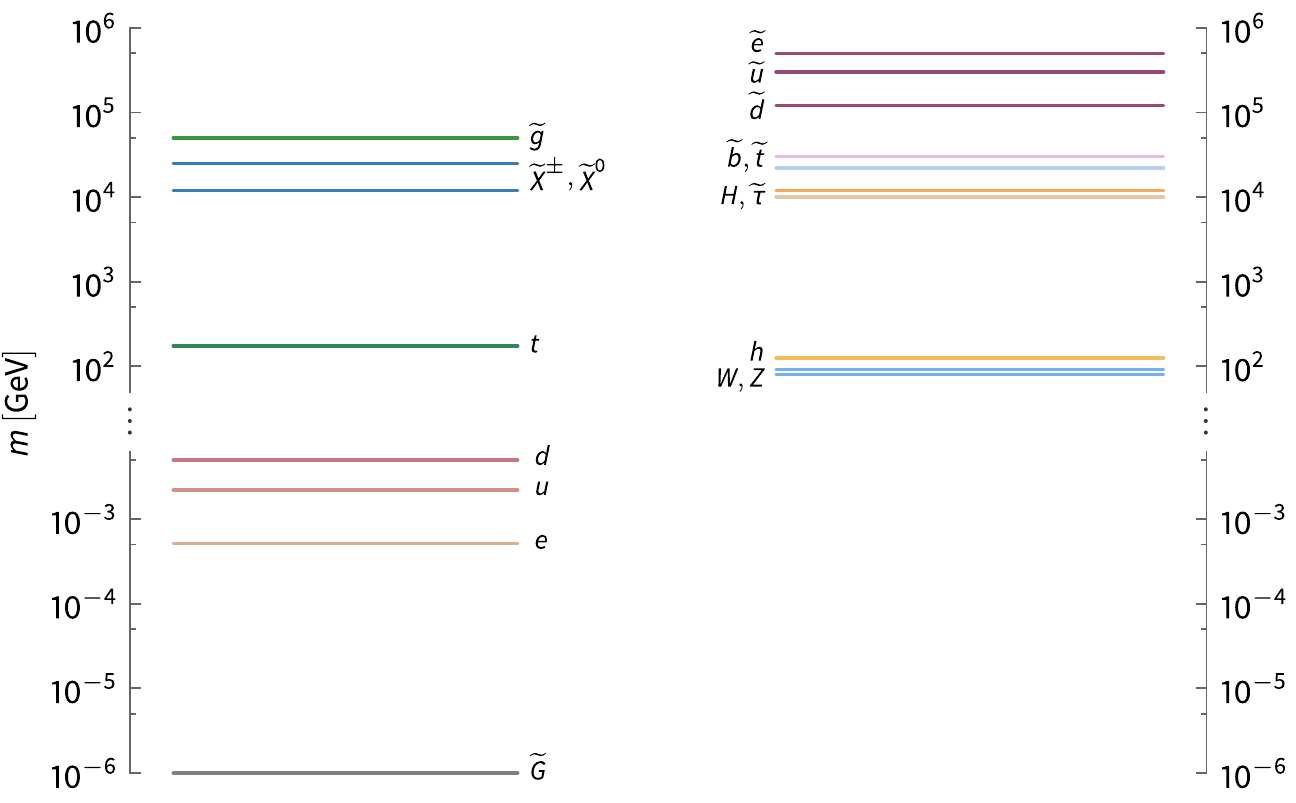}
  \caption{Schematic diagram depicting a possible particle spectrum of the partially composite supersymmetric model. The left (right) column depicts the fermions (bosons). The sfermion mass hierarchy is inversely related to the fermion mass hierarchy and the LSP is the gravitino.}
  \label{fig:spectrum-schematic}
\end{figure}

The nonperturbative nature of the strong dynamics only allows the sfermion mass spectrum to be qualitatively determined. To obtain a more detailed sfermion spectrum that is consistent with a 125 GeV Higgs boson, we use the anti-de Sitter/conformal field theory (AdS/CFT) correspondence to model the strong dynamics associated with partial compositeness in a slice of five-dimensional anti-de Sitter space (AdS$_5$)~\cite{Randall:1999ee}. The supersymmetric Higgs sector is localized on the UV brane, while supersymmetry breaking is confined to the IR brane. The supersymmetric matter fields are bulk fields with the top quark (light fermions) localized near the UV (IR) brane. In the dual, five-dimensional (5D) gravity description, the overlap of fermion profiles~\cite{Grossman:1999ra} with the UV-localized Higgs fields mimics the partial compositeness and explains the fermion mass hierarchy~\cite{Gherghetta:2000qt}. The Standard Model Yukawa couplings are used to constrain the bulk fermion profiles, which then determine the soft masses at the IR scale. Both the Higgs-sector soft masses and the soft trilinear scalar couplings arise at loop order, due to radiative corrections from the bulk that transmit the breaking of supersymmetry. Renormalization group evolution is then used to run the soft masses down to the electroweak scale and obtain the 125 GeV Higgs boson mass. Using this procedure, we analyze two benchmark scenarios: one for the case that the gaugino masses arise from a singlet spurion and the other in nonsinglet spurion case. We find that the observed Higgs boson mass naturally accommodates sparticle spectra that hierarchically suppress the masses of the stops and the other third-generation sfermions below the mass scale of the first- and second-generation sfermions. In particular, if the masses of the first- and second-generation sfermions are restricted to be above 100 TeV to additionally suppress flavor-changing neutral currents that arise in supersymmetric models, the stop masses lie in the range 20--100 TeV, while the masses of the lightest stau and neutralino may be as low as 10 TeV. Previous attempts to explain the sfermion mass hierarchy in a slice of AdS$_5$ before the Higgs boson mass was known were considered in Refs.~\cite{Gherghetta:2000kr, Gabella:2007cp}. The results obtained in this paper are the first predictions for the sfermion mass spectrum from partial compositeness that are compatible with a 125 GeV Higgs boson. Additionally, we include for the first time the full one-loop radiative corrections to the bulk scalar soft masses squared, the Higgs-sector soft terms, and the soft trilinear scalar couplings. For stops and other UV-localized sfermions, these corrections provide the dominant soft mass contributions, and accordingly have important phenomenological consequences. For the Higgs sector, they control the breaking of electroweak symmetry.

A schematic diagram of a possible mass spectrum in the partially composite supersymmetric model is depicted in Fig.~\ref{fig:spectrum-schematic}. It assumes an elementary Higgs sector interacting with an elementary top quark sector (UV localized). The first- and second-generation  fermions are composites (IR localized) of a strongly coupled sector that is also responsible for dynamically breaking supersymmetry. The strong dynamics therefore leads to a sfermion mass spectrum that is inversely related to the fermion mass spectrum. Furthermore, the graviton supermultiplet is elementary, and, since it couples gravitationally, the gravitino receives a small supersymmetry-breaking contribution, becoming the lightest supersymmetric particle (LSP) with a mass $\gtrsim 1$ keV. This differs from other split-supersymmetry models where the gravitino is usually the heaviest superpartner. Finally, even though the first two generations of matter are composite, gauge coupling unification still occurs at approximately $10^{16}$ GeV, as in the usual supersymmetric standard model (assuming any underlying strong dynamics is SU(5) symmetric). It should also be noted that the elementary matter is present at the grand unified theory (GUT) scale with order-one Yukawa couplings. This helps one to avoid the tension that occurs from Yukawa coupling unification in usual grand unification scenarios, where the lighter first two fermion generations
are elementary (do not mix with a composite sector), and consequently have tiny Yukawa couplings at the GUT scale.

The rest of this paper is summarized as follows: 
In Sec.~\ref{sec:partcomp}, we review the idea of partial compositeness in the context of the fermion mass hierarchy. This is then generalized to a supersymmetric model when supersymmetry is assumed to be broken by the strong dynamics.
This gives rise to a sfermion mass hierarchy that inverts the fermion mass ordering.
We then construct the gravitational dual of this model in a slice of AdS$_5$ in Sec.~\ref{sec:5d}. The soft masses resulting from supersymmetry breaking on the IR brane are calculated, including one-loop radiative corrections from the bulk theory to the sfermions and Higgs fields.
In Sec.~\ref{sec:spectrum-5d}, we develop the parameter space available in the 5D model and estimate the constraints arising on this space from various phenomenological and theoretical requirements. 
We select two representative benchmark scenarios and perform a full numerical analysis. We conclude by presenting the resulting spectra.
In Appendix~\ref{app:holographic-mixing}, we write down the partially composite theory below the confinement scale in terms of component fields, and then determine the mass eigenstates resulting from the mixing of the elementary and composite sectors.
Next, in Appendix~\ref{app:zero-modes}, we summarize the zero-mode profiles for fields in the bulk of a slice of AdS$_5$, including their deformations due to the presence of supersymmetry breaking on the IR brane.
Finally, in Appendix~\ref{app:radiativecorrections}, we derive the one-loop radiative corrections to scalar soft masses squared, the Higgs soft $b$-term, and the soft trilinear scalar couplings arising in the 5D theory after supersymmetry is broken on the IR brane.


\section{Partial Compositeness}\label{sec:partcomp}

\subsection{The fermion mass hierarchy}\label{sec:fermionhierarchy}

We begin by briefly reviewing how partial compositeness explains the fermion mass hierarchy~\cite{Contino:2004vy, Gherghetta:2010cj}. Consider two sectors, an elementary sector with a Weyl fermion $\psi$ and a composite sector with a (charge-conjugate) fermion operator
$\mathcal{O}^c_\psi$. The scaling dimension of the fermion operator is written as $\frac{3}{2} + \delta$, where $\delta$ denotes the deviation from the canonical scaling dimension. The Lagrangian at the UV scale, $\Lambda_{\text{UV}}$, is taken to have the form
\begin{equation}
  \label{eq:fermionLag}
  \mathcal{L}_{\psi} 
    = i \psi^{\dagger} \bar{\sigma}^{\mu} \partial_{\mu} \psi 
    - \frac{1}{\Lambda_{\text{UV}}^{\delta-1}}  \left( \psi \mathcal{O}^c_\psi + \text{H.c.} \right) \, ,
\end{equation}
where the Minkowski metric $\eta_{\mu\nu} = \operatorname{diag}({-}, {+}, {+}, {+})$ and an order-one UV coefficient has been assumed in the second term. The mixing term in \eqref{eq:fermionLag} means that after confinement at an IR scale, $\Lambda_{\text{IR}}$, the mass eigenstates are an admixture of elementary and composite states. This is analogous to $\gamma$-$\rho$ mixing in QCD.
To obtain analytic estimates of this mixing, the strong dynamics is assumed to be described by a large-$N$ gauge theory, where $N$ is the number of colors. In the large-$N$ limit, the two-point function for a composite operator, $\mathcal{O}$ can be written as $\langle \mathcal{O}(p) \, \mathcal{O}(-p) \rangle = \sum_{n} a_n^2 / (p^2 + m_n^2)$ to leading order in $\frac{1}{N}$, where $a_n = \langle 0| \mathcal{O} |n \rangle \propto \frac{\sqrt{N}}{4 \pi}$ is the matrix element for $\mathcal{O}$ to create the $n^{\text{th}}$ state from the vacuum and $m_n$ is the mass of that state~\cite{Witten:1979kh}.

Applying these results to fermions, we consider a simple three-state system containing an elementary Weyl fermion $\psi$, together with a lowest-lying composite Dirac fermion $(\psi^{(1)}, \psi^{c(1)})$, with mass, $m_{\psi}^{(1)} = g_\psi^{(1)}\Lambda_{\text{IR}}$, where $g_\psi^{(1)}$ is an order-one coupling. Note that having a composite Dirac fermion follows from assuming that the strong dynamics does not break any Standard Model gauge symmetries. The Lagrangian at the scale $\Lambda_{\text{IR}}$ is given by
\begin{align}
  \mathcal{L}_\psi 
    &= i \psi^{\dagger} \bar{\sigma}^{\mu} \partial_{\mu} \psi 
     + i \psi^{\dagger(1)} \bar{\sigma}^{\mu} \partial_{\mu} \psi^{(1)} 
     + i \psi^{\dagger c(1)} \bar{\sigma}^{\mu} \partial_{\mu} \psi^{c(1)} \nonumber \\[1ex]
    &  \qquad
     - \varepsilon_\psi \Lambda_{\text{IR}} \left( \psi \psi^{c(1)} + \text{H.c.} \right) 
     - m_{\psi}^{(1)} \left( \psi^{(1)} \psi^{c(1)} + \text{H.c.} \right) \, .
       \label{eq:fermionLag2}
\end{align}
The dimensionless coupling $\varepsilon_\psi$ is defined at the IR scale to be
\begin{equation}
  \label{eq:omegaLR}
  \varepsilon_\psi \equiv \tilde{\varepsilon}_\psi (\Lambda_{\text{IR}}) \, \frac{\sqrt{N}}{4\pi} 
    = \frac{1}{\sqrt{\smash[b]{Z_\psi}}} 
      \left( \frac{\Lambda_{\text{IR}}}{\Lambda_{\text{UV}}} \right)^{\delta-1} 
      \frac{\sqrt{N}}{4\pi}
    \simeq \frac{1}{\sqrt{\smash[b]{\zeta_\psi}}} 
           \sqrt{\frac{\delta-1}{\left( \frac{\Lambda_{\text{IR}}}{\Lambda_{\text{UV}}} \right)^{2(1 - \delta)} - 1}} \, ,
\end{equation}
where the running parameter $\tilde{\varepsilon}_\psi(\mu)$ satisfies $\mu\frac{d \tilde{\varepsilon}_\psi}{d \mu} = (\delta - 1) \tilde{\varepsilon}_\psi + \zeta_\psi \frac{N}{16\pi^2} \, \tilde{\varepsilon}_\psi^3$, $\zeta_\psi$ is an order-one constant due to the (unknown) strong dynamics~\cite{Contino:2004vy}, and the coefficient $Z_\psi$ is the wavefunction renormalization of the elementary fermion. In the final expression we have taken the large-$N$ approximation of $Z_\psi$.

The diagonalization of the Lagrangian \eqref{eq:fermionLag2} generates fermionic admixtures of the
elementary and composite states~\cite{Batell:2007jv}. In particular, the massless fermion eigenstate is given by
\begin{equation}
  |\psi_0 \rangle 
    \simeq \mathcal{N}_\psi
           \left\{ |\psi\rangle - \frac{\varepsilon_\psi}{g_\psi^{(1)}}|\psi^{(1)}\rangle \right\}
    \simeq \mathcal{N}_\psi 
           \left\{ 
             |\psi\rangle 
           - \frac{1}{g_\psi^{(1)}} 
             \frac{1}{\sqrt{\smash[b]{\zeta_\psi}}} 
             \sqrt{\frac{\delta-1}{\left(\frac{\Lambda_{\text{IR}}}{\Lambda_{\text{UV}}}\right)^{2(1 - \delta)} - 1}} \, 
             |\psi^{(1)}\rangle
            \right\} \,
      \label{eq:zerostate}
\end{equation}
where $\mathcal{N}_\psi$ is a normalization constant. This expression shows that for $\delta \geq 1$ the mass eigenstate is mostly elementary, while for $0 \leq \delta < 1$ the state has a sizeable composite admixture. These admixtures play a crucial role in determining mass hierarchies.

At the UV scale, the chiral elementary fermions $\psi_{L,R}$ are coupled to the elementary Higgs field $H$ via the Yukawa interaction $\lambda \psi_L \psi_R H$, where $\lambda$ is an order-one proto-Yukawa coupling and the Higgs field is assumed to develop a vacuum expectation value (VEV) $\langle H \rangle = v / \sqrt{2}$. Diagonalizing the fermion Lagrangian at the IR scale with the Higgs contribution gives the fermion mass expression
\begin{equation}
  \label{eq:fermionmass}
  m_{\psi} 
    \simeq \frac{\lambda}{\sqrt{\smash[b]{Z_L Z_R}}} \frac{v}{\sqrt{2}} \, \mathcal{N}_\psi^2 
    \simeq \begin{cases} 
             \; \dfrac{\lambda}{\zeta_\psi} (\delta - 1)
                \dfrac{16 \pi^2}{N} 
                \dfrac{v}{\sqrt{2}} 
           & \phantom{\; 0 \leq} \delta \geq 1 \, , \\[1em]
             \; \dfrac{\lambda}{\zeta_\psi} (1 - \delta) 
                \dfrac{16 \pi^2}{N} 
                \dfrac{v}{\sqrt{2}} 
                \left( \dfrac{\Lambda_{\text{IR}}}{\Lambda_{\text{UV}}} \right)^{2(1 - \delta)} 
           & 0 \leq \delta < 1 \, .
           \end{cases}
\end{equation}
where for simplicity we have assumed that $\delta \equiv \delta_L = \delta_R$, $g_\psi^{(1)} \equiv g_{\psi_L}^{(1)} = g_{\psi_R}^{(1)}$, $\varepsilon_\psi \equiv \varepsilon_{\psi_L} = \varepsilon_{\psi_R}$, and explicitly included the normalization factor from \eqref{eq:zerostate}. The wavefunction coefficients $Z_{L,R}$ are approximated as in \eqref{eq:omegaLR}, and for $0 \leq \delta < 1$ we have assumed $\varepsilon_\psi\lesssim g_\psi^{(1)}$. Notice that when $\delta \geq 1$ there is no power-law suppression in the Yukawa coupling since the mass eigenstates are mostly elementary and have order-one couplings to the elementary Higgs field. This, for instance, would explain the Yukawa coupling of the top quark. This contrasts with the case $0 \leq \delta < 1$, where the mass eigenstates have a sizeable composite admixture with a power-law suppressed Yukawa interaction (due to the fact that the proto-Yukawa coupling $\lambda$ in the elementary sector is divided by the large wavefunction renormalization factor, $\sqrt{\smash[b]{Z_L Z_R}}$, at the IR scale). These states describe the remaining light fermions in the Standard Model, where for each flavor $i$ there is a corresponding operator with anomalous dimension $\delta_i$. Therefore, the hierarchical Yukawa couplings result from order-one anomalous dimensions of operators. In the supersymmetric generalization, these anomalous dimensions also determine the magnitude of the corresponding sfermion masses.

\subsection{Supersymmetric partial compositeness}\label{sec:sfermionmass}

\subsubsection{Chiral supermultiplet}

Next, we consider the supersymmetric generalization of partial compositeness. Consider the elementary chiral superfield $\Phi = \phi + \sqrt{2} \theta \psi + \theta\theta F$, where $\phi$ is a complex scalar field, $\psi$ is a Weyl fermion, and $F$ is an auxiliary field. In addition, we introduce
a supersymmetric chiral operator $\mathcal{O} = \mathcal{O}_\phi + \sqrt{2} \theta \mathcal{O}_\psi + \theta \theta \mathcal{O}_F$. The scaling dimension of the scalar operator is $\operatorname{dim} \mathcal{O}_\phi = 1 + \delta_{\mathcal{O}}$, the scaling dimension of the fermion operator is $\operatorname{dim} \mathcal{O}_\psi = \frac{3}{2} + \delta_{\mathcal{O}}$, and the scaling dimension of the auxiliary operator is $\operatorname{dim} \mathcal{O}_F = 2 + \delta_{\mathcal{O}}$, where $\delta_{\mathcal{O}} \geq 0$~\cite{Cacciapaglia:2008bi}.

The supersymmetric Lagrangian contains separate elementary and composite sectors, together with linear mixing terms of the form $[ \Phi {\mathcal{O}^c} ]_F$ for each chiral superfield. The superfield Lagrangian at the scale $\Lambda_{\text{UV}}$  is given by
\begin{equation}
  \mathcal{L}_\Phi 
    = [ \Phi^{\dagger} \Phi ]_D 
    + \frac{1}{\Lambda_{\text{UV}}^{\delta-1}} \bigl( [ \Phi\mathcal{O}^c ]_F + \text{H.c.} \bigr) \, 
\end{equation}
where $\mathcal{O}^c$ is the charge-conjugate composite operator with anomalous dimension $\delta$ and we have assumed an order-one UV coefficient.\footnote{A kinetic mixing between the elementary and composite sectors of the form $\Lambda_{\text{UV}}^{- \delta_{\mathcal{O}}} \bigl[ \Phi^{\dagger} \mathcal{O} + \text{H.c.} \bigr]_D$ has been omitted in our minimal setup.} The composite sector is assumed to confine at the infrared scale $\Lambda_{\text{IR}}$, and thus, for the large-$N$ strong dynamics, the two-point function $\langle \mathcal{O} \mathcal{O}\rangle$ is again assumed to be a sum over one-particle states. The Lagrangian at the IR scale can be written as
\begin{align}
  \mathcal{L}_\Phi 
    &= [ \Phi^{\dagger} \Phi ]_D 
     + [ \Phi^{c(1) \dagger } \Phi^{c(1)} ]_D 
     + [ \Phi^{(1) \dagger } \Phi^{(1)} ]_D \nonumber \\[1ex]
    &  \qquad
     + \varepsilon_\Phi \Lambda_{\text{IR}} \bigl( [ \Phi \Phi^{c(1)} ]_F + \text{H.c.} \bigr)
     + m_{\Phi}^{(1)} \bigl( [ \Phi^{(1)} \Phi^{c(1)} ]_F + \text{H.c.} \bigr) \, , \label{eq:scalarLag}
\end{align}
where $\Phi^{(1)}$ ($\Phi^{c(1)}$) is the lowest-lying composite chiral superfield corresponding to $\mathcal{O}$ ($\mathcal{O}^c$) and $m_{\Phi}^{(1)} = g_\Phi^{(1)} \Lambda_{\text{IR}}$ is the lowest-lying resonance mass, with $g_\Phi^{(1)}$ an order-one coupling. Note that we have neglected heavier resonances and higher-order terms in \eqref{eq:scalarLag}. The Lagrangian contains mixing terms between the elementary superfield $\Phi$ and the lowest-lying composite superfield $\Phi^{c(1)}$. The dimensionless constant $\varepsilon_\Phi$ is defined at the IR scale to be
\begin{equation}
  \label{eq:epsPhi}
  \varepsilon_\Phi 
  \equiv \tilde{\varepsilon}_\Phi (\Lambda_{\text{IR}}) \, \frac{\sqrt{N}}{4\pi}
    = \frac{1}{\sqrt{\smash[b]{Z_\phi}}} 
      \left(\frac{\Lambda_{\text{IR}}}{\Lambda_{\text{UV}}} \right)^{\delta - 1} 
      \frac{\sqrt{N}}{4\pi}
    \simeq \begin{cases}
             \; \dfrac{\sqrt{\delta-1}}{\sqrt{\smash[b]{\zeta_\phi}}} 
                \left( \dfrac{\Lambda_{\text{IR}}}{\Lambda_{\text{UV}}} \right)^{\delta - 1} 
           & \phantom{\; 0 \leq} \delta > 1 \, \\[1em]
             \; \dfrac{\sqrt{1 - \delta}}{\sqrt{\smash[b]{\zeta_\phi}}} 
           & 0 \leq \delta < 1 \, .
           \end{cases}
\end{equation}
where $\tilde{\varepsilon}_\Phi$ satisfies $\mu\frac{d \tilde{\varepsilon}_\Phi}{d \mu} = (\delta - 1) \tilde{\varepsilon}_\Phi + \zeta_\Phi \frac{N}{16\pi^2} \, \tilde{\varepsilon}_\Phi^3$, with $\zeta_\Phi$ an order-one constant. Note that the supersymmetric nonrenormalization theorem guarantees that $\varepsilon_\Phi$ only depends on the wavefunction renormalization, unlike the nonsupersymmetric case, where the vertex renormalization piece in \eqref{eq:omegaLR} was neglected.

Just as for the fermions in Sec.~\ref{sec:fermionhierarchy}, the mixing terms in the Lagrangian \eqref{eq:scalarLag} cause the scalar mass eigenstates to be admixtures of elementary and composite states. The details of this mixing are given in Appendix~\ref{app:holographic-mixing}. Using the result \eqref{eq:masslesseqn} and \eqref{eq:epsPhi}, the massless scalar eigenstate $\phi_0$ is given by
\begin{equation}
  \label{eq:scalaradmix}
  |\phi_0 \rangle 
    \simeq \mathcal{N}_\Phi 
           \left\{ 
             |\phi\rangle 
           - \frac{1}{g_\Phi^{(1)}\sqrt{\smash[b]{\zeta_\phi}}}
             \sqrt{\frac{\delta-1} {\left( \frac{\Lambda_{\text{IR}}}{\Lambda_{\text{UV}}} \right)^{2(1-\delta)} -1}} \,
             |\phi^{(1)}\rangle 
           \right\} \, ,
\end{equation}
where $\phi$ is an elementary scalar, $\phi^{(1)}$ is the lowest-lying composite scalar, and $\mathcal{N}_\Phi$ is a normalization constant. Similarly to the fermion case \eqref{eq:zerostate}, the massless scalar eigenstates are mostly elementary for $\delta > 1$, whereas for $0 \leq \delta < 1$ they are an admixture of elementary and composite states. In fact, supersymmetry guarantees that the admixtures for both the fermion and scalar field in the chiral multiplet are the same.

\subsubsection{Vector supermultiplet}

Next, we consider the supersymmetric generalization of partial compositeness for gauge fields. For simplicity, we consider the case of an Abelian U(1) gauge field, and assume our discussion can be generalized to non-Abelian gauge fields. In the nonsupersymmetric case, the source field $A_\mu$ mixes with the conserved U(1) current $J_\mu$ and induces a kinetic mixing of the form $F^{\mu \nu} F_{\mu \nu}^{(1)}$~\cite{Agashe:2002jx, Batell:2007jv}, where $F_{\mu \nu}^{(1)}$ is the field strength of the lowest-lying composite field, $A_\mu^{(1)}$. To supersymmetrize this coupling we introduce the vector superfield, $V = \theta^{\dagger} \bar{\sigma}^{\mu} \theta A_{\mu} + \theta^{\dagger} \theta^{\dagger} \theta \lambda + \theta \theta\theta^{\dagger} \lambda^{\dagger} + \frac{1}{2} \theta \theta \theta^{\dagger} \theta^{\dagger} D$, with field strength superfield $W_{\alpha} = \lambda_{\alpha} + \theta_{\alpha} D + \frac{i}{2} ( \sigma^{\mu} \bar{\sigma}^{\nu} \theta )_{\alpha} F_{\mu \nu} + i\theta \theta ( \sigma^{\mu} \partial_{\mu} \lambda^{\dagger} )_{\alpha} $. The conserved current $J_\mu$ can be embedded into a linear supermultiplet ${\mathcal J}$, which satisfies the condition $D^2 \mathcal{J} = D^{\dagger 2} \mathcal{J} = 0$ and guarantees current conservation~\cite{Weinberg:2000cr}.

The supersymmetric Lagrangian at the UV scale is given by
\begin{equation}
  \label{eq:UVvectorLag}
  \mathcal{L}_V 
    = \left( \frac{1}{4} [ W^{\alpha} W_{\alpha} ]_F + \text{H.c.} \right) 
    + 2 {\tilde{\varepsilon}}_V [ V \mathcal{J} ]_D \, ,
\end{equation}
where $\tilde{\varepsilon}_V$ is the mixing parameter. Since $[ V \mathcal{J} ]_D$ can be transformed into a kinetic term such as $[ W^{\alpha} \mathcal{W}_{\alpha} ]_F + \text{H.c.}$, where $\mathcal{W}$ is the field-strength superfield operator associated with a composite vector superfield operator $\mathcal{V}$, we omit such a term in the Lagrangian.

After confinement, the IR Lagrangian for the source $V$, together with the lowest-lying composite vector $V^{(1)}$, field-strength superfield $W_\alpha^{(1)}$, and chiral adjoint superfield $\Phi_V^{(1)}$, can be written as
\begin{align}
  \label{eq:vectorsuperL}
  \mathcal{L}_V 
     = \biggl( 
         \frac{1}{4} [ W^{\alpha} W_{\alpha} ]_F 
    &  + \frac{1}{4} [ W^{(1)\alpha} W_{\alpha}^{(1)} ]_F + \text{H.c.} 
       \biggr) \nonumber \\[1ex]
    &+ \Lambda_{\text{IR}}^2 
       \left[ 
         \biggl( 
           \varepsilon_V V + g_V^{(1)}V^{(1)} 
         + \frac{\Phi_V^{(1)} 
         + \Phi_V^{(1)\dagger}}{\sqrt{2}\Lambda_{\text{IR}}} 
         \biggr)^2 
       \right]_D \, ,
\end{align}
where $g_V^{(1)}$ is the composite vector coupling, and the dimensionless constant, $\varepsilon_V \equiv {\tilde{\varepsilon}}_V(\Lambda_{\text{IR}}) \, \frac{\sqrt{N}}{4\pi}$ parametrizes the mixing at the IR scale. The running parameter $\tilde{\varepsilon}_V(\mu)$ satisfies $\mu \frac{d\tilde{\varepsilon}_V}{d\mu} = \zeta_V \frac{N}{16\pi^2}{\tilde \varepsilon}_V^{3}$, with $\zeta_V$ an order-one constant that comes from the (unknown) strong dynamics. This immediately leads to the solution
\begin{equation}
 {\tilde{\varepsilon}}_V(\Lambda_{\text{IR}})
    = \frac{1}{\sqrt{\frac{1}{\tilde{\varepsilon}_V^2} + \frac{\zeta_V N}{8\pi^2} \log \bigl( \frac{\Lambda_{\text{UV}}}{\Lambda_{\text{IR}}} \bigr)}} 
    \simeq \frac{1}{\sqrt{2\zeta_V  \log \bigl( \frac{\Lambda_{\text{UV}}}{\Lambda_{\text{IR}}} \bigr)}} 
           \frac{4\pi}{\sqrt{N}} \, ,
\end{equation}
where we have taken the large-$N$ limit in the last term.

The Lagrangian \eqref{eq:vectorsuperL} can be diagonalized to obtain the mass eigenstates (the details are given in Appendix~\ref{app:holographic-mixing}). Using \eqref{eq:gaugeboson0mode}, the massless gauge boson eigenstate is found to be
\begin{equation}
  \label{eq:A0decomp}
  |A_{\mu 0} \rangle 
    \simeq 
      \mathcal{N}_V 
      \left\{ 
        |A_{\mu} \rangle 
      - \frac{1}{ g_V^{(1)} 
        \sqrt{2\zeta_V \log \bigl( \frac{\Lambda_{\text{UV}}}{\Lambda_{\text{IR}}} \bigr)} } \, 
        | A_{\mu}^{(1)} \rangle 
      \right\} \, ,
\end{equation}
where $\mathcal{N}_V$ is the normalization constant and $A_\mu^{(1)}$ is the lowest-lying composite gauge boson. Assuming a large hierarchy ($\Lambda_{\text{UV}}\gg \Lambda_{\text{IR}}$) and order-one values for the couplings, this expression shows that the massless eigenstate is mostly elementary, but with a sizeable composite admixture. The corresponding gauge coupling for this massless state is given in \eqref{eq:zerogc}.

The gaugino part of the Lagrangian \eqref{eq:vectorsuperL} is similarly given in Appendix~\ref{app:holographic-mixing}, where it is shown to be a special case $(\delta = 1)$ of the fermion Lagrangian
\eqref{eq:fermionLag2}. After diagonalizing the mass terms (see Appendix~\ref{app:holographic-mixing}), the massless gaugino eigenstate becomes
\begin{equation}
  \label{eq:gauginoadmix}
  |\lambda_0 \rangle 
    \simeq 
      \mathcal{N}_V 
      \left\{ 
        |\lambda \rangle 
      - \frac{1}{g_V^{(1)} 
        \sqrt{2\zeta_V \log \bigl( \frac{\Lambda_{\text{UV}}}{\Lambda_{\text{IR}}} \bigr)} } \, 
        | \lambda^{(1)} \rangle 
      \right\} \, ,
\end{equation}
where we have used \eqref{eq:masslessgaugino} and $\lambda^{(1)}$ represents the lowest-lying composite gaugino. By supersymmetry, the overall normalization constant, $\mathcal{N}_V$, and composite admixture are the same as in \eqref{eq:A0decomp}.

\subsubsection{Gravity supermultiplet}\label{sec:gravity}

Next, we consider the supersymmetric generalization of partial compositeness for gravity fields. In the nonsupersymmetric case the graviton $h_{\mu \nu}$ mixes with the conserved energy-momentum tensor $T_{\mu \nu}$, inducing a kinetic mixing between the graviton and massive spin-2 resonances~\cite{Batell:2007jv} (which is completely analogous to graviton-$f_2$ mixing in QCD). In the supersymmetric extension, the graviton $h_{\mu \nu}$ and gravitino $\psi_{\mu}$ can be embedded into the real supergravity field,
\begin{equation}
  H_{\mu} 
    = - \frac{1}{\sqrt{2}} \bigl( \theta^{\dagger} \bar{\sigma}^{\nu} \theta \bigr) h_{\mu \nu} 
      - i \theta \theta \theta^{\dagger} \lambda_{\mu}^{\dagger} 
      + i \theta^{\dagger} \theta^{\dagger} \theta \lambda_{\mu} + \dotsb \, ,
\end{equation}
where we have assumed the gauge $\frac{1}{2} \psi_{\mu} \equiv \lambda_{\mu} + \frac{1}{3} \sigma_{\mu} \bar{\sigma}^{\rho} \lambda_{\rho} $ and only written the graviton and gravitino parts. The conserved energy-momentum tensor $T_{\mu \nu}$ can be embedded into the supercurrent $\Theta_{\mu}$, which satisfies the condition $- i \bar{\sigma}^{\mu} D \Theta_{\mu} = D^{\dagger} X$, where $X$ is an antichiral superfield, and guarantees current conservation: $\partial_{\mu} T^{\mu \nu} = 0$~\cite{Weinberg:2000cr}. The supersymmetric Lagrangian at the UV scale is given by
\begin{equation}
  \label{eq:UVgravityLag}
  \mathcal{L}_H 
    = \frac{4}{3} [ H_{\mu} E^{\mu} ]_D 
    + \frac{2\tilde{\varepsilon}_H}{\Lambda_{\text{UV}}}  [ H_{\mu} \Theta^{\mu} ]_D \, ,
\end{equation}
where $\tilde{\varepsilon}_H$ is the mixing parameter and $E^{\mu}$ is the Einstein superfield~\cite{Ferrara:1977mv}. In an analogous fashion to the gauge boson case, the IR Lagrangian after confinement can be written as
\begin{equation}
  \label{eq:IRgravityLag}
  \mathcal{L}_H 
    = \frac{4}{3} [ H_{\mu} E^{\mu} ]_D 
    + \frac{4}{3} [ H_{\mu}^{(1)} E^{(1) \mu} ]_D 
    + 2 \Lambda_{\text{IR}}^2 
      \left[ \left( \varepsilon_H H^{\mu} + g_H^{(1)} H_{\mu}^{(1)} \right)^2 \right]_D \, ,
\end{equation}
where $\varepsilon_H = \tilde{\varepsilon}_H(\Lambda_{\text{IR}}) \, \frac{\sqrt{N}}{4\pi}$ and $H_{\mu}^{(1)}$ is a composite real superfield with corresponding Einstein superfield $E^{(1)\mu}$. The running parameter $\tilde{\varepsilon}_H (\mu) \equiv \frac{\tilde{\varepsilon}_H}{\sqrt{Z_H}}\frac{\mu}{\Lambda_{\text{UV}}}$ satisfies $\mu \frac{d\tilde{\varepsilon}_H}{d\mu} = \tilde{\varepsilon}_H + \zeta_H \frac{N}{16\pi^2} {\tilde \varepsilon}_H^3$, with $\zeta_H$ an order-one constant that arises from the (unknown) strong dynamics. The solution is then given by 
\begin{equation}
  \tilde{\varepsilon}_H (\Lambda_{\text{IR}}) 
    = \frac{\Lambda_{\text{IR}}}{\Lambda_{\text{UV}}} 
      \frac{1}{\sqrt{\frac{1}{\tilde\varepsilon_H^2} + \frac{\zeta_H N}{16\pi^2} \left( 1 - \left( \frac{\Lambda_{\text{IR}}}{\Lambda_{\text{UV}}} \right)^2 \right) }} 
    \simeq \frac{1}{\sqrt{\smash[b]{\zeta_H}}} 
           \frac{\Lambda_{\text{IR}}}{\Lambda_{\text{UV}}} 
           \frac{4\pi}{\sqrt{N}} \, ,
\end{equation}
where, in the last term, we have taken the large-$N$ limit. Since the graviton couples to the energy-momentum tensor with strength $1/M_P$ (where $M_P=2.4 \times 10^{18}$ GeV is the reduced Planck mass), we obtain the matching condition $\frac{1}{M_P} \equiv \frac{\varepsilon_H}{\Lambda_{\text{IR}}}$ at the IR scale, which leads to the relation
\begin{equation}
  \label{eq:planckmass-4d}
  M_P \simeq \sqrt{\smash[b]{\zeta_H}} \, 
             \Lambda_{\text{UV}} 
             \sqrt{1 - \Bigl( \frac{\Lambda_{\text{IR}}}{\Lambda_{\text{UV}}} \Bigr)^2}
      \simeq \sqrt{\smash[b]{\zeta_H}} \, \Lambda_{\text{UV}}\,.
\end{equation}
The Lagrangian \eqref{eq:IRgravityLag} contains mass mixing terms, and the diagonalization details are given in Appendix~\ref{app:holographic-mixing}. Using \eqref{eq:graviton0mode}, the massless graviton eigenstate is found to be
\begin{equation}
  \label{eq:h0decomp}
  |h_{\mu \nu 0} \rangle 
    \simeq 
      \mathcal{N}_H 
      \left\{ 
        |h_{\mu \nu} \rangle 
      - \frac{1}{g_H^{(1)} 
        \sqrt{\smash[b]{\zeta_H}}} \frac{\Lambda_{\text{IR}}}{\Lambda_{\text{UV}}} \, 
        | h_{\mu \nu}^{(1)} \rangle 
      \right\} \, ,
\end{equation}
Assuming $\Lambda_{\text{IR}} \ll \Lambda_{\text{UV}}$, this shows that the massless eigenstate is mostly elementary with a tiny composite admixture.

The gravitino part of the Lagrangian \eqref{eq:IRgravityLag} is given in Appendix~\ref{app:holographic-mixing}, where it is shown to be a special case $(\delta = 2)$ of the fermion Lagrangian \eqref{eq:fermionLag2}. After diagonalizing the mass terms (see Appendix~\ref{app:holographic-mixing}), the massless gravitino eigenstate becomes
\begin{equation}
  \label{eq:gravitinoadmix}
  |\psi_{\mu 0} \rangle 
    \simeq 
      \mathcal{N}_H 
      \left\{ 
        |\psi_{\mu} \rangle 
      - \frac{1}{g_H^{(1)} 
        \sqrt{\smash[b]{\zeta_H}}} \frac{\Lambda_{\text{IR}}}{\Lambda_{\text{UV}}} \, 
        | \psi_{\mu}^{(1)} \rangle 
      \right\} \, ,
\end{equation}
where we have used \eqref{eq:masslessgravitino} and $\psi_{\mu}^{(1)}$ represents the lowest-lying composite gravitino. By supersymmetry, the composite admixture is the same as in \eqref{eq:h0decomp}, and thus the gravitino is again mostly elementary.

\subsection{Supersymmetry breaking}\label{sec:susybreaking-4d}

Supersymmetry is assumed to be broken in the composite sector by the strong dynamics and will be parameterized by a composite spurion, chiral superfield, ${\mathcal{X}} = \theta\theta \mathcal{F}$, where the vacuum expectation value $ \langle \mathcal{F} \rangle \equiv F_{\mathcal{X}} \neq 0$. In the large-$N$ limit, we have $F_{\mathcal{X}} \propto \frac{\sqrt{N}}{4\pi} \Lambda^2_{\text{IR}}$~\cite{Gherghetta:2011na}.

\subsubsection{Sfermion masses}

Supersymmetry-breaking scalar masses are only generated for the composite sector fields, since there is no direct coupling of the supersymmetry-breaking spurion to the elementary fields. For instance, the massive chiral superfield $\Phi^{(1)}$ directly couples to the supersymmetry breaking in the composite sector via the following $D$-term:
\begin{equation}
  \xi_4 \frac{g_{\Phi}^{(1)2}}{\Lambda_{\text{IR}}^2}  
  [ \mathcal{X}^\dagger \mathcal{X} \Phi^{(1)\dagger} \Phi^{(1)} ]_D 
    = \xi_4 g_{\Phi}^{(1)2} \frac{|F_{\mathcal{X}}|^2}{\Lambda_{\text{IR}}^2} \phi^{(1)\dagger} \phi^{(1)} \, ,
\end{equation}
where $\xi_4$ is a dimensionless parameter, which for a large-$N$ gauge theory is proportional to $\frac{16\pi^2}{N}$~\cite{Witten:1979kh}. This $D$-term interaction gives a supersymmetry-breaking mass to the composite scalar field $\phi^{(1)}$. Given the scalar admixture \eqref{eq:scalaradmix}, the corresponding sfermion mass squared becomes:
\begin{equation}
  \label{eq:sfermionmass}
  \widetilde{m}^2 
    \simeq
      \mathcal{N}_\Phi^2 \varepsilon_\Phi^2
      \xi_4 \frac{|F_{\mathcal{X}}|^2}{\Lambda_{\text{IR}}^2} 
    \simeq
      \begin{cases}
        \; \dfrac{(\delta - 1)}{\zeta_\Phi}
           \left( \dfrac{\Lambda_{\text{IR}}}{\Lambda_{\text{UV}}} \right)^{2(\delta - 1)} 
           \xi_4 
           \dfrac{|F_{\mathcal{X}}|^2}{\Lambda_{\text{IR}}^2} 
      & \phantom{\; 0 \leq} \delta \geq 1 \, , \\[1em]
        \; \dfrac{(1 - \delta)}{\zeta_\Phi} 
           \xi_4 
           \dfrac{|F_{\mathcal{X}}|^2}{\Lambda_{\text{IR}}^2} 
      & 0 \leq \delta < 1 \, ,
      \end{cases}
\end{equation}
where we have assumed $\varepsilon_\Phi \lesssim g_\Phi^{(1)}$. Note that the sfermion mass is power-law suppressed for $\delta > 1$. This is because the mass eigenstate is mostly elementary and therefore can only obtain a supersymmetry-breaking mass via mixing with the composite sector. This contrasts with the case $0 \leq \delta < 1$, where the mass eigenstate has a sizeable composite admixture and therefore directly feels the supersymmetry breaking from the composite sector without any power-law suppression. In this way, a sfermion mass hierarchy can be explained by anomalous dimensions of supersymmetric operators.

Note, however, that as $\delta$ is increased beyond one, the scalar mass squared becomes increasingly
small, since the scalar is becoming more elementary [using \eqref{eq:scalaradmix}]. Eventually, radiative corrections are sufficiently large that they provide the dominant contribution to the sfermion mass squared. For instance, the sfermion-fermion-gaugino interaction with a massive gaugino leads to the one-loop contribution
\begin{equation}
  \label{eq:gauginocont}
  \delta\widetilde{m}^2 \simeq \frac{g_i^2}{16\pi^2} M_{\lambda_i}^2 \, ,
\end{equation}
where $g_i$ $(i = 1,2,3)$ are the Standard Model $\text{U(1)}_Y \times \text{SU(2)}_L \times \text{SU(3)}$ gauge couplings, respectively.
Note that this radiative correction is assumed to be finite. Therefore, there is a maximum value of the anomalous dimension $\delta^\ast$, beyond which the gaugino-mediated contribution \eqref{eq:gauginocont} dominates.

\begin{figure}[t]
  \centering
  \includegraphics[width=0.75\textwidth]{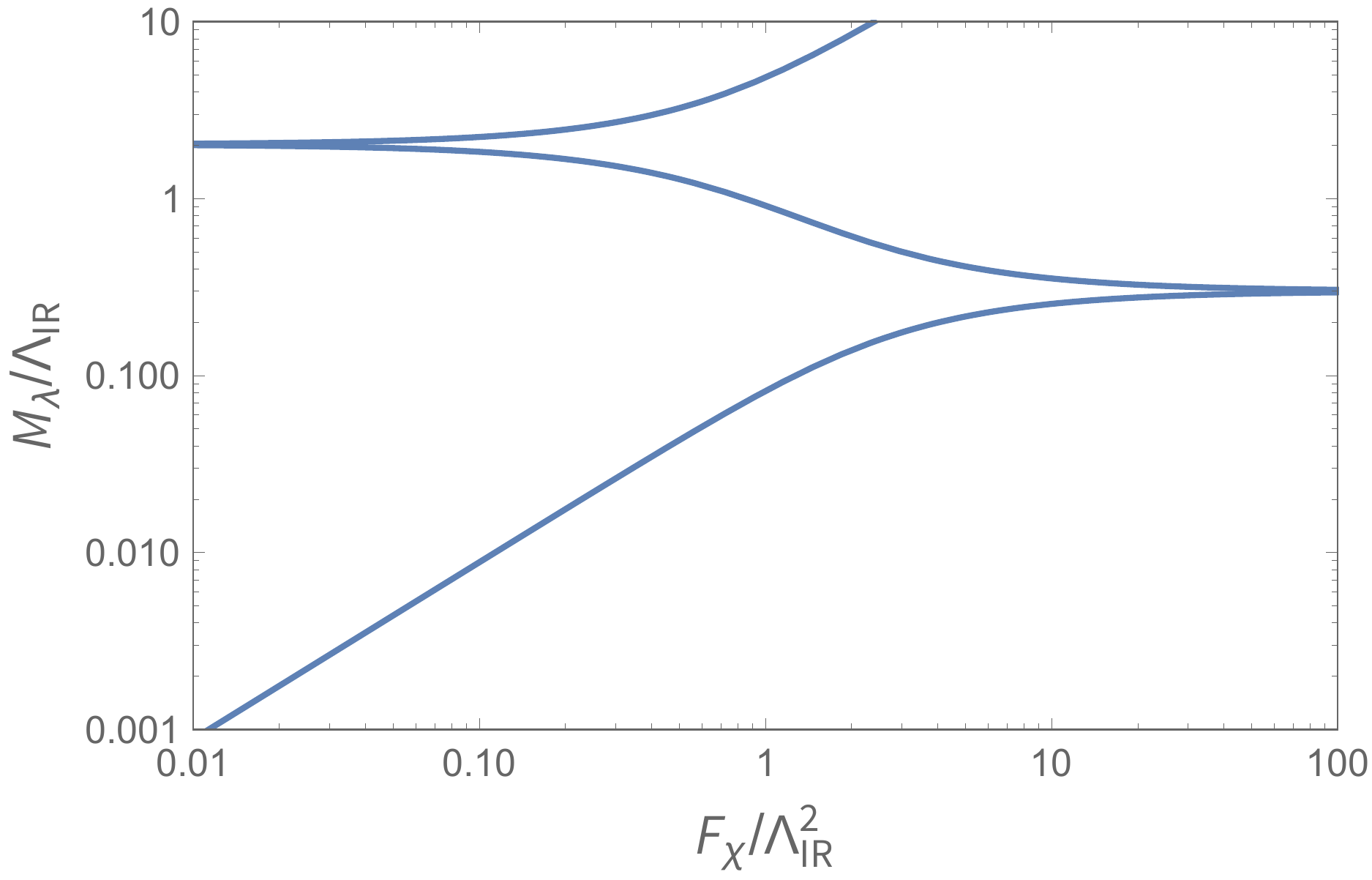}
  \caption{Gaugino mass eigenvalues for $g_V^{(1)} = 2$, $\varepsilon_V = 0.3$, and $\xi_3 = 1$ as a function of the supersymmetry-breaking order parameter $F_{\mathcal{X}}$.}
  \label{fig:gaugino10^-2}
\end{figure}

\subsubsection{Gaugino masses}

The spurion superfield $\mathcal{X}$ can also be used to generate gaugino masses.
Since the massless gaugino \eqref{eq:gauginoadmix} contains a $\lambda^{(1)}$ admixture, supersymmetry breaking in the composite sector is transmitted by the following interactions:
\begin{equation}
  \label{eq:gauginosusybreaking}
  \frac{ \xi_3}{2} \frac{g_V^{(1)2}}{\Lambda_{\text{IR}}}
  \left( [ \mathcal{X}W^{\alpha (1)} W_\alpha^{(1)} ]_F + \text{H.c.} \right) \, ,
\end{equation}
where $\xi_3$ is proportional to $\frac{4\pi}{\sqrt{N}}$ for a large-$N$ gauge theory. The supersymmetry-breaking interaction \eqref{eq:gauginosusybreaking}, adds the Majorana mass term $\xi_3 g_V^{(1)2} \frac{F_{\mathcal{X}}}{2\Lambda_{\text{IR}}} \lambda^{(1)} \lambda^{(1)}+ \text{H.c.}$ to the gaugino mass terms in \eqref{eq:gauginoLag}. Note that there is no $\chi^{(1)}$ mass term because the interaction $[ \mathcal{X}V^{(1)2} ]_D$ is not gauge invariant. Diagonalization of the mass matrix then leads to a gaugino mass
\begin{equation}
  \label{eq:gauginomass}
  M_{\lambda} 
    = \mathcal{N}_V^2 {\varepsilon}_V^2 \xi_3 \frac{F_{\mathcal{X}}}{\Lambda_{\text{IR}}}
    \simeq g^2\xi_3 \frac{F_{\mathcal{X}}}{\Lambda_{\text{IR}}} \, ,
\end{equation}
where we have used the gauge coupling relation \eqref{eq:zerogc} and ${\varepsilon}_V \simeq g_s( \Lambda_{\text{IR}})$. The result \eqref{eq:gauginomass} is consistent with simply using \eqref{eq:gauginoadmix} and the fact that the massless mode only has a $\lambda^{(1)}$ admixture. Notice that the gaugino mass is suppressed by a $\log(\Lambda_{\text{UV}}/\Lambda_{\text{IR}})$ factor due to the fact that the elementary gaugino mixes with the composite sector via a marginal coupling. This causes the gauginos to be generically lighter than the heavy composite sfermions with $0 \leq \delta < 1$, leading to a ``split'' spectrum.

However when $\delta$ is greater than the critical value $\delta^*$, the gaugino mass radiative correction \eqref{eq:gauginocont} gives the dominant contribution to the scalar masses. The critical value $\delta^*$ occurs when $\delta\widetilde{m}^2 \simeq \widetilde{m}^2$, and using \eqref{eq:gauginomass} it takes the value
\begin{equation}
  \delta^* 
    \simeq 
      1 
    + \frac{\log \left[ \frac{4\pi}{g_i} \log \bigl( \frac{\Lambda_{\text{UV}}}{\Lambda_{\text{IR}}} \bigl) \right]}
           {\log \bigl( \frac{\Lambda_{\text{UV}}}{\Lambda_{\text{IR}}} \bigl)} \, ,
\end{equation}
where we have included only the dominant gauge-coupling contribution. Thus, for $\delta > \delta^*$, the gaugino is heavier than the corresponding sfermion mass by a factor of $\sqrt{\smash[b]{\alpha_i / (4 \pi)}}$.

Note that when $\sqrt{\smash[b]{F_{\mathcal{X}}}} \gg \Lambda_{\text{IR}}$, the lightest gaugino becomes a Dirac fermion with a mass 
\begin{equation}
  M_{\lambda} 
    \simeq \varepsilon_V \Lambda_{\text{IR}} 
    \simeq \frac{\Lambda_{\text{IR}}}{\sqrt{ 2 \zeta_V \log \bigl( \frac{\Lambda_{\text{UV}}}{\Lambda_{\text{IR}}} \bigr) }}
    \, .
\end{equation}
This value parametrically matches the gaugino mass that is found in a five-dimensional (5D) supersymmetric standard model where supersymmetry breaking arises from ``twisted" boundary conditions~\cite{Gherghetta:2000kr}. The Dirac limit can also be seen in Fig.~\ref{fig:gaugino10^-2}, which shows the elementary state $\lambda$ marrying the composite Weyl fermion $\chi^{(1)}$ as $\sqrt{\smash[b]{F_{\mathcal{X}}}}$ becomes larger, giving rise to a Dirac mass ${\sim} g_s (\Lambda_{\text{IR}}) \,  \Lambda_{\text{IR}}$ that is smaller than the mass \eqref{eq:Diracgauginomass} that applies when $\sqrt{\smash[b]{F_{\mathcal{X}}}}\ll \Lambda_{\text{IR}}$.

\subsubsection{Gravitino masses}

The supersymmetry breaking from the composite sector gives rise to a positive vacuum energy.
This contribution can be cancelled by introducing a constant superpotential $W$, which induces a mass term for the elementary gravitino
\begin{equation}
  - \frac{1}{4} \frac{W}{M_P^2} \, \psi_{\rho} \left[ \sigma^{\mu}, \bar{\sigma}^{\rho} \right] \psi_{\mu} + \text{H.c.}
\end{equation}
At the IR scale, this term becomes
\begin{equation}
  \label{eq:elemgravitinomass}
  - \frac{1}{4Z_H} 
    \frac{W}{M_P^2} \,
    \psi_{\rho} 
    \left[ \sigma^{\mu}, \bar{\sigma}^{\rho} \right] 
    \psi_{\mu} + \text{H.c.} 
    \simeq - \frac{1}{4\zeta_H \tilde{\varepsilon}_H^2} 
             \frac{4\pi}{\sqrt{N}} 
             \frac{F_{\mathcal{X}}}{\sqrt{3}M_P} \,
             \psi_{\rho} 
             \left[ \sigma^{\mu}, \bar{\sigma}^{\rho} \right] 
             \psi_{\mu} + \text{H.c.} \,,
\end{equation}
where $Z_H$ is defined from $\tilde{\varepsilon}_H(\Lambda_{\text{IR}})$ in Sec.~\ref{sec:gravity} and in the second equation, we have tuned $|F_{\mathcal{X}}|^2 \frac{N}{16\pi^2} \simeq 3 \frac{|W|^2}{M_P^2}$ so that the cosmological constant vanishes. Note that this means the elementary sector is supersymmetric AdS$_4$, with Minkowski space obtained after the ``uplift" from the supersymmetry-breaking composite sector. Diagonalization of the mass matrix [given in \eqref{eq:diaggravitino}] with the inclusion of \eqref{eq:elemgravitinomass} then leads to a gravitino mass
\begin{equation}
  \label{eq:dualgravitinomass}
  m_{3/2} \simeq \xi_{3} \frac{F_{\mathcal{X}}}{\sqrt{3} M_P} \, ,
\end{equation}
where $F_{\mathcal{X}} \ll \Lambda_{\text{IR}} M_P$, $\mathcal{N}_H \simeq 1$, and $\xi_{3} = \frac{1}{\zeta_H \tilde{\varepsilon}_H^2} \frac{4\pi}{\sqrt{N}}$. The result \eqref{eq:dualgravitinomass} is consistent with simply using \eqref{eq:gravitinoadmix} and the fact that the massless mode is mostly elementary. Since $\Lambda_{\text{IR}} \ll M_P$, the gravitino is generically much lighter than the heavy composite sfermions with $0 \leq \delta <1$. There is also a heavy Dirac
gravitino mostly comprised of $\psi_\mu^{(1)}$ and $\chi_\mu^{(1)}$ with mass ${\sim} ( g_H^{(1)2} + \varepsilon_H^2 )^{1/2} \Lambda_{\text{IR}}$.

In the opposite limit $F_{\mathcal{X}} \gg \Lambda_{\text{IR}} M_P$, one gravitino Weyl state decouples, and the elementary state $\psi_\mu$ marries the composite state $\psi_\mu^{(1)}$~\cite{Gherghetta:2000kr} such that the lightest gravitino then becomes a Dirac fermion with a mass $m_{3/2} \simeq g_H^{(1)} \Lambda_{\text{IR}}$. This is similar to what occurs for the gaugino, except that since the mixing is very small, the Dirac limit is reached much more slowly for the gravitino.

\subsubsection{Higgs soft mass}\label{sec:Higgs soft mass}

There is no direct coupling of the supersymmetry-breaking spurion to the elementary fields, and therefore the Higgs soft mass can only be generated radiatively, via loops of fields with a composite admixture. In particular, gaugino and sfermion loops can transmit the supersymmetry-breaking effects to the elementary sector.

The first type of sfermion contributions are due to Yukawa interactions. These lead to the Higgs soft
mass squared
\begin{align}
  (\Delta m_H^2)_{y} 
    &\simeq 
       \frac{y_{\delta}^2(\Lambda_{\text{IR}})}{16\pi^2} \,
       \widetilde{m}^2 
       \log \Bigl( \frac{\Lambda_{\text{IR}}}{\text{TeV}} \Bigr) \nonumber \\[1ex]
    &\simeq 
       \frac{\lambda^2}{16\pi^2} 
       \frac{\mathcal{N}_{\Phi}^2}{Z_{\Phi}^2} \,
       \widetilde{m}^2 
       \log \Bigl( \frac{\Lambda_{\text{IR}}}{\text{TeV}} \Bigr) \nonumber \\[1ex]
    &\simeq 
       \frac{16\pi^2 \lambda^2}{N^2} 
       \frac{\mathcal{N}_{\Phi}^4}{\zeta_{\Phi}^3}
       \left( \frac{\delta - 1}{1 - \left( \frac{\Lambda_{\text{IR}}}{\Lambda_{\text{UV}}} \right)^{2(\delta - 1)} } \right)^3 
       \left( \frac{\Lambda_{\text{IR}}}{\Lambda_{\text{UV}}} \right)^{2(\delta -1)} 
       \xi_4 
       \frac{|F_{\mathcal{X}}|^2}{\Lambda_{\text{IR}}^2} 
       \log \Bigl( \frac{\Lambda_{\text{IR}}}{\text{TeV}} \Bigr) \, ,
       \label{eq:mH2Yukawa}
\end{align}
where $y_{\delta}(\Lambda_{\text{IR}}) \equiv m_\psi/  \langle H \rangle$ is the Yukawa coupling from Sec.~\ref{sec:fermionhierarchy} with $\delta = \delta_L = \delta_R$ chosen for simplicity, $\lambda$ is the proto-Yukawa coupling,  $\mathcal{N}_{\Phi}$ is a normalization constant given in \eqref{eq:masslesseqn}, and $Z_\Phi$ is defined in \eqref{eq:epsPhi}.

The second type of sfermion corrections are due to $D$-term gauge interactions and the mixing term $V - \Phi_V^{(1)}$. Since the scalar and the auxiliary field component of $\Phi_V^{(1)}$ mediates the supersymmetry breaking, the easiest way to estimate the $D$-term contribution is to approximately solve for $\Phi_V^{(1)}$ in order to determine how the elementary $D$-term is coupled to the composite sfermions. This procedure generates the following correction:
\begin{align}
  (\Delta m_H^2)_{D} 
    &\simeq 
       Y(H) \, Y(\phi) \,
       \frac{g_1(\Lambda_{\text{IR}}) }{8\pi^2} 
       \varepsilon_V 
       \left( 
         \log \Bigl( \frac{\Lambda_{\text{IR}}}{\text{TeV}} \Bigr) 
       + \frac{\mathcal{N}_{\Phi}^2}{2\varepsilon_{\Phi}^2 \left( 1 + \frac{m_{\Phi_1}^2}{\Lambda_{\text{IR}}^2} \right)} \right) 
       \widetilde{m}^2 \, \nonumber \\[1ex]
    &\simeq 
       Y(H) \, Y(\phi) \,
       \frac{g_1( \Lambda_{\text{IR}}) }{8\pi^2} 
       \varepsilon_V 
       \mathcal{N}_{\Phi}^2 \nonumber \\[1ex]
    &  \qquad \times 
       \left( 
         \frac{ \log \bigl( \frac{\Lambda_{\text{IR}}}{\text{TeV}} \bigr)}{\zeta_{\Phi}} 
         \frac{\delta-1}{ \bigl( \frac{\Lambda_{\text{IR}}}{\Lambda_{\text{UV}}} \bigr)^{2(1 - \delta)} - 1 } 
       + \frac{\mathcal{N}_{\Phi}^2}{2 \left( 1 + \frac{m_{\Phi_1}^2}{\Lambda_{\text{IR}}^2} \right)} 
       \right) 
       \xi_4 
       \frac{|F_{\mathcal{X}}|^2}{\Lambda_{\text{IR}}^2} \, ,
       \label{eq:mH2Dterm}
\end{align}
where $g_{1}$ is the U(1)$_Y$ gauge coupling, $Y$ denotes the hypercharge, and $m_{\Phi_1}^2 = ( \varepsilon_{\Phi}^2 + g_{\Phi}^{(1)2} ) \Lambda_{\text{IR}}^2$.

\begin{figure}[t]
  \centering
  \includegraphics[width=0.75 \textwidth]{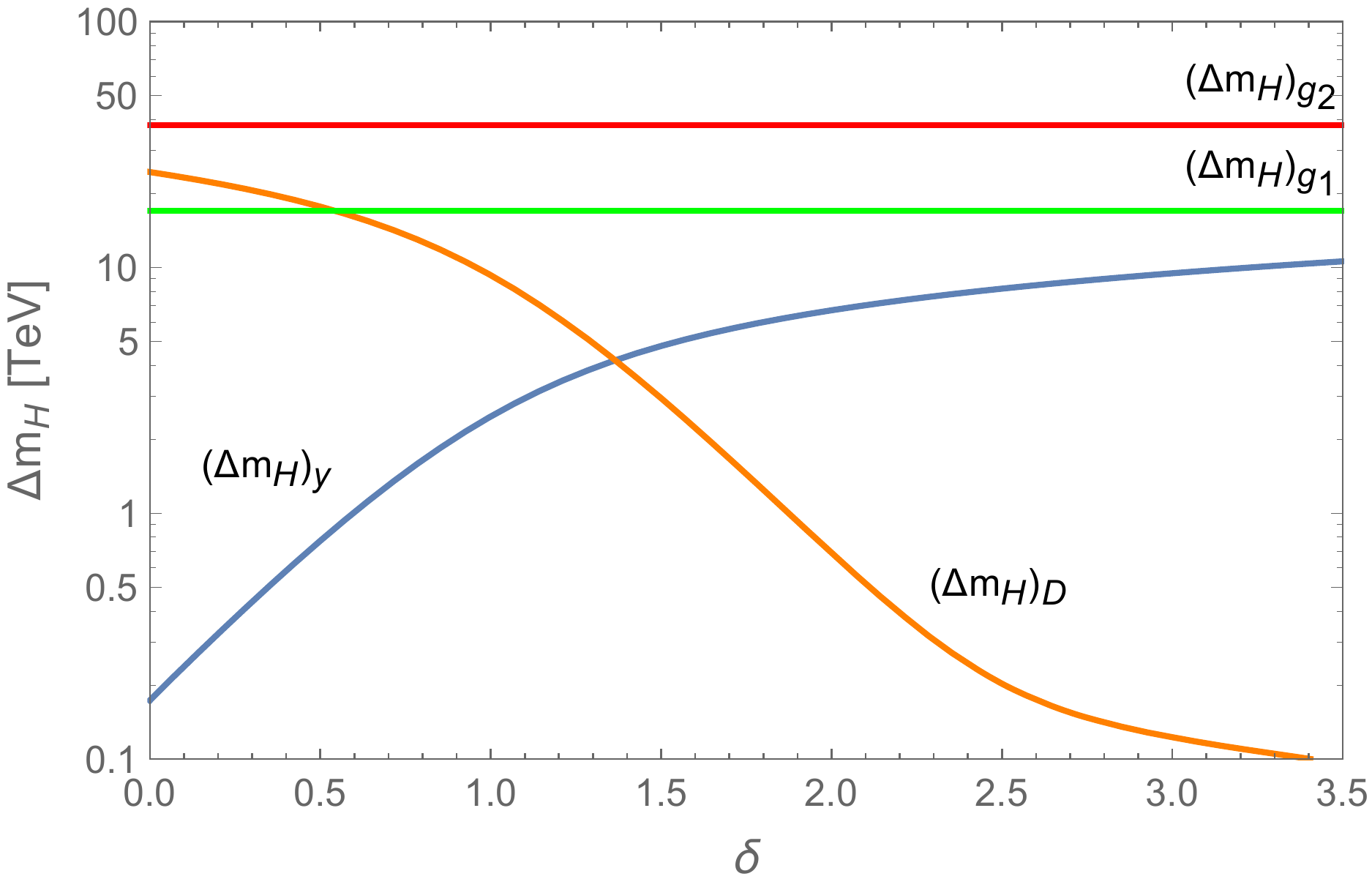}
  \caption{One-loop corrections to the Higgs mass parameter that arise from gauge, Yukawa, and $D$-term interactions due to a single sfermion or gaugino as a function of the anomalous dimension $\delta$.}
  \label{fig:correctionstoHiggsmass}
\end{figure}

Finally, the corrections that involve gauginos arise from gauge interactions. The dominant contribution is given by
\begin{equation}
  \label{eq:mH2gaugino}
  (\Delta m_H^2)_{g} 
    \simeq 
      4 C(R_H) \, 
      \mathcal{N}_V^2 \,
      \frac{g_2^2(\Lambda_{\text{IR}})}{8\pi^2} 
      M_{\lambda}^2 
      \log \Bigl( \frac{\Lambda_{\text{IR}}}{\text{TeV}} \Bigr) \, ,
\end{equation}
where $g_{2}$ is the SU(2) gauge coupling and $C(R)$ is the quadratic Casimir of the representation $R$. In Fig.~\ref{fig:correctionstoHiggsmass}, we plot the approximate expressions \eqref{eq:mH2Yukawa}, \eqref{eq:mH2Dterm}, and \eqref{eq:mH2gaugino} for a single sfermion (with $Y(\phi) = 1$) and gaugino, assuming 
$\sqrt{\smash[b]{\xi_4 / \zeta_{\Phi}}} \, F_{\mathcal{X}} = \xi_3 F_{\mathcal{X}}= ( 4.75 \times 10^{10}~\text{GeV} )^2$, 
$\Lambda_{\text{IR}} = 2 \times 10^{16}~\text{GeV}$, 
$\Lambda_{\text{IR}} / \Lambda_{\text{UV}} \simeq 0.026$, 
$g_{\Phi_L}^{(1)} g_{\Phi_R}^{(1)} \frac{16\pi^2}{N} \lambda \simeq 1$, 
$g_V^{(1)2} \simeq g_5^2 k$, and 
$\zeta_{\Phi_i,V} \simeq 1 / g_{\Phi_i,V}^{(1)2}$, where $i = L,R$. 
Furthermore, for later comparison with the 5D calculation, we use values for $m_{\Phi_1}^2$ and $m_{V_1}^2$ corresponding to the 5D Kaluza-Klein masses. Only the maximal contribution arising from the Yukawa interaction (corresponding to $\delta_L = \delta$ and $\delta_R =1$) is shown in the figure. All contributions are qualitatively similar to the precise 5D results that are given in Sec.~\ref{sec:susybreaking-5d} and calculated in Appendix~\ref{app:radiativecorrections}.

\subsection{Higgsino mass}\label{sec:Higgsinomass}

Since the Higgs supermultiplet is elementary, there is no direct coupling to the supersymmetry-breaking spurion in the composite sector, and therefore the generation of a $\mu$-term via a Giudice-Masiero mechanism~\cite{Giudice:1988yz} is forbidden. Instead, to generate a sizeable $\mu$-term, we consider the Kim-Nilles mechanism~\cite{Kim:1983dt} and introduce an elementary Standard Model singlet $S$, together with a global U(1) Peccei-Quinn symmetry, where $H_u, H_d,S$ have the charges $+1,+1,-1$, respectively.\footnote{To allow for Yukawa interactions with the Higgs fields, the fermions $Q,L,{\bar u},{\bar d},{\bar e}$ must have charges $-1,-1,0,0,0$, respectively.} This global symmetry then allows the following nonrenormalizable superpotential term:
\begin{equation}
  W_{KN} = \frac{\kappa_{\mu}}{2 M_P} S^2 H_u H_d \, ,
\end{equation}
where $\kappa_{\mu}$ is a dimensionless coupling. Assuming that the Peccei-Quinn symmetry is spontaneously broken by a nonzero vacuum expectation value $\langle S \rangle \sim f$, an effective $\mu$-term of size
\begin{equation}
  \label{eq:muterm}
  \mu \simeq \frac{\kappa_\mu f^2}{2M_P}
\end{equation}
is then generated. Since the global symmetry is anomalous (and assuming all other sources of breaking are small), the pseudo Nambu-Goldstone boson associated with the spontaneous symmetry breaking can be identified with the axion. This axion is of the invisible DFSZ type, and is consistent with the present astrophysical constraints provided that $10^9~\text{GeV} \lesssim f \lesssim 10^{12}~\text{GeV}$. For this range of $f$, the $\mu$-term \eqref{eq:muterm} can easily accommodate values in the range $0.1~\text{TeV} \lesssim \mu \lesssim 100$~TeV. Thus, using the Kim-Nilles mechanism, we can solve the strong-$CP$ problem and generate the required values of the $\mu$ term.

\subsection{The sfermion mass hierarchy}\label{sec:numexample}

\begin{figure}[t]
  \centering
  \includegraphics[clip, width=0.55\textwidth]{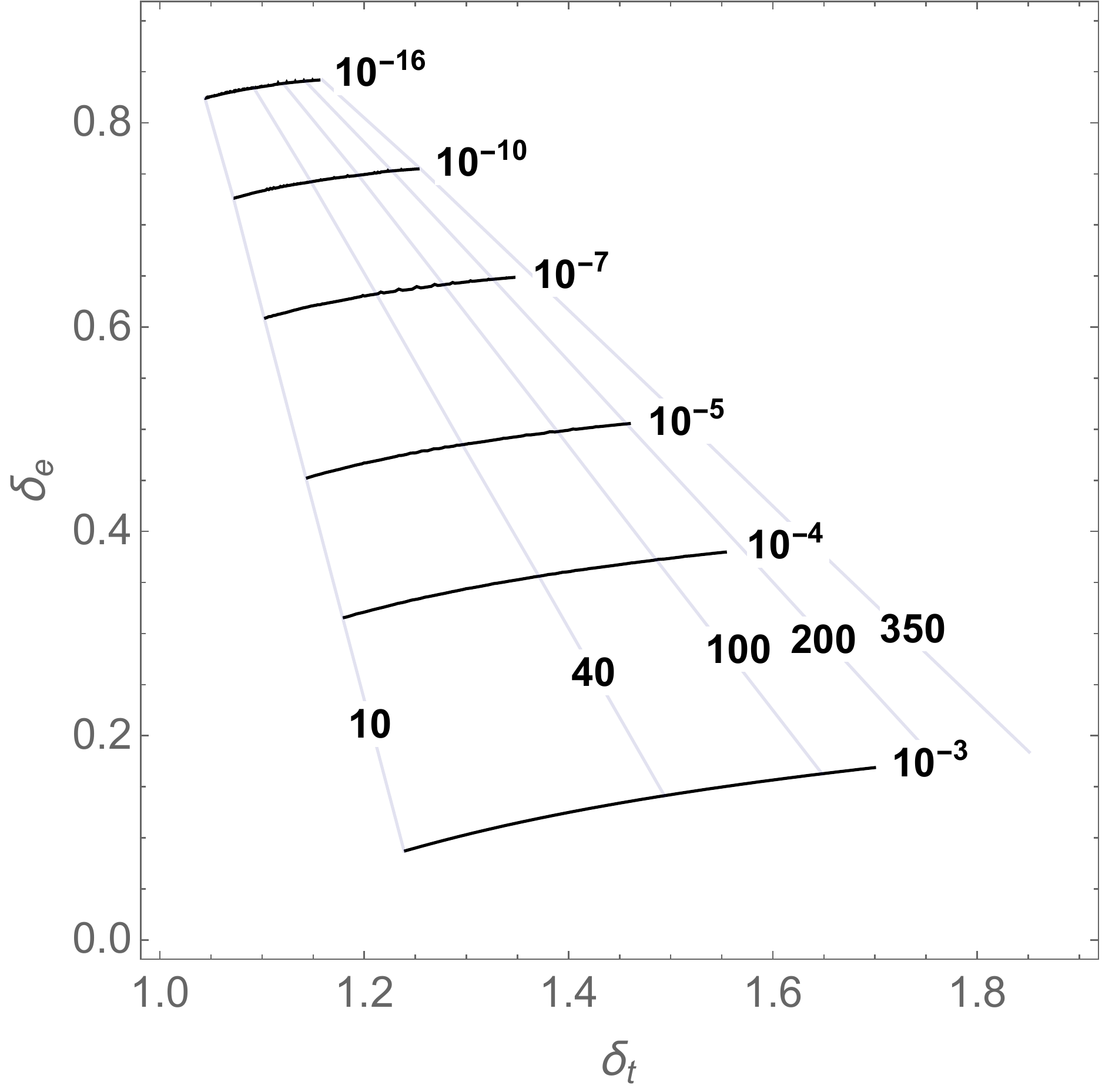}
  \caption{Values of the anomalous dimensions $\delta_e$ and $\delta_t$ that give rise to the electron to top-quark Yukawa coupling ratio. The approximately horizontal lines are contours of the ratio $\Lambda_{\text{IR}}/\Lambda_{\text{UV}}$, while the approximately vertical lines are contours of the tree-level sfermion mass ratio $m_{\tilde{e}}/m_{\tilde{t}}$. The running of the Yukawa couplings has been included, assuming $\Lambda_{\text{UV}} = 10^{18}$~GeV, $\tan \beta = 3$, and a supersymmetric mass threshold at 50 TeV.}
  \label{fig:fermionratio}
\end{figure}

In a supersymmetric model, partial compositeness relates the fermion and sfermion mass hierarchies.  To explicitly see this relation we consider a numerical example involving the electron and the top quark. In order to explain the fermion mass hierarchy, the top quark must be mostly elementary with $\delta_t > 1$ (i.e., irrelevant mixing), while the electron must have a sizeable composite admixture with $ 0< \delta_e < 1$ (i.e., relevant mixing). Using the fermion mass expressions \eqref{eq:fermionmass}, the ratio of the Yukawa couplings at the IR scale is then given by
\begin{equation}
  \label{eq:fermionratio}
  \frac{y_e}{y_t} 
    \simeq 
      \frac{1 - \delta_e}{\delta_t - 1} 
      \left( \frac{\Lambda_{\text{IR}}}{\Lambda_{\text{UV}}} \right)^{ 2(1-\delta_e)} \, .
\end{equation}
The ratio of the electron to the top-quark Yukawa coupling is determined in terms of $\delta_{e,t}$, and for a sufficiently large hierarchy between $\Lambda_{\text{IR}}$ and $\Lambda_{\text{UV}}$, depends sensitively on the anomalous dimension $\delta_e$. It is plotted in Fig.~\ref{fig:fermionratio} for various values of $\Lambda_{\text{IR}} / \Lambda_{\text{UV}}$, assuming that the Yukawa coupling ratio is approximately $y_e / y_t \simeq 10^{-5}$ over most IR scales. Note that \eqref{eq:fermionratio} is a simplified expression for $\delta$ values not very close to one, but exact expressions are used in the figure.

Using the sfermion mass expressions \eqref{eq:sfermionmass} and \eqref{eq:gauginocont}, we similarly obtain the ratio of the selectron mass squared to the stop mass squared at the IR scale:
\begin{equation}
  \label{eq:scalarratio}
  \frac{m^2_{\tilde{e}}}{m^2_{\tilde{t}}} 
    = \begin{cases}
        \; \dfrac{1- \delta_e}{\delta_t -1} 
           \left( \dfrac{\Lambda_{\text{IR}}}{\Lambda_{\text{UV}}} \right)^{ 2(1 - \delta_t)}
      & \delta_t < \delta_t^* \, , \\[1em]
        \; \dfrac{4\pi}{\alpha_3} 
           (1 - \delta_e) 
           \log^2 \left( \dfrac{\Lambda_{\text{UV}}}{\Lambda_{\text{IR}}} \right)
      & \delta_t \geq \delta_t^* \, .
     \end{cases}
\end{equation}
Note that this expression is separately valid for left- and right-handed scalar masses and only the gluino contribution is included in \eqref{eq:gauginocont} for the gaugino-mediated contribution to the stop mass squared. Interestingly, the two expressions \eqref{eq:fermionratio} and \eqref{eq:scalarratio} differ only in the exponents for $\delta_t < \delta_t^*$, while for $\delta_t\geq \delta_t^*$ the ratio no longer depends on $\delta_t$.  As shown in Fig.~\ref{fig:fermionratio}, the allowed region is $0 \lesssim \delta_e\lesssim 0.9$ and $1 \lesssim \delta_t \lesssim 1.8$,
depending on the value of $\Lambda_{\text{IR}}/\Lambda_{\text{UV}}$. The largest value of the ratio
$m_{\tilde{e}}/m_{\tilde{t}}$ is approximately 140 (390) for $\Lambda_{\text{IR}}/\Lambda_{\text{UV}}\simeq 10^{-3}~(10^{-16})$. Note that 
the Yukawa coupling ratio contours in Fig.~\ref{fig:fermionratio}, end at $\delta_t^*$ because the sfermion mass ratio  becomes approximately 
constant as seen in (\ref{eq:scalarratio}).

Although these are na\"{i}ve tree-level results obtained at the IR scale, they clearly reveal an inverted mass hierarchy for the sparticle spectrum where $m_{\tilde{e}} / m_{\tilde{t}} \sim 10 \text{--} 350$. To obtain a physical mass spectrum, this range is further restricted due to renormalization group (RG) effects. We next include these effects, as well as a number of phenomenological constraints, such as the 125 GeV Higgs boson. Since the strong dynamics is nonperturbative, the AdS/CFT correspondence is used to calculate the mass spectrum
in a slice of five-dimensional AdS.


\section{The Five-Dimensional Picture}\label{sec:5d}

\subsection{Supersymmetry in a slice of AdS}\label{sec:ads5-susy}

We consider a warped five-dimensional spacetime $(x^\mu, y)$, where $\mu=0,1,2,3$ are the 4D coordinates and $-\pi R \le y \le \pi R$ is the coordinate of an extra dimension compactified on a $S^1/\mathbb{Z}_2$ orbifold of radius $R$. The spacetime metric is anti-de Sitter, given by
\begin{equation} \label{eq:ads5-metric}
  ds^2 = e^{-2 k|y|} \eta_{\mu\nu} \, dx^{\mu} \, dx^{\nu} + dy^2 \equiv g_{MN} \, dx^M \, dx^N \, ,
\end{equation}
where $k$ is the AdS curvature scale and capital Latin indices $M = (\mu, 5)$ label the 5D coordinates. The 5D spacetime is a slice of AdS\textsubscript{5} geometry, bounded by two 3-branes located at the orbifold fixed points: a UV brane at $y = 0$ and an IR brane at $y = \pi R$.\footnote{We do not specify a particular mechanism to stabilize the extra dimension. In a supersymmetric theory, one possibility is supersymmetrization of the Goldberger-Wise mechanism~\cite{Goldberger:1999uk,Goh:2003yr,Gherghetta:2011wc}.}

The cutoff scale of the UV brane is $\Lambda_{\text{UV}} = M_5$, where $M_5$ is the 5D Planck scale, while the scale of the IR brane is $\Lambda_{\text{IR}} = \Lambda_{\text{UV}} e^{-\pi k R}$. The 4D reduced Planck mass, $M_P$, is given by~\cite{Randall:1999ee}
\begin{equation}
  M_P^2 = \frac{M_5^3}{k} \left(1 - e^{-2 \pi k R}\right) \simeq \frac{M_5^3}{k} \, ,
\end{equation}
where we are assuming $\pi k R \gg 1$. Note that this expression is consistent with the result \eqref{eq:planckmass-4d} derived from partial compositeness. In order for the classical metric solution to be valid, the AdS curvature must be small enough compared to the 5D Planck scale so that higher-order curvature terms in the 5D gravitational action can be neglected. This requires $k/M_5 \lesssim 2$~\cite{Agashe:2007zd}, but, in the following, we have taken $k$ to be generically smaller, choosing $k / M_5 = 0.1$.

Besides gravity, we introduce the matter and gauge field content of the minimal supersymmetric standard model (MSSM) in the bulk. Since only Dirac fermions are allowed by the 5D Lorentz algebra, the bulk supersymmetry has eight supercharges, corresponding to $\mathcal{N} = 2$ supersymmetry (SUSY) from the 4D perspective. All fields that propagate in the AdS bulk are thus in $\mathcal{N} = 2$ representations of supersymmetry, but the orbifold compactification breaks this to an $\mathcal{N} = 1$ supersymmetry at the massless level, preserving four supercharges~\cite{Gherghetta:2000qt,Gherghetta:2000kr}. The massless modes which form this 4D MSSM are the zero-mode solutions in the Kaluza-Klein (KK) decompositions of the 5D fields.

The zero-mode profiles for the bulk fields are summarized in Appendix~\ref{app:zero-modes}. In a bulk hypermultiplet, the fermion and scalar zero-mode profiles depend on the bulk fermion mass parameter $c$. By the AdS/CFT correspondence, the scaling dimension of fermionic operators in the 4D dual theory is $\operatorname{dim} \mathcal{O}_\psi = \frac{3}{2} + |c \pm \frac{1}{2}|$, and for scalar operators it is $\operatorname{dim} \mathcal{O}_\phi = 1 + | c \pm \frac{1}{2}|$,
where the upper (lower) sign is used for left-handed (right-handed) fields. Thus, there is direct relation $\delta_i = |c_i \pm \frac{1}{2}|$ between the anomalous dimensions $\delta_i$ in the 4D dual theory (introduced in Sec.~\ref{sec:sfermionmass}) and the bulk fermion mass parameters $c_i$. Furthermore, for a bulk vector supermultiplet, the zero-mode profiles of the gauge field and gaugino field are flat, such that the bulk mass parameter of the Majorana fermion gaugino is $c = \frac{1}{2}$. Similarly, the gravity supermultiplet contains a graviton with the UV-localized zero mode ($\propto e^{-k y}$) and a spin-$\frac{3}{2}$ Rarita-Schwinger fermion (the gravitino), whose bulk mass parameter is fixed to be $c = \frac{3}{2}$.

\begin{figure}[t]
  \centering
  \includegraphics{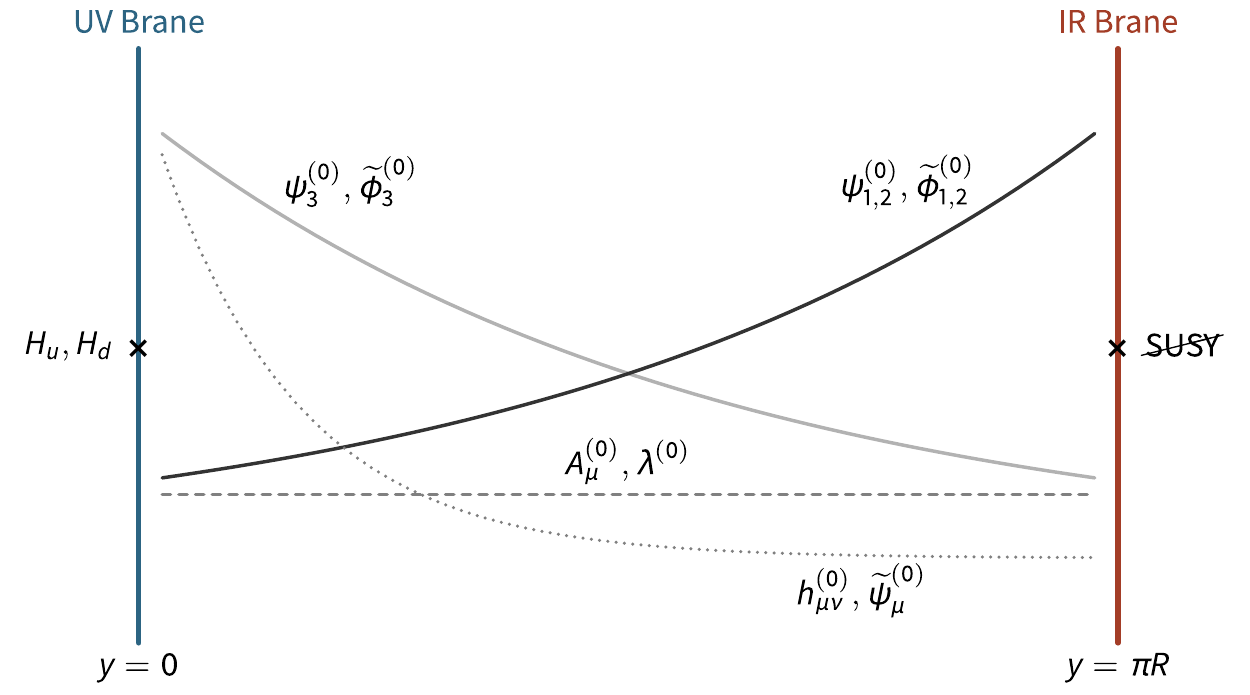}
  \caption{Schematic diagram of the 5D model with the Higgs fields, $H_{u,d}$, localized on the
  UV brane and supersymmetry broken on the IR brane. The Yukawa coupling hierarchy requires that the lighter
  first- and second-generation fermions $\psi_{1,2}^{(0)}$ (and their superpartners $\widetilde{\phi}_{1,2}^{(0)}$) are IR localized (dark gray line) and the heavier, third-generation fermions $\psi_3^{(0)}$ (and their superpartners $\widetilde{\phi}_3^{(0)}$) are UV localized (light gray line). The vector supermultiplet, $A_\mu^{(0)}$ and $\lambda^{(0)}$, (dashed line) is not localized, while the gravity multiplet, $h_{\mu\nu}^{(0)}$ and $\widetilde{\psi}_\mu^{(0)}$, (dotted line) is UV localized.}
\label{fig:adsusy-fields}
\end{figure}

The warped geometry naturally generates a separation of scales that can be used to explain the hierarchy between the scale of supersymmetry breaking and the Planck scale. The IR brane is therefore identified with the scale where supersymmetry is broken; the bulk and UV brane remain supersymmetric. The Higgs fields are localized on the UV brane, while the rest of the MSSM fields propagate in the bulk and couple to the Higgs fields with brane-localized Yukawa couplings. As discussed in Sec.~\ref{sec:fermionhierarchy-5d}, the degree of overlap between a particular bulk fermion zero-mode profile and a UV-localized Higgs field determines its effective 4D Yukawa coupling and thus the size of the corresponding fermion mass. In this setup, third-generation fermions are therefore UV localized, whereas the lighter first- and second-generation fermions are more IR localized. Including the SM gauge fields, the gauginos, and gravity, a full schematic diagram of the 5D model is depicted in Fig.~\ref{fig:adsusy-fields}.

Since the localization of the fermion in each chiral supermultiplet determines the corresponding scalar localization, this 5D fermion geography has distinctive consequences in the SUSY sector: the third-generation sfermions are generally UV localized and the first- and second-generation sfermions are IR localized. Due to the properties of localization, the effective coupling strength of each superfield on the IR brane is inversely related to its coupling strength on the UV brane. Therefore, when supersymmetry is broken on the IR brane, as discussed in Sec.~\ref{sec:susybreaking-5d}, the localization of the sfermions induced by the fermion mass spectrum results in an inverted scalar soft mass spectrum: light fermions have heavy superpartners, while heavy fermions have light superpartners. We next construct the details of this distinctive supersymmetric particle spectrum.

\subsection{The fermion mass hierarchy}\label{sec:fermionhierarchy-5d}

Consider first the generation of the fermion mass hierarchy~\cite{Gherghetta:2000qt,Huber:2000ie}.
In our 5D spacetime, the SM fermion mass hierarchy is determined from the overlap of the bulk SM fermion
zero modes with the UV-localized Higgs fields. The Yukawa interactions take the form
\begin{align}
  S_5 
   &= \int d^5 x \, 
      \sqrt{-g} \, 
      Y^{(5)}_{ij}
      \Bigl[ \, \bar{\Psi}_{iL}(x^\mu, y) \, \Psi_{jR}(x^\mu, y) + \text{H.c.} \Bigr]
      H(x^\mu) \, \delta(y) \nonumber \\[1ex]
   &\equiv 
      \int d^4 x \,
      \Bigl[
      \, y_{ij} \, \bar{\psi}_{iL}^{(0)}(x^\mu) \, \psi_{jR}^{(0)}(x^\mu) \, H(x^\mu) + h.c + \dotsb \,
      \Bigr] \text{ ,}
      \label{eq:yukawa-5d}
\end{align}
where the $Y^{(5)}_{ij}$ (with flavor indices $i,j=1,2,3$) are dimensionful (inverse mass) 5D Yukawa couplings, $\Psi_{L(R)}$ is a Dirac spinors that contains an SU(2)\textsubscript{\textit{L}} doublet (singlet) of the MSSM as its zero mode, and $H$ is the appropriate Higgs field. Using the 5D fermion zero-mode\footnote{Use of the zero-mode approximation for the profiles, where the backreaction of the boundary Higgs-generated fermion mass is neglected, is a valid approximation provided $v^2/\Lambda_{\text{UV}}^2 \ll 1$.} profiles \eqref{eq:zeromode-fermion}, the effective 4D SM Yukawa couplings $y_{ij}$ are then given by~\cite{Gherghetta:2000qt}
\begin{equation}
  y_{ij}
    = Y^{(5)}_{ij} \tilde{f}_{iL}^{(0)}(0) \, \tilde{f}_{jR}^{(0)}(0)
    = Y^{(5)}_{ij} k
      \sqrt{\frac{\frac{1}{2} - c_{iL}}{e^{2(\frac{1}{2} - c_{iL})\pi k R} - 1}}
      \sqrt{\frac{\frac{1}{2} + c_{jR}}{e^{2(\frac{1}{2} + c_{jR})\pi k R} - 1}}~.
      \label{eq:yukawa-4d}
\end{equation}
By assuming that the dimensionless 5D couplings $Y^{(5)}_{ij} k$ are of order one, and since
$\pi k R \gg 1$, the hierarchy in the 4D Yukawa couplings $y_{ij}$ is generated by the order
one bulk mass parameters $c_i$ of the fermions. Recall that in the 4D dual theory, this
is equivalent to choosing the anomalous dimensions $\delta_i$.

After electroweak symmetry breaking, the neutral components of each MSSM Higgs doublet acquire VEVs, $v_u = \langle H_u^0 \rangle$ and $v_d = \langle H_d^0 \rangle$, which are related by $\tan \beta \equiv v_u / v_d$, and the fermions obtain masses
\begin{subequations}
\label{eq:yukawa-mass}
\begin{align}
    (m_e)_{ij} &= (y_e)_{ij} \, v \cos \beta \, , \\
    (m_d)_{ij} &= (y_d)_{ij} \, v \cos \beta \, , \\
    (m_u)_{ij} &= (y_u)_{ij} \, v \sin \beta \, ,
  \end{align}
\end{subequations}
where $(m_{u,d,e})_{ij}$ and $(y_{u,d,e})_{ij}$ are the SM fermion mass and Yukawa coupling matrices, respectively, and $v^2 \equiv v_u^2 + v_d^2 \simeq (174 \, \text{GeV})^2$ is the SM Higgs VEV. In the mass basis, these matrices are diagonal. Neglecting quark mixing (see Refs.~\cite{Casagrande:2008hr,Bauer:2009cf} for a fuller treatment) the interaction basis coincides with the mass basis, resulting (given values for $\tan \beta$ and $\Lambda_{\text{IR}}$) in a system with 24 free parameters $(Y^{(5)}_{e,d,u})_{ii}$ and $c_{L_i,e_i,Q_i,d_i,u_i}$ and nine constraint equations following from \eqref{eq:yukawa-mass}. If we take a universal value for the 5D Yukawa couplings such that $(Y^{(5)}_{e,d,u})_{ii} = Y^{(5)} k$, there remains one parameter degree of freedom within each generation of leptons and within each generation of quarks, which we choose as the doublet $c$ parameters $c_{L_i,Q_i}$, without loss of generality.

\begin{figure}[t]
  \centering
  \includegraphics{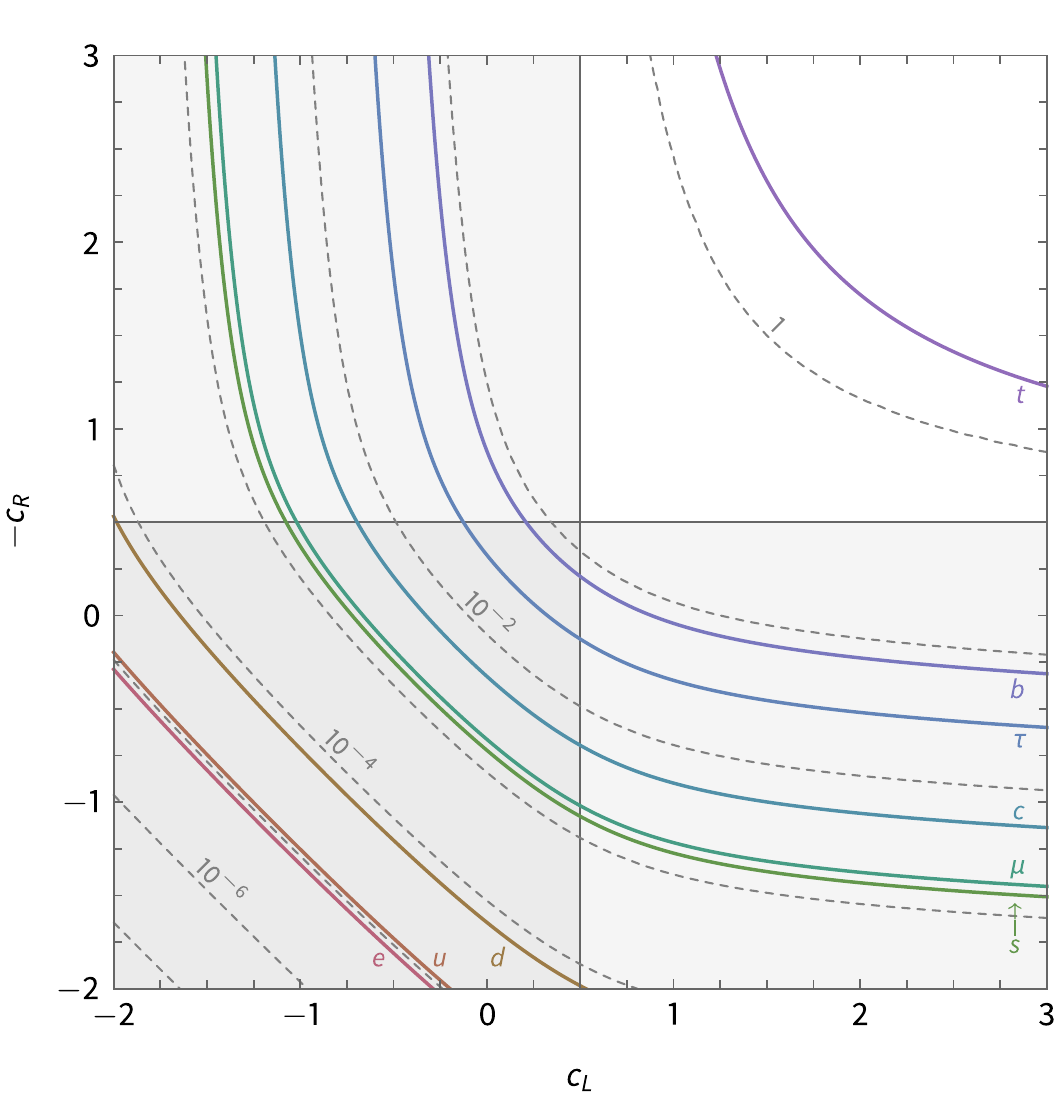}
  \caption{Contours of the effective 4D Yukawa coupling \eqref{eq:yukawa-4d} at the IR-brane scale as a function of the localization parameters $c_L$ and $c_R$ of the bulk fermion fields for $\tan \beta = 3$, $\Lambda_{\text{IR}} = 2 \times 10^{16}$~GeV, and $Y^{(5)} k = 1$. The dashed gray lines give contours of the Yukawa coupling strength. In color are contour lines corresponding to the coupling strengths of the SM Yukawa couplings at the IR-brane scale. The region in which each field is IR localized is shaded light gray and the region where both fields are IR-localized is darker gray.}
  \label{fig:yukawa-matching}
\end{figure}

The relations \eqref{eq:yukawa-mass} hold for the running masses. Ultimately, we are interested in the bulk mass parameters at the IR-brane scale, where they determine the soft masses received by the sfermions when supersymmetry is broken. Therefore, before we perform the 4D-5D Yukawa matching, we first evolve the 4D Yukawa couplings to the IR-brane scale. We discuss the procedure we use to consistently perform the renormalization in Sec.~\ref{sec:numerical-spectrum-renormalization}. In Fig.~\ref{fig:yukawa-matching}, we give an example of the resulting matching, showing the allowed range of localizations ($c$ parameters) of the SM leptons and quarks when the 5D Yukawa coupling takes the universal value $Y^{(5)} k = 1$ and $\Lambda_{\text{IR}} = 2 \times 10^{16}$~GeV. Generally, we see that the largeness of the third-generation Yukawa couplings requires the third-generation fermions for both leptons and quarks to be UV-localized (white region), while the smaller Yukawa couplings of the first and second generations lead to IR localization (darker gray region). Although it is always possible to make one of the chiral fermions in each generation UV-localized, only for the third generation can this be done without making at least one of the other chiral fermions IR localized.

We note in passing that in the quark sector, since both singlet fields in a given generation must be separately matched to the same doublet field according to \eqref{eq:yukawa-4d}, the asymmetry between the 4D couplings $y_{u_i}$ and $y_{d_i}$ precludes the solution $c_{Q_i} = - c_{d_i} = - c_{u_i}$ unless $y_{u_i} / y_{d_i} = Y^{(5)}_{u_i} / Y^{(5)}_{d_i}$. In this case, the 4D Yukawa coupling hierarchies are simply moved into the 5D couplings. This behavior is an indication of the universality of the warped extra dimension: while it can explain the magnitude of the Yukawa coupling hierarchies, since it is flavor-blind, the underlying flavor structure remains as an order-one feature. In practice, we take $Y^{(5)} k$ to have a universal value and absorb the flavor structure completely into the quark $c$ parameters. A similar situation does not arise in the lepton sector as we do not include Yukawa couplings for the neutrinos. Neutrinos can be naturally incorporated in a warped extra dimension to generate the required neutrino masses~\cite{Grossman:1999ra,Goldberger:2002pc,Huber:2003sf,Gherghetta:2003hf,Agashe:2015izu}.

\subsection{Supersymmetry breaking}\label{sec:susybreaking-5d}

Supersymmetry is assumed to be broken on the IR brane and is parameterized by the introduction of a spurion superfield $X = \theta \theta F_X$ that couples to the sfermions and the gauginos. The sfermions and gauginos acquire tree-level soft masses with a characteristic scale $F / \Lambda_{\text{IR}}$, where $F = F_X e^{- 2 \pi k R}$, modulated by their overlap with the IR brane. A gravitino mass of order $F / M_P$ is generated by the super-Higgs effect. The Higgs fields receive no tree-level soft masses, as they are confined to the UV brane and do not couple directly to the supersymmetry-breaking spurion. We do not include any mechanism to generate tree-level trilinear soft scalar couplings, although, like the Higgs-sector soft terms, they arise radiatively.

Contributions to the soft masses also generically arise from anomaly mediation. Since the gravitino mass is Planck-scale suppressed (as opposed to the other soft masses, which are suppressed by the IR-brane scale), the anomaly-mediated contribution typically is subdominant to the effects (both at tree level and loop level) of the supersymmetry-breaking sector on the IR brane. An additional source of supersymmetry breaking arises due to the stabilization of the radion of the extra dimension, which generically requires a nonzero $F$-term for the radion superfield (equivalently, the introduction of a constant superpotential on the IR brane). The scale of the radion-mediated contribution to the soft masses depends on the details of the stabilization model. We are interested in the regime where such effects are subdominant to the effects of the supersymmetry-breaking sector on the IR brane. In the model of Ref.~\cite{Gherghetta:2011wc}, this can be accomplished if the Goldberger-Wise bulk hypermultiplet is sufficiently UV localized.

\subsubsection{Gravitino mass}\label{sec:susybreaking-5d-gravitino}

When supersymmetry is spontaneously broken on the IR boundary, the effective 4D cosmological constant receives a positive contribution from the VEV of $F_X$. In the 5D warped geometry, this contribution can be canceled by the addition of a constant superpotential $W$ on the UV brane~\cite{Randall:1998uk, Luty:2000ec, Luty:2002ff, Gherghetta:2002nr, Chacko:2003tf, Itoh:2006fv, Gherghetta:2011wc}, which introduces a boundary mass term for the gravitino:
\begin{equation} \label{eq:susybreaking-action-gravitino}
  S_5
    = \int d^5 x \, \sqrt{-g}
      \left[ \,
        \frac{1}{4}
        \frac{W}{M_5^3}
        \psi_{\mu} \left[ \sigma^{\mu}, \bar{\sigma}^{\nu} \right] \psi_{\nu} \,
      + \text{H.c.} \,
      \right]
      \delta(y) \, .
\end{equation}
The cosmological constant vanishes when
\begin{equation}
  |F|^2 \simeq 3 \frac{|W|^2}{M_P^2} \, ,
\end{equation}
such that the lightest gravitino obtains a Majorana mass:
\begin{equation}
  \label{eq:KKgrav}
  m_{3/2} \simeq \frac{F}{\sqrt{3} M_P} \, .
\end{equation}
This is the super-Higgs effect.\footnote{Note that a constant superpotential can also be introduced on the IR brane, as is generically expected in the context of radion stabilization. However, such a superpotential provides a positive contribution to the cosmological constant, and so it cannot be the sole source for the gravitino mass.} Again, higher-order terms can be included to account for the backreaction of the gravitino boundary mass, although in practice this is not necessary in the relevant regions of our parameter space. As expected, due to the universality of gravity, this matches the usual 4D result. Since the gravitational coupling is Planck-scale suppressed, the gravitino mass is lower than the characteristic soft mass scale $F/\Lambda_{\text{IR}}$ by a warp factor, and the gravitino is therefore always the LSP in the relevant regions of parameter space. This is consistent with the partial compositeness result \eqref{eq:dualgravitinomass} in the 4D dual theory, where the gravitino is mostly an elementary state.

\subsubsection{Gaugino masses}\label{sec:susybreaking-5d-gaugino}

For a field strength superfield $W^a$ of a vector supermultiplet $V^a$ containing a standard model gauge field $A^a_{\mu}$ and its Majorana fermion gaugino superpartner $\lambda_a$ (where $a$ is the gauge index), we introduce the interaction
\begin{equation} \label{eq:susybreaking-action-gaugino}
  S_5 = \int d^5 x \, \sqrt{-g}
        \int d^2 \theta
        \left[ \, \frac{1}{2} \frac{X}{\Lambda_{\text{UV}} k} W^{\alpha a} W^a_\alpha \, + \text{H.c.} \right] \delta(y - \pi R) \, .
\end{equation}
This term gives rise to a boundary Majorana mass for the gaugino field and breaks supersymmetry, shifting the masses of the Kaluza-Klein modes up such that there is no longer a massless gaugino zero-mode solution.
At tree level, the lightest KK mass is
\begin{equation}
  \label{eq:zeromodemass-gaugino-tree-x}
  M_{\lambda}
    \simeq \frac{g_5^2 k}{2\pi k R} \frac{F}{\Lambda_{\text{IR}}}
    = g^2 \frac{F}{\Lambda_{\text{IR}}} \, ,
\end{equation}
where $g_5^2 k = (2 \pi k R) g^2$. This mass expression \eqref{eq:zeromodemass-gaugino-tree-x} assumes the zero-mode approximation for the profiles, where the backreaction of the boundary Majorana mass is neglected, an approximation that is valid provided $\sqrt{F}/\Lambda_{\text{IR}} \lesssim 1$. In practice, we include terms higher order in $\sqrt{F}/\Lambda_{\text{IR}}$. The mass for arbitrary $F$ can be determined by solving the full KK mass quantization condition (see Refs.~\cite{Gherghetta:2000qt,Gherghetta:2000kr}). Note that the gaugino masses are suppressed relative to $F/\Lambda_{\text{IR}}$ by $g^2 \sim g_5^2/\pi k R$, the square of the 4D gauge coupling \cite{Chacko:2003tf},\footnote{The gauge-coupling dependence arises since we assume a generic GUT symmetry that is broken by the Higgs mechanism on the UV brane, separated from the supersymmetry-breaking sector on the IR boundary.} and hence the gauginos in general obtain masses suppressed below those of sfermions with flat profiles ($\pm c = \frac{1}{2}$). This suppression matches that found in \eqref{eq:gauginomass}, as expected from the AdS/CFT dictionary.

If the supersymmetry-breaking sector does not contain any singlets with large $F$-terms, the interaction \eqref{eq:susybreaking-action-gaugino} is forbidden. In this case, with a nonsinglet spurion $X$, the leading contribution to the gaugino masses is~\cite{Heidenreich:2014jpa}
\begin{equation}
  S_5 = \int d^5 x \, \sqrt{-g}
        \int d^4 \theta
        \left[ \, \frac{1}{2} \frac{X^{\dag} X}{\Lambda^3_{\text{UV}} k} W^{\alpha a} W^a_\alpha \, + \text{H.c.} \right] \delta(y - \pi R) \, ,
\end{equation}
such that
\begin{equation}
  \label{eq:zeromodemass-gaugino-tree-xx}
  M_{\lambda}
    \simeq \frac{g_5^2 k}{2\pi k R} \frac{F^2}{\Lambda_{\text{IR}}^3}
    = g^2 \frac{F^2}{\Lambda_{\text{IR}}^3} \, .
\end{equation}
Except in the regime $\sqrt{F} \sim \Lambda_{\text{IR}}$, this mass is highly suppressed, and other supersymmetry-breaking contributions such as radion mediation may dominate.

\subsubsection{Sfermion masses}\label{sec:susybreaking-5d-sfermion}

\begin{figure}[t]
  \centering
  \includegraphics{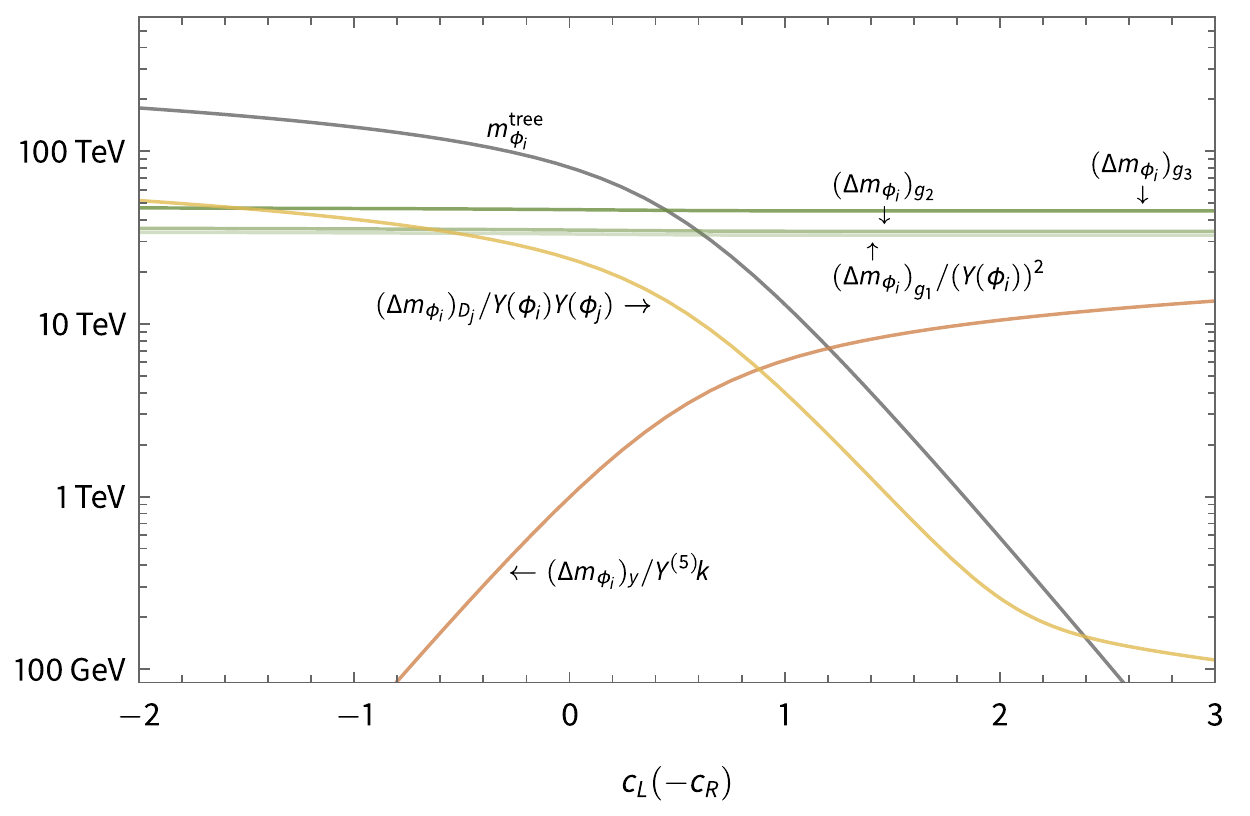}
  \caption{Plot of the magnitude of the tree-level scalar soft mass and the gauge and Yukawa one-loop radiative corrections as a function of hypermultiplet localization when the gaugino masses are given by \eqref{eq:zeromodemass-gaugino-tree-x}. We take $\Lambda_{\text{IR}} = 2 \times 10^{16}$~GeV, $\sqrt{F} = 4.75 \times 10^{10}$~GeV, and $\tan \beta = 3$.}
  \label{fig:loopcontributions-scalar-x}
\end{figure}

\begin{figure}[t]
  \centering
  \includegraphics{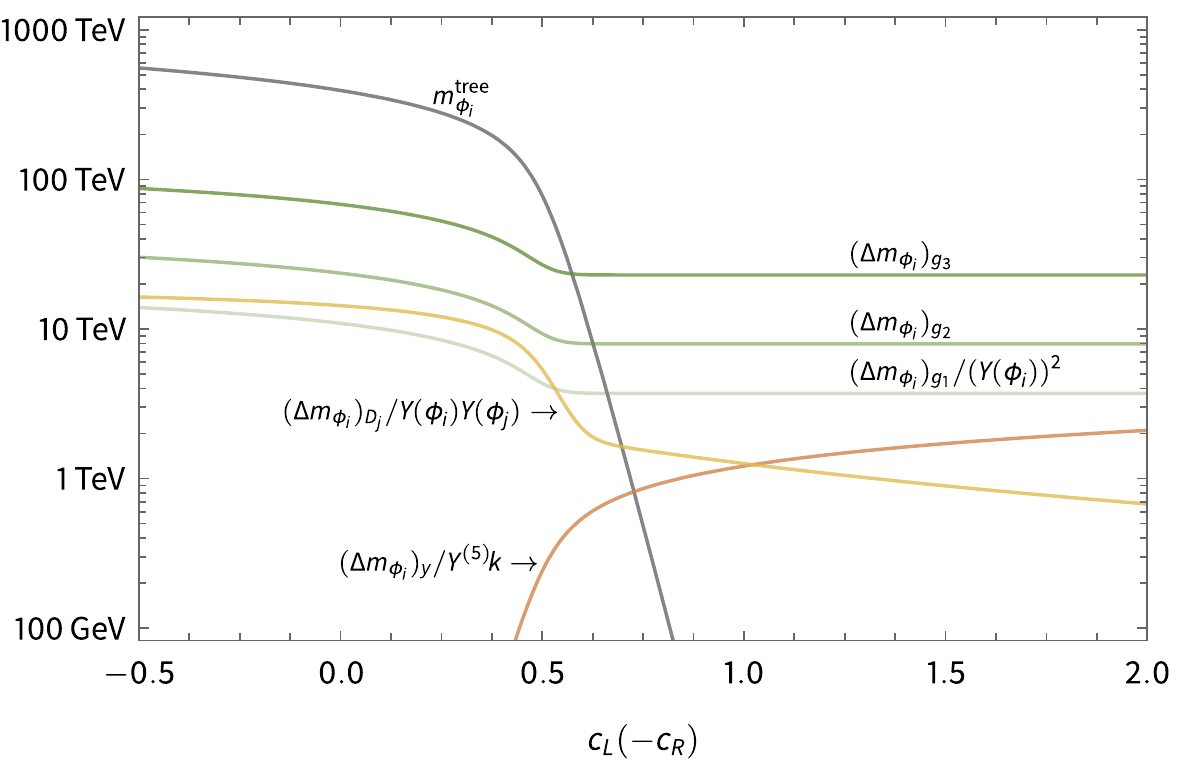}
  \caption{Plot of the magnitude of the tree-level scalar soft mass and the gauge and Yukawa one-loop radiative corrections as a function of hypermultiplet localization when the gaugino masses are given by \eqref{eq:zeromodemass-gaugino-tree-xx}. We take $\Lambda_{\text{IR}} = 6.5 \times 10^6$~GeV, $\sqrt{F} = 2 \times 10^6$~GeV, and $\tan \beta = 5$.}
  \label{fig:loopcontributions-scalar-xx}
\end{figure}

For a chiral supermultiplet $\Phi$ containing a Weyl fermion $\psi$ and its complex scalar superpartner $\phi$, we introduce the interaction
\begin{equation} \label{eq:susybreaking-action-sfermion}
  S_5 \supset \int d^5 x \, \sqrt{-g} \int d^4 \theta \, \frac{X^{\dag} X}{\Lambda_{\text{UV}}^2 k} \Phi^\dag \Phi \, \delta(y - \pi R) \, .
\end{equation}
As with the gauginos, adding this boundary mass breaks supersymmetry. At tree level, the lightest KK mass is
\begin{equation} \label{eq:zeromodemass-scalar-tree}
  m_{\phi_{L,R}}^{\text{tree}}
    \simeq
      \frac{F}{\Lambda_{\text{IR}}} \,
      \sqrt{\frac{\frac{1}{2} \mp c}{e^{2 (\frac{1}{2} \mp c) \pi k R} - 1}} \,
      e^{(\frac{1}{2} \mp c) \pi k R}
    \sim \begin{cases}
           \; (\pm c - \frac{1}{2})^{1/2} \, \dfrac{F}{\Lambda_{\text{IR}}} e^{(\frac{1}{2} \mp c) \pi k R}
         & \pm c > \frac{1}{2} \, , \\[1em]
           \; (\frac{1}{2} \mp c)^{1/2} \, \dfrac{F}{\Lambda_{\text{IR}}}
         & \pm c < \frac{1}{2} \, ,
         \end{cases}
\end{equation}
where the upper (lower) signs refer to the $L$ $(R)$ states. As with the gauginos, the scalar mass is valid in the limit $\sqrt{F}/\Lambda_{\text{IR}} \lesssim 1$, and in our numerical calculations we include terms higher order in $F/\Lambda_{\text{IR}}^2$ to account for the backreaction of the sfermion boundary mass. For simplicitly, we have taken these interactions to be flavor-diagonal, although this assumption can be relaxed. Note that the UV-localized $(\pm c > \frac{1}{2})$ sfermion masses are suppressed by a warp factor relative to the IR-localized sfermion masses $(\pm c < \frac{1}{2})$ because supersymmetry is broken on the IR brane. This behavior is illustrated in Figs.~\ref{fig:loopcontributions-scalar-x} and \ref{fig:loopcontributions-scalar-xx}. Using the relations $\Lambda_{\text{IR}}/\Lambda_{\text{UV}} = e^{-\pi kR}$ and $\delta_i = |c_i \pm \frac{1}{2}|$, the expressions \eqref{eq:zeromodemass-scalar-tree} are seen to be consistent with the masses \eqref{eq:sfermionmass} obtained in the 4D dual theory.

Since the soft masses generated at tree level by \eqref{eq:susybreaking-action-sfermion} can be exponentially small for UV-localized bulk scalar fields, quantum corrections become significant when $\pm c$ is sufficiently large. At the one-loop level, supersymmetry breaking is transmitted to the bulk scalars via interactions with other bulk scalars and gauginos. In Appendix~\ref{app:radiativecorrections-scalar}, we derive the resulting contributions to the bulk scalar masses squared in the bulk theory. From the 4D perspective, these appear as one-loop threshold corrections to the scalar soft masses squared at the IR-brane scale, arising when the KK modes of the theory are integrated out. Parametrized in terms of the gaugino and sfermion tree-level soft masses, the corrections take the forms
\begin{subequations}
  \begin{align}
    16 \pi^2 (\Delta m^2_{\tilde{Q}_i})_{\text{1-loop}}
      &= \begin{aligned}[t]
            \frac{32}{3} r^{\tilde{Q}_i}_{g_3} g_3^2 M_3^2
         &+ 6            r^{\tilde{Q}_i}_{g_2} g_2^2 M_2^2
          + \frac{2}{15} r^{\tilde{Q}_i}_{g_1} g_1^2 M_1^2 \\[1ex]
         &- 2 r^{\tilde{Q}_i}_{y_{u_i}} y_{u_i}^2 m_{\tilde{u}_i}^2
          - 2 r^{\tilde{Q}_i}_{y_{d_i}} y_{d_i}^2 m_{\tilde{d}_i}^2
          - \frac{1}{5} g_1^2 \Delta_{\mathcal{S}} \, ,
         \end{aligned} \\[1ex]
    16 \pi^2 (\Delta m^2_{\tilde{u}_i})_{\text{1-loop}}
      &= \frac{32}{3}  r^{\tilde{u}_i}_{g_3} g_3^2 M_3^2
       + \frac{32}{15} r^{\tilde{u}_i}_{g_1} g_1^2 M_1^2
       - 4             r^{\tilde{u}_i}_{y_{u_i}} y_{u_i}^2  m_{\tilde{Q}_i}^2
       + \frac{4}{5} g_1^2 \Delta_{\mathcal{S}} \, , \\[1ex]
    16 \pi^2 (\Delta m^2_{\tilde{d}_i})_{\text{1-loop}}
      &= \frac{32}{3} r^{\tilde{d}_i}_{g_3} g_3^2 M_3^2
       + \frac{8}{15} r^{\tilde{d}_i}_{g_1} g_1^2 M_1^2
       - 4            r^{\tilde{d}_i}_{y_{d_i}} y_{d_i}^2 m_{\tilde{Q}_i}^2
       - \frac{2}{5} g_1^2 \Delta_{\mathcal{S}} \, , \\[1ex]
    16 \pi^2 (\Delta m^2_{\tilde{L}_i})_{\text{1-loop}}
      &= 6           r^{\tilde{L}_i}_{g_2} g_2^2 M_2^2
       + \frac{6}{5} r^{\tilde{L}_i}_{g_1} g_1^2 M_1^2
       - r^{\tilde{L}_i}_{y_{e_i}} y_{e_i}^2  m_{\tilde{e}_i}^2
       + \frac{3}{5} g_1^2 \Delta_{\mathcal{S}} \, , \\[1ex]
    16 \pi^2 (\Delta m^2_{\tilde{e}_i})_{\text{1-loop}}
      &= \frac{24}{5} r^{\tilde{e}_i}_{g_1} g_1^2 M_1^2
       - 4            r^{\tilde{e}_i}_{y_{e_i}} y_{e_i}^2 m_{\tilde{L}_i}^2
       - \frac{6}{5} g_1^2 \Delta_{\mathcal{S}} \, ,
  \end{align}
\end{subequations}
where $\Delta_{\mathcal{S}}$, defined in \eqref{eq:scalar-correction-dterm-r}, is
\begin{equation}
  \label{eq:dterm-trace-ads}
  \Delta_{\mathcal{S}}
    = \sum_i Y(\phi_i) \, r^D_{\phi_i} m^2_{\phi_i}
    = \operatorname{Tr}
      \left[ \,
          r^D_{\tilde{Q}_i} m_{\tilde{Q}_i}^2
      - 2 r^D_{\tilde{u}_i} m_{\tilde{u}_i}^2
      +   r^D_{\tilde{d}_i} m_{\tilde{d}_i}^2
      -   r^D_{\tilde{L}_i} m_{\tilde{L}_i}^2
      +   r^D_{\tilde{e}_i} m_{\tilde{e}_i}^2 \,
      \right] \, ,
\end{equation}
and the sum is over the sfermions. The coefficients $r^{\phi_i}_g$, $r^{\phi_i}_y$, and $r^D_{\phi_i}$, defined in \eqref{eq:scalar-r-gauge}, \eqref{eq:scalar-r-yukawa}, and \eqref{eq:scalar-r-dterm}, respectively, are numerical parameters which encode the effect of the extra dimension.

The radiative corrections to the bulk scalar soft masses can be divided into three types of contributions: gauge corrections arising from loops involving bulk vector supermultiplets, Yukawa corrections arising from loops of bulk hypermultiplets and boundary Higgs fields, and $D$-term corrections arising from the Fayet-Iliopoulos $D$-term for weak hypercharge. In Fig.~\ref{fig:loopcontributions-scalar-x} we plot the magnitudes of the various contributions as functions of the hypermultiplet localization when the gaugino mass is given by \eqref{eq:zeromodemass-gaugino-tree-x} (singlet spurion), and the magnitudes when the supersymmetry-breaking sector does not contain any singlets with large $F$-terms and the gaugino mass takes the form \eqref{eq:zeromodemass-gaugino-tree-xx} in Fig.~\ref{fig:loopcontributions-scalar-xx}. In each case, the tree-level sfermion mass is plotted in grey, the one-loop U(1), SU(2), and SU(3) gauge radiative corrections are plotted in green, and the magnitude of the maximal contribution from a single Yukawa coupling (this corresponds to $c_L = \frac{1}{2}$ or $- c_R = \frac{1}{2}$ for the corresponding doublet and singlet hypermultiplets), neglecting all color multiplicity factors and modulo the 5D Yukawa coupling, is plotted in orange. Also shown is the magnitude of the one-loop $D$-term radiative contribution due to a single scalar mode (yellow), modulo hypercharge factors.\footnote{As seen in Appendix~\ref{app:radiativecorrections-scalar-dterm}, one-loop $D$-term corrections are independent of the localization of the external scalar, and so here we plot the contribution as a function of the localization of the scalar contributing in the loop (unlike the other radiative contributions, which depend on the localization of the external scalar).}

Due to the conformal flatness of the vector supermultiplet, the gauge corrections take a universal value for UV-localized sfermions that is of order of the gaugino masses. These contributions are positive and set the characteristic scale of the radiative corrections. Since the tree-level contribution is dominant for IR-localized sfermions, which accordingly receive a mass of order of the characteristic soft mass scale $\sqrt{F}/\Lambda_{\text{IR}}$, the sfermion sector thus accommodates a hierarchy $m_{\phi}^{\text{UV}} / m_{\phi}^{\text{IR}} \sim M_a \Lambda_{\text{IR}} / F$. When the gaugino mass is given by \eqref{eq:zeromodemass-gaugino-tree-x} as in Fig.~\ref{fig:loopcontributions-scalar-x}, the sfermion hierarchy is $m_{\phi}^{\text{UV}} / m_{\phi}^{\text{IR}} \sim g_a^2 \sim \mathcal{O}(0.4 \text{--} 1)$, which may be increased modestly in individual families with the inclusion of $D$-term and Yukawa radiative contributions. The moderate size of this hierarchy is an important result of the inclusion of radiative corrections, which wash out the exponential localization dependence of the sfermion soft masses that is a tree-level feature of the extra dimension.

A larger hierarchy can be achieved if the supersymmetry-breaking sector does not contain any singlets with large $F$-terms as in Fig.~\ref{fig:loopcontributions-scalar-xx}. In this case, the gaugino masses take the form \eqref{eq:zeromodemass-gaugino-tree-xx}, and the characteristic sfermion hierarchy is $m_{\phi}^{\text{UV}} / m_{\phi}^{\text{IR}} \sim g_a^2 F / \Lambda_{\text{IR}}$. The maximum splitting that can be accommodated between the two sfermion mass scales is limited by the requirement that no sfermions receive negative soft masses at the IR-brane scale or in the subsequent RG evolution of the theory to lower scales. The constraints this condition imposes on the pattern of sfermion mass spectrum at the IR-brane scale are discussed in Sec.~\ref{sec:parameterspace-tachyon}.

In the sfermion sector, localization in the extra dimension thus distinguishes between two scales: a tree-level mass scale associated with IR-localized sfermions and a lower mass scale arising from radiative corrections for UV-localized sfermions. When the localizations of the matter hypermultiplets are chosen to explain the SM fermion mass hierarchy (i.e., the third-generation fermions must be predominantly UV-localized, while the lighter generations are mostly IR-localized), the result is a split sfermion spectrum, with the third-generation sfermions hierarchically lighter than the first two generations. We note that the sfermion spectrum inverts the ordering of the fermion spectrum, a consequence of the separation of the supersymmetry-breaking sector and the Higgs sector on opposite orbifold fixed points. 
Additionally, although both are explained by the same localization mechanism, the sfermion hierarchy is necessarily less split than the fermion hierarchy. This is because the Yukawa couplings only receive wave function renormalization, while the scalar masses are soft parameters and can receive large radiatve corrections from the extra dimension and MSSM running.

\subsubsection{Higgs sector}\label{sec:susybreaking-5d-higgs}

\begin{figure}[t]
  \centering
  \includegraphics{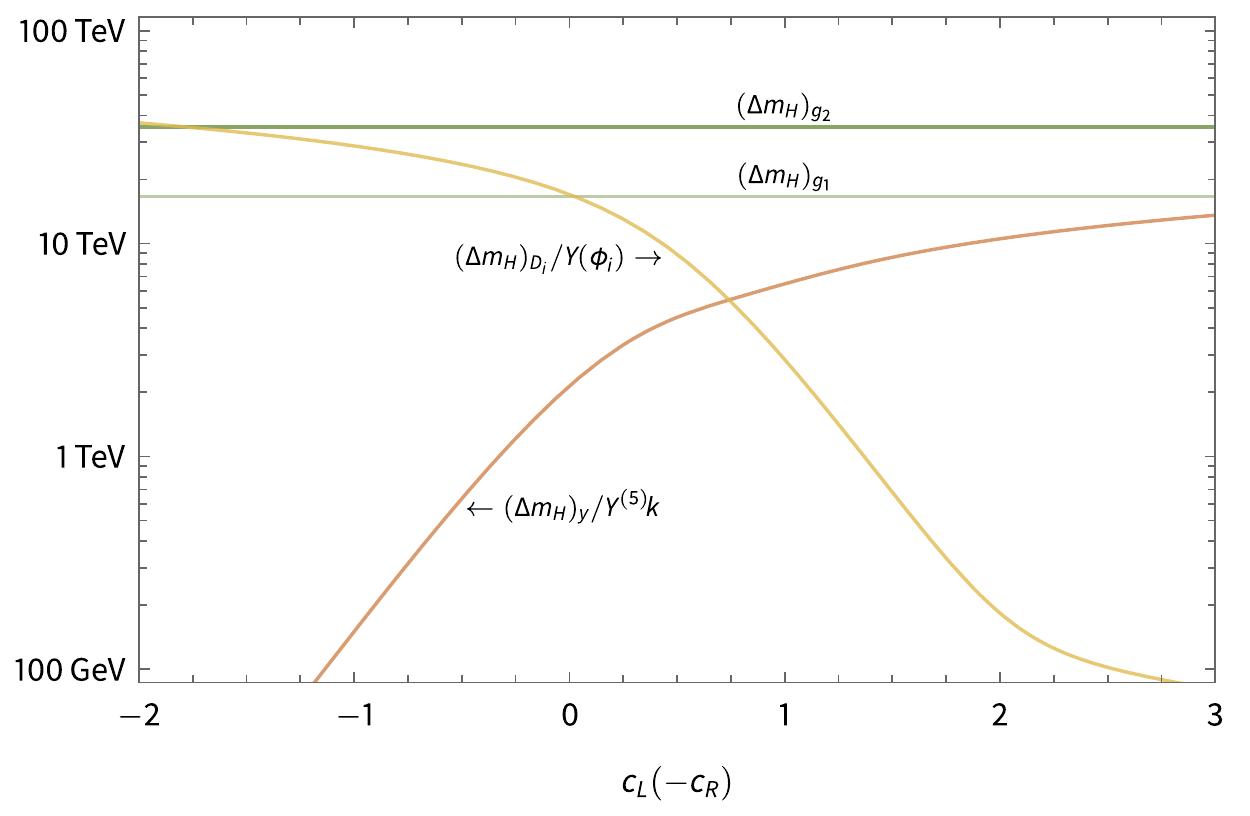}
  \caption{Plot of the magnitude of the one-loop radiative corrections that generate the Higgs soft masses as a function of hypermultiplet localization in the singlet spurion case. We take $\Lambda_{\text{IR}} = 2 \times 10^{16}$~GeV, $\sqrt{F} = 4.75 \times 10^{10}$~GeV, and $\tan \beta = 3$.}
  \label{fig:loopcontributions-higgs-x}
\end{figure}

The Higgs sector, confined to the UV brane, does not couple directly to the supersymmetry-breaking sector, and thus the Higgs soft terms at the IR-brane scale are zero at tree level. Nevertheless, as with the sfermions, the breaking is transmitted to the Higgs fields at the quantum level. We derive the one-loop corrections to $m_{H_u}^2$, $m_{H_d}^2$, and $b$ from the bulk theory in Appendix~\ref{app:radiativecorrections-higgs}. Parametrized in terms of the tree-level gaugino and sfermion soft masses, the one-loop Higgs masses take the form
\begin{subequations}
  \label{eq:softhiggs}
  \begin{align}
    16 \pi^2 m_{H_u}^2
      &= 6           r^H_{g_2} g_2^2 M_2^2
       + \frac{6}{5} r^H_{g_1} g_1^2 M_1^2
       - 6 \operatorname{Tr} \left[ \, r^H_{y_{u_i}} y_{u_i}^2 \left( m_{\tilde{Q}_i}^2 + m_{\tilde{u}_i}^2 \right) \right]
       - \frac{3}{5} g_1^2 \Delta_{\mathcal{S}} \, , \label{eq:softmhu2} \\[1ex]
    16 \pi^2 m_{H_d}^2
      &= \begin{aligned}[t]
            6           r^H_{g_2} g_2^2 M_2^2
          + \frac{6}{5} r^H_{g_1} g_1^2 M_1^2
         &- 6 \operatorname{Tr} \left[ \, r^H_{y_{d_i}} y_{d_i}^2 \left( m_{\tilde{Q}_i}^2 + m_{\tilde{d}_i}^2 \right) \right] \\[1ex]
         &- 2 \operatorname{Tr} \left[ \, r^H_{y_{e_i}} y_{e_i}^2 \left( m_{\tilde{L}_i}^2 + m_{\tilde{e}_i}^2 \right) \right]
          + \frac{3}{5} g_1^2 \Delta_{\mathcal{S}} \, ,
         \end{aligned} \label{eq:softmhd2} \\[1ex]
    16 \pi^2 b &= - \mu \left( 6 r^b_{\lambda_1} g_2^2 M_2 + \frac{6}{5} r^b_{\lambda_2} g_1^2 M_1 \right) \, , 
                    \label{eq:softb}
  \end{align}
\end{subequations}
where $r^H_g$, $r^H_y$, and $r^b_{\lambda}$ are defined in \eqref{eq:higgs-r-gauge}, \eqref{eq:higgs-r-yukawa}, and \eqref{eq:softb-r-gaugino}, respectively. The origin of the $\mu$-term on the UV brane is assumed to arise from the Kim-Nilles mechanism, as discussed in Sec.~\ref{sec:Higgsinomass}; its magnitude is determined as necessary to ensure that electroweak symmetry is broken, along with the value of $\tan \beta$, as described in Sec.~\ref{sec:ewsb-5d}.

\begin{figure}[t]
  \centering
  \includegraphics{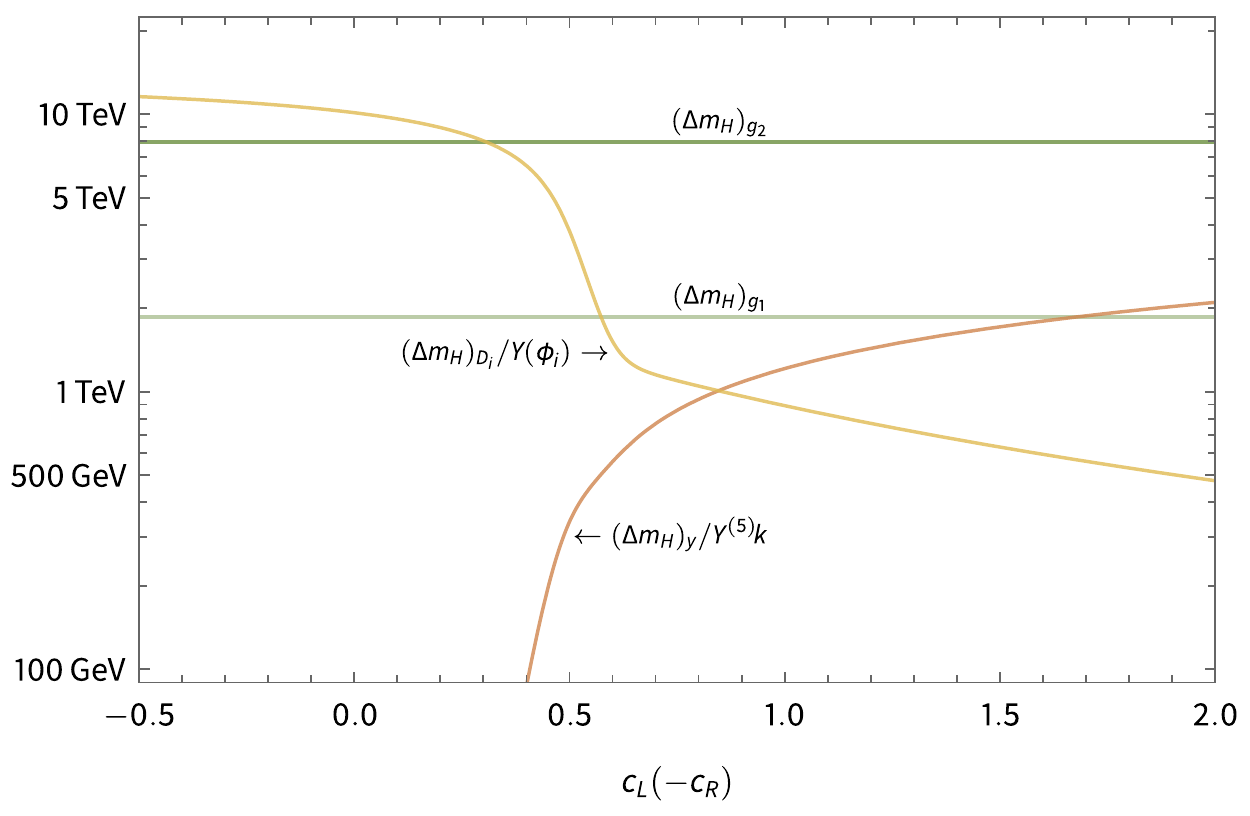}
  \caption{Plot of the magnitude of the one-loop radiative corrections that generate the Higgs soft masses as a function of hypermultiplet localization in the nonsinglet spurion case. We take $\Lambda_{\text{IR}} = 6.5 \times 10^6$~GeV, $\sqrt{F} = 2 \times 10^6$~GeV, and $\tan \beta = 5$.}
  \label{fig:loopcontributions-higgs-xx}
\end{figure}

In Figs.~\ref{fig:loopcontributions-higgs-x} and \ref{fig:loopcontributions-higgs-xx}, we plot the magnitudes of the various one-loop contributions to the Higgs soft masses as functions of hypermultiplet localization in the singlet spurion and nonsinglet spurion cases, respectively. The U(1) (lighter green) and SU(2) (darker green) gauge-sector contributions are independent of localization. In orange, we give the maximal contribution from a single Yukawa coupling (this corresponds to $c_L = \frac{1}{2}$ or $- c_R = \frac{1}{2}$ for the corresponding doublet and singlet hypermultiplets), neglecting all color multiplicity factors and modulo the 5D Yukawa coupling. These contributions are negative. In yellow is the $D$-term contribution from a single bulk scalar, modulo the scalar hypercharge. These individual contributions can be either positive or negative depending on the relative sign between the hypercharge of the Higgs field and the hypercharge of the scalar.

\subsubsection{Trilinear soft scalar couplings (\textit{a}-terms)}

Soft $a$-terms are also generated radiatively. We derive the one-loop corrections in Appendix~\ref{app:radiativecorrections-aterms}. Parametrized in terms of the tree-level gaugino masses, these take the forms:
\begin{subequations}
  \begin{align}
    16 \pi^2 a_{u_i}
      &= \begin{aligned}[t]
         - y_{u_i}
           \bigg(
      &      \frac{32}{3} (r^a_{\lambda_3})_{Q_i u_i} \, g_3^2 M_3
           + 6            (r^a_{\lambda_2})_{Q_i} \, g_2^2 \, M_2 \\[1ex]
      &      \qquad
           + \left[ \,
             - \frac{2}{5}  (r^a_{\lambda_1})_{Q_i}
             + \frac{8}{5}  (r^a_{\lambda_1})_{u_i}
             + \frac{8}{15} (r^a_{\lambda_1})_{Q_i u_i} \,
             \right] g_1^2 M_1
           \bigg) \, ,
         \end{aligned} \\[1ex]
    16 \pi^2 a_{d_i}
      &= \begin{aligned}[t]
         - y_{d_i}
           \bigg(
      &      \frac{32}{3} (r^a_{\lambda_3})_{Q_i d_i} \, g_3^2 M_3
           + 3            (r^a_{\lambda_2})_{Q_i} \, g_2^2 M_2 \\[1ex]
      &      \qquad
           + \left[ \,
               \frac{2}{5}  (r^a_{\lambda_1})_{Q_i}
             + \frac{4}{5}  (r^a_{\lambda_1})_{d_i}
             - \frac{4}{15} (r^a_{\lambda_1})_{Q_i d_i} \,
             \right] g_1^2 M_1
           \bigg) \, ,
         \end{aligned} \\[1ex]
    16 \pi^2 a_{e_i}
      &= - y_{e_i}
           \left(
             6 (r^a_{\lambda_2})_{L_i} \, g_2^2 M_2
           + \left[ \,
             - \frac{6}{5} (r^a_{\lambda_1})_{L_i}
             + \frac{12}{5} (r^a_{\lambda_1})_{e_i}
             + \frac{12}{5} (r^a_{\lambda_1})_{L_i e_i} \,
             \right] g_1^2 M_1
           \right) \, ,
  \end{align}
\end{subequations}
where the $r^a_{\lambda}$ are defined in \eqref{eq:softa-r-gaugino-1} and \eqref{eq:softa-r-gaugino-2}.

\subsection{Electroweak symmetry breaking}\label{sec:ewsb-5d}

In the MSSM, the tree-level scalar potential has a minimum breaking electroweak symmetry if the following two equations are satisfied:
\begin{subequations} \label{eq:ewsb-conditions}
  \begin{align}
    m^2_{H_u} + |\mu|^2 - b \cot \beta - \frac{1}{8} (g_1^2 + g_2^2) v^2 \cos 2 \beta &= 0 \, , \\[1ex]
    m^2_{H_d} + |\mu|^2 - b \tan \beta + \frac{1}{8} (g_1^2 + g_2^2) v^2 \cos 2 \beta &= 0\, .
  \end{align}
\end{subequations}
In our model, $m^2_{H_u}$, $m^2_{H_d}$, and $b$, given by \eqref{eq:softmhu2}, \eqref{eq:softmhd2}, and \eqref{eq:softb}, respectivelt, are radiatively generated at the IR-brane scale when the extra dimension is integrated out. Solving \eqref{eq:ewsb-conditions} determines two parameters: the magnitude of the Higgsino mass parameter $|\mu|$ and the ratio of Higgs VEVs $\tan \beta$. In the limit that the scale of supersymmetry breaking is much larger than the scale of electroweak symmetry breaking the physical solutions are
\begin{subequations} \label{eq:ewsb-parameters}
  \begin{align}
    \tan \beta &\simeq \frac{(m_{H_d}^2 - m_{H_u}^2) + \sqrt{(m_{H_d}^2 - m_{H_u}^2)^2 + 4 b^2}}{2 b} + \mathcal{O}\left( \frac{v^2}{b} \right) \, , \\[1ex]
    |\mu|^2 &\simeq \frac{m_{H_d}^2 - m_{H_u}^2 \tan^2 \beta}{\tan^2 \beta - 1} + \mathcal{O}(v^2) \, ,
  \end{align}
\end{subequations}
where we require $\operatorname{sign}(\mu) = {-1}$ and $m_{H_u}^2 < \operatorname{min} ( m_{H_d}^2, m_{H_d}^2 / \tan^2 \beta)$. Solutions with $\operatorname{sign}(\mu) = {+1}$ are excluded at tree level since $b$ is constrained to have the opposite sign from $\mu$ at the IR-brane scale according to \eqref{eq:softb} (and this typically remains true under RG evolution). Although these equations are modified when loop corrections to the Higgs scalar potential are included, they remain practical constraints since the iterative method employed to determine electroweak symmetry breaking (EWSB) in the numerical renormalization procedure (see Sec.~\ref{sec:numerical-spectrum-renormalization}) requires an initial tree-level solution. It is not guaranteed that the Higgs soft terms \eqref{eq:softmhu2}, \eqref{eq:softmhd2}, and \eqref{eq:softb} permit a tree-level solution at the IR-brane scale, in which case electroweak symmetry must be broken radiatively.

The conditions \eqref{eq:ewsb-parameters} for EWSB strongly favor $m_{H_u}^2 < 0$. In order for this to occur, the combined radiative corrections to $m_{H_u}^2$ both from the KK modes at the IR-brane scale and from the MSSM running to lower scales must be overall negative. We discuss the constraints this imposes on the parameter space of our theory in Sec.~\ref{sec:parameterspace-tachyon}.


\section{Superpartner Mass Spectrum}\label{sec:spectrum-5d}

\subsection{Model parameter space}\label{sec:parameterspace}

As we have seen in generating the fermion mass hierarchy and breaking supersymmetry, the parameter space available for our partially composite supersymmetric model is in general quite large. The overall mass scale of our sparticle spectrum is jointly determined by $\Lambda_{\text{IR}}$, the scale of the IR brane, and $\sqrt{F}$, the scale of supersymmetry breaking. Together these two parameters fill the roles associated with $M_{\text{GUT}}$, $m_{1/2}$, and $m_0$ in classic universal supergravity (SUGRA) models. As we discuss in Sec.~\ref{sec:ewsb-5d}, we do not have the usual freedom in $\tan \beta$ and the sign of $\mu$, which are in this case determined by electroweak symmetry breaking. In addition to these universal parameters, our model features nonuniversal IR-scale boundary conditions for the sfermion soft masses, which we specify in a flavor-dependent way by choosing field localizations to explain the SM fermion mass spectrum, as described in Sec.~\ref{sec:fermionhierarchy-5d}. We choose a universal value $Y^{(5)} k$ for all 5D Yukawa couplings, such that this specification requires fixing six additional free parameters, which we take as the doublet $c$ parameters $c_{L_i}$ and $c_{Q_i}$. In this section we discuss various phenomenological and theoretical constraints that impose limits on the set of model parameters.

\subsubsection{Phenomenological considerations}\label{sec:parameterspace-pheno}

We first consider five phenomenological desiderata that constrain our model.

\begin{enumerate}[label=\textbf{\arabic*.},labelindent=0pt,labelwidth=1em,leftmargin=!]

  \item \textbf{Gravitino dark matter}

  \noindent
  Since the gravitino mass is Planck-scale suppressed, it is the LSP throughout our parameter space. In the absence of $R$-parity violation, the LSP is absolutely stable, and as such, the gravitino makes an attractive dark matter candidate.
  However, the stability of a gravitino LSP can lead to cosmological problems, as the thermal gravitino mass density arising from freeze-out is sufficient to overclose the universe unless the gravitino is very light ($\mathcal{O}(100)$~eV)~\cite{Pagels:1981ke,Feng:2010ij}. In this case, observations of the matter power spectrum at small cosmological scales limit the free-streaming length of the gravitino, further requiring $m_{3/2} < 4.7$~eV in order for the gravitino to be adequately cold~\cite{Osato:2016ixc}, and gravitinos in this scenario cannot therefore account for all of the observed dark matter density. 
  
  Throughout the relevant parameter space of our model, the gravitino is sufficiently heavy that we require inflation to dilute the initial thermal population~\cite{Ellis:1982yb} and must restrict the reheating temperature so that the gravitino does not subsquently come back into thermal equilibrium. In this case, structure-formation constraints are weaker, requiring $m_{3/2} \gtrsim \mathcal{O}(1)$~keV for generic warm dark matter candidates~\cite{Viel:2005qj,Irsic:2017ixq,Yeche:2017upn}, such the gravitino may be the dominant component of dark matter. In this scenario, gravitinos may still be produced from the scattering of particles in thermal equilibrium with the plasma. The largest contribution arises from gluinos, such that the thermal gravitino density takes the form
  \begin{equation}
    \label{eq:thermal-gravitino-density}
    \Omega^{\text{thermal}}_{3/2} h^2 
      \sim 0.3 \left( \frac{100 \, \text{GeV}}{m_{3/2}} \right)
               \left( \frac{m_{\tilde{g}}}{1 \, \text{TeV}} \right)^2
               \left( \frac{T_R}{10^{10} \, \text{GeV}} \right) \, .
  \end{equation}
  For $m_{3/2} \lesssim 1 \, \text{keV}$, thermal scattering production of gravitinos cannot supply all of the observed dark matter density unless $T_R$ is high enough to bring the gravitino into thermal equilibrium. 

  \begin{figure}[t]
    \centering
    \includegraphics{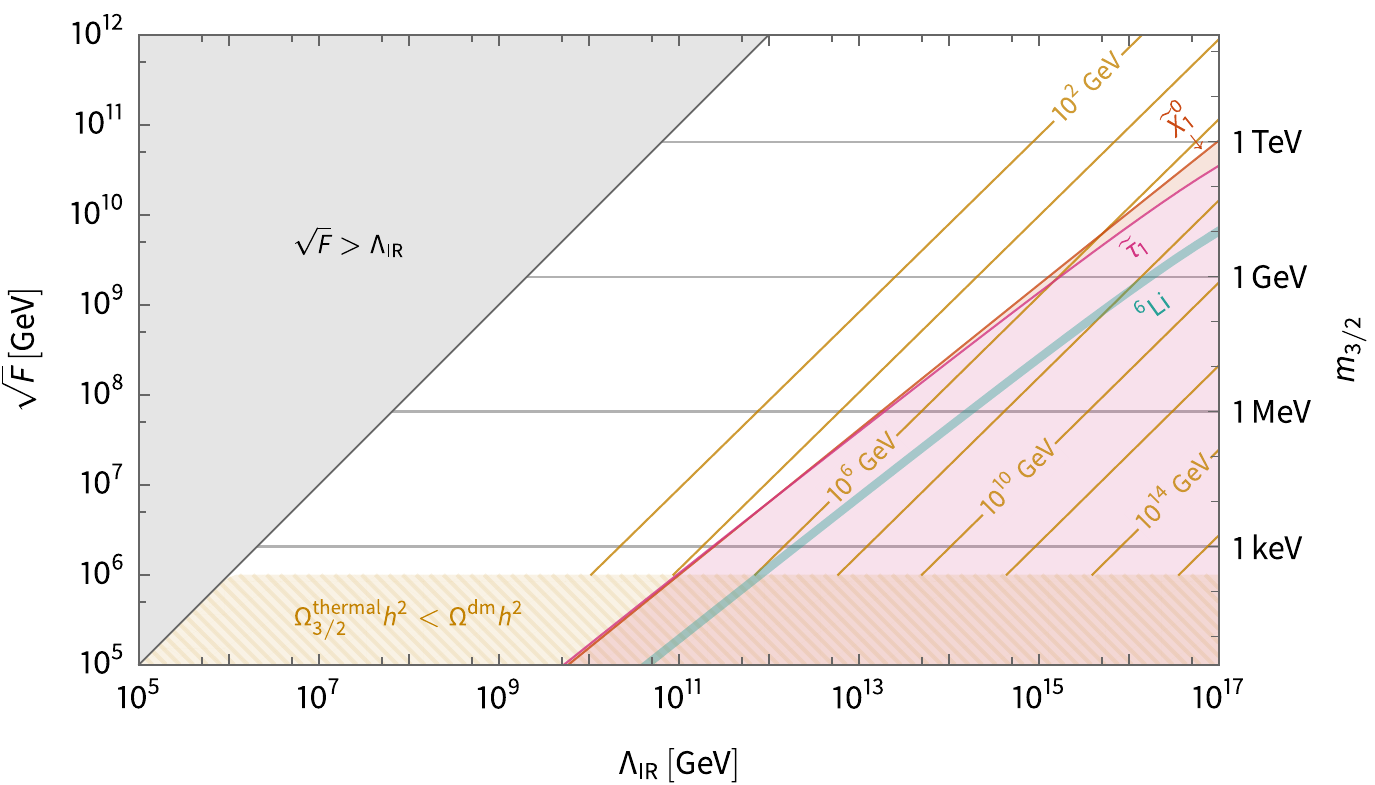}
    \caption{Plot of the gravitino dark matter constraints on the parameter space of our model in the ($\Lambda_{\text{IR}}, \sqrt{F}$) plane in the singlet spurion case for $Y^{(5)} k = 1$.}
    \label{fig:parameterspace-darkmatter-x}
  \end{figure}
  
  \begin{figure}[t]
    \centering
    \includegraphics{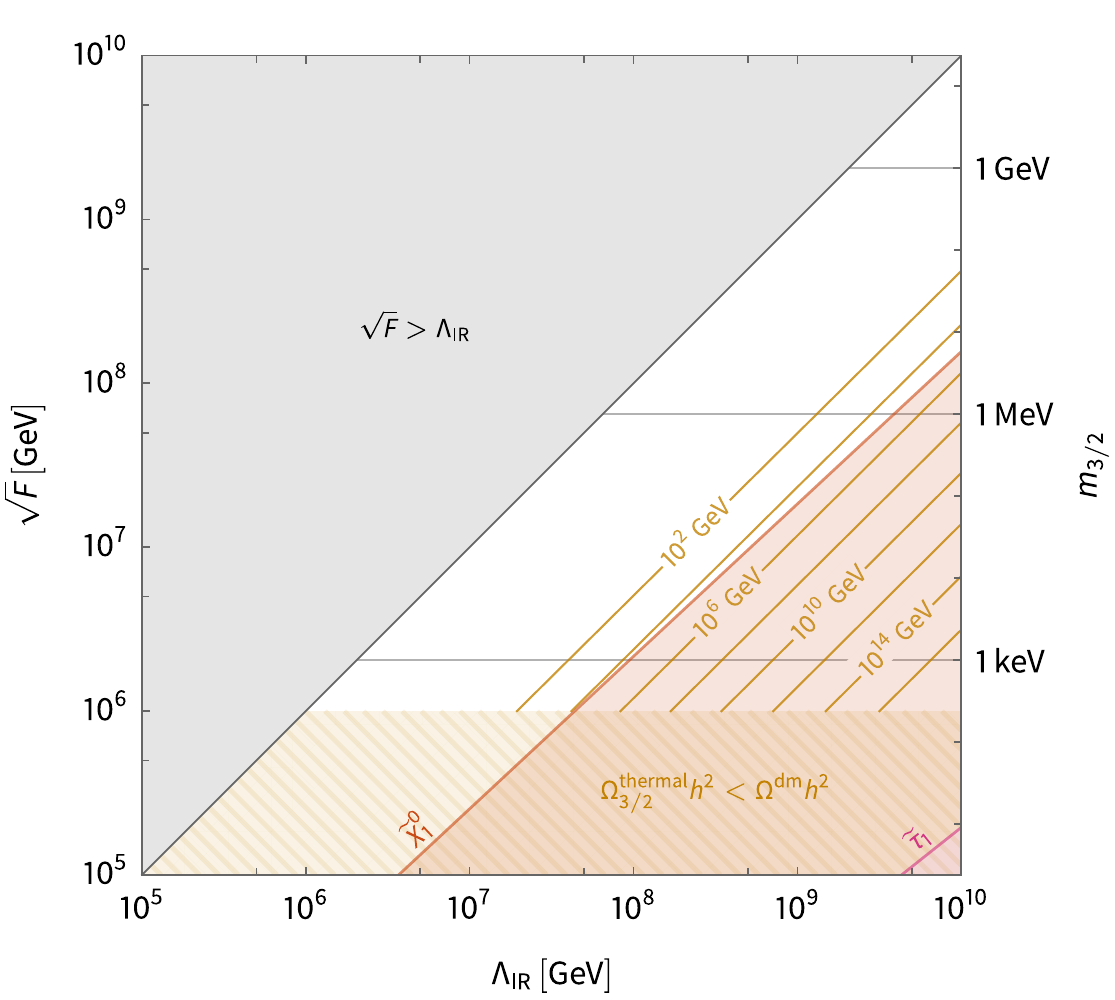}
    \caption{Plot of the gravitino dark matter constraints on the parameter space of our model in the ($\Lambda_{\text{IR}}, \sqrt{F}$) plane in the nonsinglet spurion case for $Y^{(5)} k = 1$.}
    \label{fig:parameterspace-darkmatter-xx}
  \end{figure}
    
  Gravitinos are also produced nonthermally, via decays of the next-to-lightest supersymmetric particle (NLSP), which in our model is typically either a (Bino-like or Higgsino-like) neutralino or a (mainly right-handed) stau. 
  In both cases, the NLSP is sufficiently long-lived throughout our parameter space to decay after freeze-out, such that the resulting nonthermal gravitino population takes the form (superWIMP scenario~\cite{Feng:2003xh}):
  \begin{equation}
    \label{eq:superwimp-gravitino-density}
    \Omega^{\text{nonthermal}}_{3/2} h^2 
      = \frac{m_{3/2}}{m_{\text{NLSP}}} \Omega_{\text{NLSP}} h^2 \, .
  \end{equation}
  The initial NLSP population that contributes to $\Omega_{\text{NLSP}} h^2$ is moderated by $T_R$. In particular, if the reheating temperature is low enough that the NLSP never comes into thermal equilibrium after inflation ($T_R \lesssim m_{\text{NLSP}} / 20$), the initial NLSP population is Boltzmann-suppressed.

  The observed dark matter abundance ($\Omega_{\text{dm}} h^2 = 0.1186 \pm 0.0020$~\cite{Tanabashi:2018oca}) thus places an upper limit on the reheating temperature. 
  We show an estimate of these limits in the space of $(\Lambda_{\text{IR}},\sqrt{F})$ in Figs.~\ref{fig:parameterspace-darkmatter-x} and \ref{fig:parameterspace-darkmatter-xx}. In each case, the yellow contours give the reheating temperature necessary for the thermal gravitino relic density \eqref{eq:thermal-gravitino-density} to provide the dominant component of dark matter. If nonthermal production is significant, the reheating temperature must be lowered to suppress the contribution from thermal production. 
  In the hatched yellow regions, the thermal relic density is insufficient to provide all of the dark matter, and some level of nonthermal production is required to obtain the observed dark matter density. 

  Further parameter space constraints arise from big bang nucleosynthesis (BBN), which strongly limits the energy density of the NLSP if it is long-lived enough to decay to the gravitino during or after the formation of the light elements. To avoid altering the successful predictions of the standard BBN scenario, decays to the gravitino must be prompt, limiting the lifetime of the NLSP to $\tau_{\text{NLSP}} < \mathcal{O}(0.1 \text{--} 100)$~s, or, for $m_{3/2} \ll m_{\text{NLSP}}$,
  \begin{equation}
    m_{\text{NLSP}} > \mathcal{O}(1\text{--}4)~\text{TeV} \left( \frac{m_{3/2}}{10~\text{GeV}} \right)^{2/5} \, ,
  \end{equation}
  as a conservative estimate.
  This condition places constraints in the space of $\Lambda_{\text{IR}}$ and $\sqrt{F}$. We show the regions where $\tau_{\text{NLSP}} < 0.1$~s in the neutralino case (dark orange) and in stau case (magenta) in Figs.~\ref{fig:parameterspace-darkmatter-x} and \ref{fig:parameterspace-darkmatter-xx}. Evading these limits restricts the reheating temperature for our model to the range $T_R \sim \mathcal{O}(10^2 \text{--} 10^6)$~GeV, necessitating an alternative to thermal leptogenesis to generate the baryon asymmetry.
 
  If the NLSP is the stau and is sufficiently stable to survive into BBN, the formation of bound states with nuclei can catalyze the production of light elements~\cite{Pospelov:2006sc}. In particular, if $10^3~\text{s} \lesssim \tau_{\tilde{\tau}_1} \lesssim 5 \times 10^3~\text{s}$, the catalytic enhancement for ${}^6$Li can solve the lithium problem in the standard BBN scenario~\cite{Pradler:2006hh}. We show an estimate of the region in which the stau lifetime falls within these limits in teal in Fig.~\ref{fig:parameterspace-darkmatter-x}. We note however, that this solution to the lithium problem is ruled out for our model, as the entirety of the catalytic region is excluded by current LHC limits (see Fig.~\ref{fig:parameterspace-x}). In the nonsinglet spurion case, the stau is not sufficiently long-lived to survive into BBN anywhere in the relevant parameter space.

  \item \textbf{The supersymmetric flavor problem}

  \noindent
  It is well known that the additional couplings and degrees of freedom introduced in supersymmetry can generate flavor-changing neutral currents (FCNCs) and $CP$-violation at levels above current experimental limits. This problem can be alleviated if the sfermions of the first and second generations are heavy, $\mathcal{O} (100)$~TeV~\cite{Dine:1990jd,Dimopoulos:1995mi,Pomarol:1995xc,Cohen:1996vb}, or even heavier~\cite{Gabbiani:1996hi,ArkaniHamed:1997ab,Moroi:2013sfa}, a solution that has gained popularity in the absence of any low-energy experimental signatures of supersymmetry. Such a solution is a natural and elegant choice in our model, as the inverted hierarchy in the sfermion soft mass spectrum naturally separates the scale of the first two generations from that of the third, which can remain light enough to explain the Higgs mass and offer the possibility of experimental detection.

  Thus, to ameliorate the flavor problem, we restrict the masses of the first- and second-generation  sfermions to be at least 100 TeV. As a constraint, this condition places an upper (lower) limit on the $c$ parameters of the first- and second-generation left-handed (right-handed) bulk hypermultiplet fields. This is a further limit on the hypermultiplet localizations beyond the structure necessary to explain the SM fermion mass spectrum. In Figs.~\ref{fig:parameterspace-x} and \ref{fig:parameterspace-xx}, we show an estimate of the region in the space of $(\Lambda_{\text{IR}}, \sqrt{F})$ where this limit is incompatible with the SM fermion mass for at least one field (the strictest constraint typically comes from the muon, the heaviest of the first- and second-generation fermions).

  \item \textbf{125 GeV Higgs mass and collider exclusion limits}

  \noindent
  In the MSSM, the tree-level mass of the neutral scalar Higgs boson is bounded from above by $m_h^{\text{tree}} < m_Z$. Radiative corrections, the largest arising from top and stop loops, can raise $m_h$ considerably, and accordingly, the observed value of $125.18 \pm 0.16$~GeV~\cite{Tanabashi:2018oca,Sirunyan:2017exp,Aad:2015zhl} offers an important constraint on the sparticle spectrum, particularly on the masses of the stops. In the MSSM, the observed Higgs mass constrains the general stop mass scale $\sqrt{\smash[b]{m_{\tilde{t}_1} m_{\tilde{t}_2}}}$ and is broadly compatible with stop masses ranging from $\mathcal{O}(1)$~TeV for $\tan \beta \sim 50$ to $\mathcal{O}(100)$~TeV for $\tan \beta \sim 3$~\cite{Giudice:2011cg}. In our model, the stop masses depend critically (at least $\mathcal{O}(1)$) on the localizations of the bulk hypermultiplets, and hence the Higgs mass principally induces constraints on the $c$ parameters of the theory. The precise calculation of the Higgs mass requires a complete numerical analysis (see Sec.~\ref{sec:numerical-spectrum}). To obtain a conservative evaluation of the constraint induced by the Higgs mass on our model we can exclude the region where the stop masses are always greater than 100 TeV. We show an estimate of this region in the space of $(\Lambda_{\text{IR}}, \sqrt{F})$ in Figs.~\ref{fig:parameterspace-x} and \ref{fig:parameterspace-xx}.

  Further limits on sparticle masses are set by experiments searching for direct sparticle production at colliders. In the context of our model, the strictest of constraints arise from the exclusion limits on the masses of the gluino and the squarks, which must be heavier than $\mathcal{O}(1)$~TeV when the LSP is the gravitino~\cite{Khachatryan:2016fll,Khachatryan:2016ojf,Sirunyan:2017yse,Aad:2015tin}.\footnote{These limits are weakly model-dependent: see the review~\cite{Tanabashi:2018oca}.} Qualitatively, these limits place an effective lower bound on the soft mass scale of our model, restricting the ratio $\sqrt{F}/\Lambda_{\text{IR}}$. We show an estimate of the excluded region in Figs.~\ref{fig:parameterspace-x} and \ref{fig:parameterspace-xx}.

  \item \textbf{Gauge coupling unification}

  \noindent
  Gauge coupling unification is a generic feature of the minimal supersymmetric model.
  The renormalization of gauge couplings depends on the number of degrees of freedom present in the theory at a given energy scale; in the MSSM, unification is most sensitive to the Higgsino mass $\mu$ as well as the ratio of the Wino mass to the gluino mass~\cite{Carena:1993ag}, and it can be spoiled if the magnitude of $\mu$ is larger than a few hundred TeV~\cite{Arvanitaki:2012ps}. As discussed in Sec.~\ref{sec:ewsb-5d}, $\mu$ is determined as necessary to achieve electroweak symmetry breaking. Generically, this implies that the scale of $\mu$ is of the same order of magnitude as the soft masses in the Higgs sector, i.e., $|\mu|^2 \sim |b| \sim |m^2_{H_u}| \sim |m^2_{H_d}|$. Since the Higgs soft masses are generated radiatively (and therefore characteristically of the scale of the gaugino masses) a first-order estimate of this constraint is $ |\mu| \sim M_2 \lesssim 100$~TeV. A more precise constraint can be obtained by solving the tree-level EWSB equations \eqref{eq:ewsb-conditions}. We show an estimate of the excluded region in the space of $(\Lambda_{\text{IR}},\sqrt{F})$ where $|\mu^{\text{tree}}| \gtrsim 100$~TeV in Figs.~\ref{fig:parameterspace-x} and \ref{fig:parameterspace-xx}.
  
  Note that for the case where $\Lambda_{\text{IR}}$ is much below ${\sim} 10^{16}$~GeV, we are implicitly assuming 
  that the gauge boson Kaluza-Klein states form complete SU(5) multiplets so that there is a universal shift in the running of the gauge couplings. In the warped extra dimension this can be modeled by considering the full SU(5) gauge symmetry in the bulk, although there are no Kaluza-Klein states for the UV-localized Higgsino. For simplicity, we will not consider the full SU(5) extension here, since it does not affect the details of our low-energy spectrum.\footnote{We note that in a theory with IR-localized bulk hypermultiplets, higher-dimension operators may only be suppressed by the IR-brane scale, which, in the context of grand unification, can lead to proton decay constraints \cite{Gherghetta:2000qt}. These may be addressed by the introduction of an additional global symmetry in the bulk, such as a U(1)$_B$ baryon number symmetry as in Refs.~\cite{Agashe:2004bm,Frigerio:2011zg}, or by an orbifold GUT scenario \cite{Goldberger:2002pc}.}

  \item \textbf{Minimal supersymmetric particle content}

  \noindent
  In the construction of our model, we are motivated to explain the observed Higgs mass using only the minimal supersymmetric particle content at low energy. While the orbifold compactification allows us to recover this particle content as the zero modes of the 5D $\mathcal{N} = 1$ supersymmetric theory, the essentially Dirac nature of fermions in five dimensions is a nontrivial feature of the model and can have phenomenological implications when the scale of $\mathcal{N} = 1$ supersymmetry breaking on the IR brane, $\sqrt{F}$, approaches the local compactification scale, $\Lambda_{\text{IR}}$. In this case, the backreaction of the supersymmetry-breaking boundary mass on the wavefunction profiles of the gaugino and sfermion fields cannot be neglected. In particular for the gauginos, the effect of larger $\sqrt{F}/\Lambda_{\text{IR}}$ is to increase the zero-mode mass, but at the same time to decrease the magnitude of the mass of the next-to-lightest KK mode, $m^{(1)}_{\lambda}$. This behavior smoothly approaches the twisted limit ($\sqrt{F}/\Lambda_{\text{IR}} \gg 1$), where the magnitudes of the masses of the lowest two gaugino KK modes meet at a common value and the two states form a Dirac spinor. For $\sqrt{F}/\Lambda_{\text{IR}} \lesssim  1$, the gaugino mass is only approximately Dirac, but the first KK mode is light enough that it must be included in the spectrum. While the presence of such modes in the theory at low energy can be helpful to achieve a more natural model, we leave the exploration of this region of parameter space for the future. Under this criterion, we exclude the region in which $m_{\lambda_1}^{(1)} < k e^{- \pi k R}$, shown in Figs.~\ref{fig:parameterspace-x} and \ref{fig:parameterspace-xx}.

 \end{enumerate}

\subsubsection{Charge- and color-breaking minima}\label{sec:parameterspace-tachyon}

One of the primary features of our model is the presence of significant hierarchies in the soft mass parameters---both within the sfermion sector and between the heavier sfermions and the gauginos---resulting from the structure imposed on the matter bulk mass hypermultiplets in order to explain the SM fermion mass spectrum. Although such hierarchies have desirable phenomenological features, they can also be the source of considerable constraints in the renormalization of the spectrum, as radiative corrections from heavier scalars may be large enough compared to the lighter scalar mass scale to destabilize the running masses. While, this may be favorable in the Higgs sector for electroweak symmetry breaking, for the sfermions it results in phenomenologically unacceptable charge- and color-breaking minima.

In our model, large negative radiative corrections to the scalars can arise both in the 5D bulk theory and in the effective 4D MSSM below the scale of compactification. As discussed in Secs.~\ref{sec:susybreaking-5d-sfermion} and \ref{sec:susybreaking-5d-higgs}, we calculate the bulk contributions as threshold corrections to the scalar soft masses squared at the IR brane scale. Due to the flavor structure imposed on the matter bulk hypermultiplets that explains the SM fermion mass hierarchy, the trace \eqref{eq:dterm-trace-ads} is generically nonzero.\footnote{This is ultimately a consequence of the asymmetry between the 4D Yukawa couplings $y_{u_i}$ and $y_{d_i}$, which, as mentioned in Sec.~\ref{sec:fermionhierarchy-5d}, precludes the solution $c_{Q_i} = - c_{u_i} = - c_{d_i}$. The trace $\Delta_{\mathcal{S}}$ can be tuned to zero, but this requires some additional intergenerational or interfamilial correlation.} If the spectrum also contains sufficiently IR-localized scalars, the bulk $D$-term corrections \eqref{eq:scalar-correction-dterm} may provide the dominant radiative contributions to the scalar masses. These corrections are negative (positive) for scalars with hypercharge of the same (opposite) sign as $\Delta_{\mathcal{S}}$. In order to avoid negative sfermion soft masses squared at the IR-brane scale in this case, the localizations of the matter bulk hypermultiplets must be correlated. The corresponding restrictions on the allowed $c$-parameter ranges have a distinctive structure that depends on hypercharge. In particular, upper (lower) limits arise on the $c$ parameters of left-handed (right-handed) sfermions with hypercharge of the same sign as the sign of $\Delta_{\mathcal{S}}$, which is determined by the heaviest (most IR-localized) sfermions.

Scalars in our theory also receive negative Yukawa corrections of the form \eqref{eq:scalar-correction-yukawa} from the bulk. As a result of the bulk hypermultiplet $c$-parameter structure necessary to explain the SM fermion mass hierarchy, the magnitude of the Yukawa contribution to a left-handed (right-handed) field grows as that field becomes more UV-localized (see Figs.~\ref{fig:loopcontributions-scalar-x} and \ref{fig:loopcontributions-scalar-xx}).
These corrections can become large, particularly for the third-generation fields, such that upper (lower) $c$-parameter limits for each left-handed (right-handed) field must be imposed in order to avoid any tachyonic masses. In some cases, the combined $D$-term and Yukawa limits for one or more of the third-generation fields may exclude all solutions compatible with the SM fermion mass spectrum.

Further contributions from heavy scalars arise in the MSSM running below the IR-brane scale. At the one-loop level, the $\beta$ function of each scalar soft mass squared $m_{\phi_i}^2$ includes a contribution from the tree-level Fayet-Iliopoulos (FI) $D$-term for weak hypercharge~\cite{Dimopoulos:1995mi,Cohen:1996vb}
\begin{equation} \label{eq:scalar-correction-dterm-mssm}
  16 \pi^2 (\beta_{m^2_{\phi_i}})_{\text{1-loop}}
    \supset \frac{6}{5} g_1^2 \, Y(\phi_i) \operatorname{Tr} \big[ \, Y(\phi_j) \, m_{\phi_j}^2 \, \big]
    \equiv  \frac{6}{5} g_1^2 \, Y(\phi_i) \, \mathcal{S} \, ,
\end{equation}
where $\mathcal{S}$ is the trace
\begin{equation} \label{eq:dterm-trace-mssm}
  \mathcal{S} = m^2_{H_u}
              - m^2_{H_d}
              + \operatorname{Tr}
                \left[ \,
                    \mathbf{m}^{\mathbf{2}}_{\tilde{\mathbf{Q}}}
                -   \mathbf{m}^{\mathbf{2}}_{\tilde{\mathbf{L}}}
                - 2 \mathbf{m}^{\mathbf{2}}_{\tilde{\mathbf{u}}}
                +   \mathbf{m}^{\mathbf{2}}_{\tilde{\mathbf{d}}}
                +   \mathbf{m}^{\mathbf{2}}_{\tilde{\mathbf{e}}} \,
                \right] \, .
\end{equation}
The scalar soft masses squared also receive negative contributions from scalars at the two-loop level. In the MSSM, the dominant contributions take the form
\begin{equation} \label{eq:scalar-correction-twoloop-mssm}
  (16\pi^2)^2 (\beta_{m^2_{\phi_i}})_{\text{2-loop}}
    \supset 4 \sum_{a} g_a^4 \, C_a(R_{\phi_i}) \, \sigma_a \, ,
\end{equation}
where
\begin{subequations}
  \begin{align}
    \sigma_1 &= \frac{1}{5}
                \left(
                  3 m^2_{H_u}
                + 3 m^2_{H_d}
                + \operatorname{Tr}
                   \left[ \,
                       \mathbf{m}^{\mathbf{2}}_{\tilde{\mathbf{Q}}}
                   + 3 \mathbf{m}^{\mathbf{2}}_{\tilde{\mathbf{L}}}
                   + 8 \mathbf{m}^{\mathbf{2}}_{\tilde{\mathbf{u}}}
                   + 2 \mathbf{m}^{\mathbf{2}}_{\tilde{\mathbf{d}}}
                   + 6 \mathbf{m}^{\mathbf{2}}_{\tilde{\mathbf{e}}} \,
                   \right]
                \right)\, , \\
    \sigma_2 &= m^2_{H_u}
              + m^2_{H_d}
              + \operatorname{Tr}
                \left[ \,
                  3 \mathbf{m}^{\mathbf{2}}_{\tilde{\mathbf{Q}}}
                +   \mathbf{m}^{\mathbf{2}}_{\tilde{\mathbf{L}}} \,
                \right] \, ,\\
    \sigma_3 &= \operatorname{Tr}
                \left[ \,
                  2 \mathbf{m}^{\mathbf{2}}_{\tilde{\mathbf{Q}}}
                +   \mathbf{m}^{\mathbf{2}}_{\tilde{\mathbf{u}}}
                +   \mathbf{m}^{\mathbf{2}}_{\tilde{\mathbf{d}}} \,
                \right] \, .
  \end{align}
\end{subequations}
These terms are loop-suppressed compared to \eqref{eq:scalar-correction-dterm-mssm}, but cannot be reduced by tuning the scalar masses to obtain cancellations between the various masses as can be done for \eqref{eq:scalar-correction-dterm-mssm}.\footnote{Note that if some symmetry or universality in the soft mass boundary conditions are assumed such that \eqref{eq:scalar-correction-dterm-mssm} is zero, it remains zero at all scales.}

In the context of the supersymmetric flavor problem and high-scale supersymmetry breaking, it was noted in Ref.~\cite{ArkaniHamed:1997ab} that the two-loop contributions from heavy scalars provide considerable tachyonic constraints on the allowable hierarchy among the scalar soft mass parameters unless the effect can be balanced by positive contributions from the gauginos. This analysis assumed a common mass for the heavy scalars. When this assumption is lifted, we note that the presence of nonuniversality among the soft scalar masses generically induces a nonzero value for the trace $\mathcal{S}$. The MSSM $D$-term corrections have the same hypercharge dependence as the $D$-term corrections from the bulk. Thus, the resulting contributions to scalars with hypercharge of sign opposite to that of the trace $\mathcal{S}$ are positive, and consequently may ameliorate the effect of the negative two-loop contributions. This comes, however, at the cost of increasing the limits for scalars with hypercharge of the same sign as that of $\mathcal{S}$, which receive negative corrections in the running. We provide an analytical estimate that illustrates this behavior in Appendix~\ref{app:tachyon-analytics}.

For our model, the MSSM corrections further restrict the accessible $c$-parameter ranges of the matter bulk hypermultiplets. Since the sign of $\mathcal{S}$ is typically the same as that of $\Delta_{\mathcal{S}}$, the MSSM limits generally reinforce the bulk limits, further restricting the viable IR-brane hierarchies for sfermions (particularly squarks) with hypercharge of the same sign as $\mathcal{S}$. Large hierarchies in the high-scale spectrum are only allowed for sfermions with hypercharge sign opposite to the sign of $\mathcal{S}$. For example, if $\mathcal{S} > 0$, then $\widetilde{u}_3$ (hypercharge $-\frac{2}{3}$) may receive large negative corrections, and therefore cannot have a mass at the IR-brane scale significantly lower than the scale of the other sfermions.
Such high-scale sfermion structure is a generic signature of a nonuniversal split sfermion spectrum.

A similar analysis can be performed in the Higgs sector, where, conversely, large negative corrections are typically favorable for electroweak symmetry breaking.
Negative corrections to the Higgs soft masses squared $m_{H_u}^2$ and $m_{H_d}^2$ may arise from both the bulk, in the form of Yukawa contributions \eqref{eq:higgs-correction-yukawa} and $D$-term contributions \eqref{eq:higgs-correction-dterm} at one loop, or from the MSSM, in the form of Yukawa corrections
\begin{subequations}
  \begin{align}
    16 \pi^2 (\beta_{m_{H_u}^2})_{\text{1-loop}}
      &\supset 6 |y_t|^2 \left( m_{H_u}^2 + m^2_{\tilde{Q}_3} + m^2_{\tilde{u}_3} \right)~, \\[1ex]
    16 \pi^2 (\beta_{m_{H_d}^2})_{\text{1-loop}}
      &\supset 6 |y_d|^2 \left( m_{H_d}^2 + m^2_{\tilde{Q}_3} + m^2_{\tilde{d}_3} \right)
             + 2 |y_{\tau}|^2 \left( m_{H_d}^2 + m^2_{\tilde{L}_3} + m^2_{\tilde{e}_3} \right) \,,
  \end{align}
\end{subequations}
and $D$-term corrections \eqref{eq:scalar-correction-dterm-mssm} at one loop and the corrections \eqref{eq:scalar-correction-twoloop-mssm} at two loops.
As discussed in Sec.~\ref{sec:ewsb-5d}, EWSB in our model requires $m_{H_u}^2 < \operatorname{min} (m_{H_d}^2, m_{H_d}^2 / \tan^2 \beta)$. If the sfermion hierarchy is relatively modest, such that at least one of $m^2_{\tilde{Q}_3}$ or $m^2_{\tilde{u}_3}$ is relatively heavy, this may be achieved in the familiar way in the MSSM through Yukawa radiative corrections. If, however, the sfermion splitting is large, the MSSM Yukawa contributions may be suppressed, and successful EWSB may rely on the presence of $D$-term contributions to destabilize the Higgs VEV, setting a lower limit on the net correction ($\Delta_{\mathcal{S}}$ and $\mathcal{S}$ together). The predicted Higgs boson mass in this case is also correlated with the $D$-term corrections, and consistency with the observation may require additional limits on $\Delta_{\mathcal{S}}$ and $\mathcal{S}$. Together, these requirements introduce a global constraint on the $c$ parameters of the heavy (typically, first- and second-generation) sfermions which give the dominant contributions to $\Delta_{\mathcal{S}}$ and $\mathcal{S}$ and can also induce additional tachyonic limits on the matter bulk hypermultiplet $c$ parameters, primarily constraining the families $Q$, $d$, and $e$.  

In some regions of the parameter space of our model, the union of all limits imposed on the bulk hypermultiplet $c$ parameters by these effects excludes all solutions compatible with the SM fermion mass spectrum.When the gaugino mass is given by \eqref{eq:zeromodemass-gaugino-tree-x} (singlet spurion), the splitting between the masses of the third-generation sfermions and the heavier mass scale of the first and second generations that results from the explanation of the SM fermion masses is small enough throughout the parameter space that tachyonic constraints are not significant, but when the gaugino mass is given by \eqref{eq:zeromodemass-gaugino-tree-xx} (nonsinglet spurion), larger hierarchies arise, excluding some areas of the parameter space. We show an estimate of the excluded region in the space of $(\Lambda_{\text{IR}}, \sqrt{F})$ in Fig.~\ref{fig:parameterspace-xx}.

Hierarchies in the sfermion mass spectrum (and separations of scale in general) also complicate the numerical renormalization procedure, since they necessitate a careful account of particle decoupling if precision in mass spectrum calculations is to be obtained. In mass-independent renormalization schemes such as $\xoverline[1]{\text{DR}}$, the effects of heavy particles do not decouple, and hence, at renormalization scales small compared to the particle masses, finite quantum corrections may involve terms with large logarithms of the masses of these particles (see~\cite{Pierce:1996zz} for MSSM expressions). In order for mass calculations to be precise when large hierarchies in the soft mass parameters are present, such large logarithmic corrections need to be resummed, a process which is most naturally accomplished by the use of an effective theory, or of a tower of effective theories~\cite{Tamarit:2012ry}. Precision in the case of scalar hierarchies is especially critical, since the light scalar masses depend crucially on the heavy scalar masses through the factors such as \eqref{eq:scalar-correction-dterm-mssm} and \eqref{eq:scalar-correction-twoloop-mssm}. It is important to note as well that the scale of supersymmetry breaking in our model, the IR brane, which can be significantly lower than the Planck scale or the GUT scale due to the warped 5D geometry, is the natural cutoff for the IR-localized (or composite) part of the 4D MSSM. Thus, the effects of the heavy scalars, which may receive masses very near (or even above, depending on the choice of $F$ and $k$) the IR-brane scale, may decouple after a little running, minimizing the effect of the heavy scalar contributions. In an effective field theory (EFT) approach, these heavy scalars are integrated out, introducing threshold corrections to the lighter scalar masses. At the one- and two-loop level, such corrections can be large and negative as a result of the effects mentioned above, but overall, the decoupling procedure may substantially relax the tachyonic bounds indicated in purely MSSM $\xoverline[1]{\text{DR}}$ renormalization~\cite{Tamarit:2012ie,Tamarit:2012yg}. Our renormalization procedure, discussed in Sec.~\ref{sec:numerical-spectrum-renormalization} does not implement decoupling; hence, we expect that some regions of our parameter space with light scalars may be unnecessarily excluded on tachyonic grounds, solely as an artifact of the numerical renormalization method.

\subsubsection{Parameter space constraints}\label{sec:parameterspace-constraints}

\begin{figure}[t]
  \centering
  \includegraphics{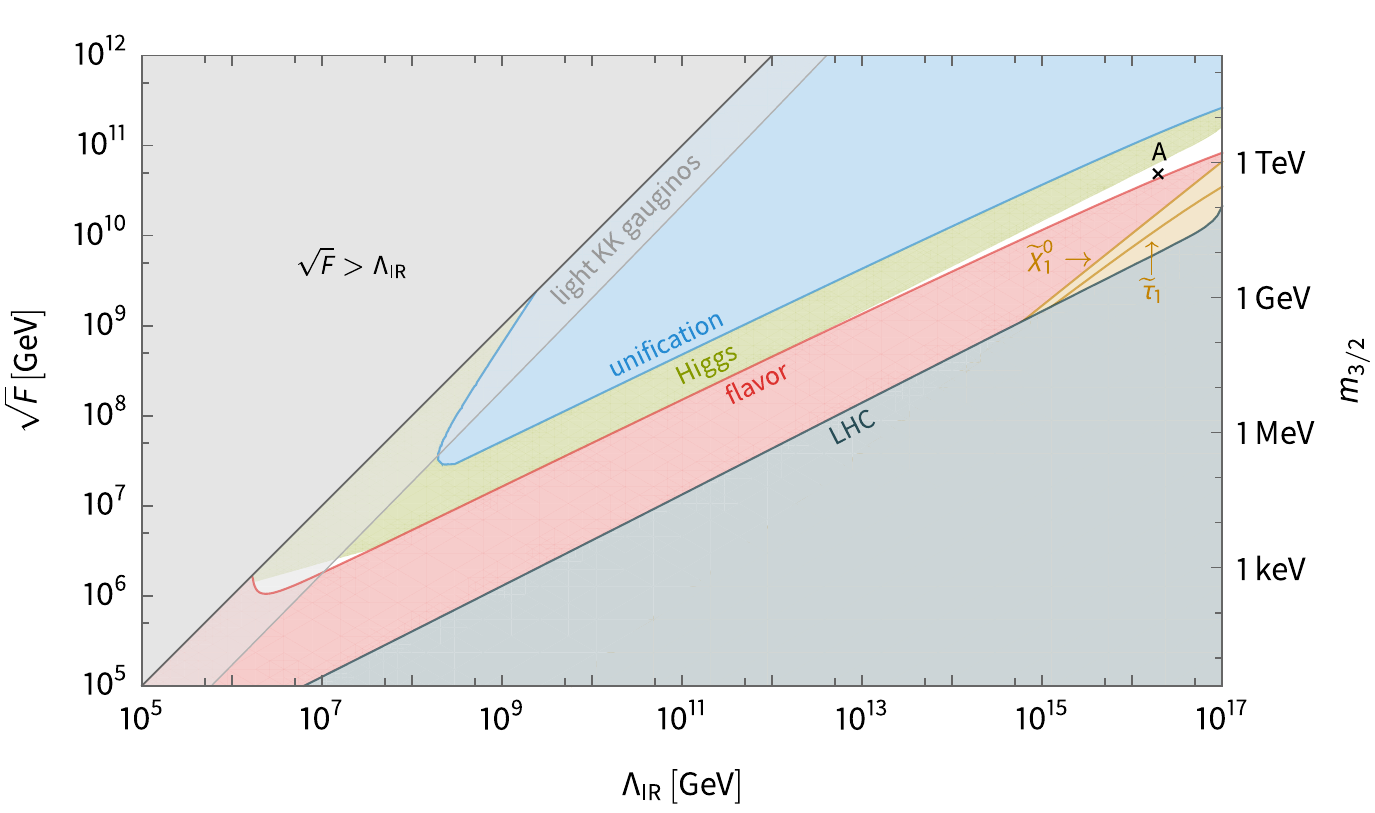}
  \caption{Plot of the constraints on the parameter space of our model in the ($\Lambda_{\text{IR}}, \sqrt{F}$) plane in the singlet spurion case for $Y^{(5)} k = 1$.}
  \label{fig:parameterspace-x}
\end{figure}

\begin{figure}[t]
  \centering
  \includegraphics{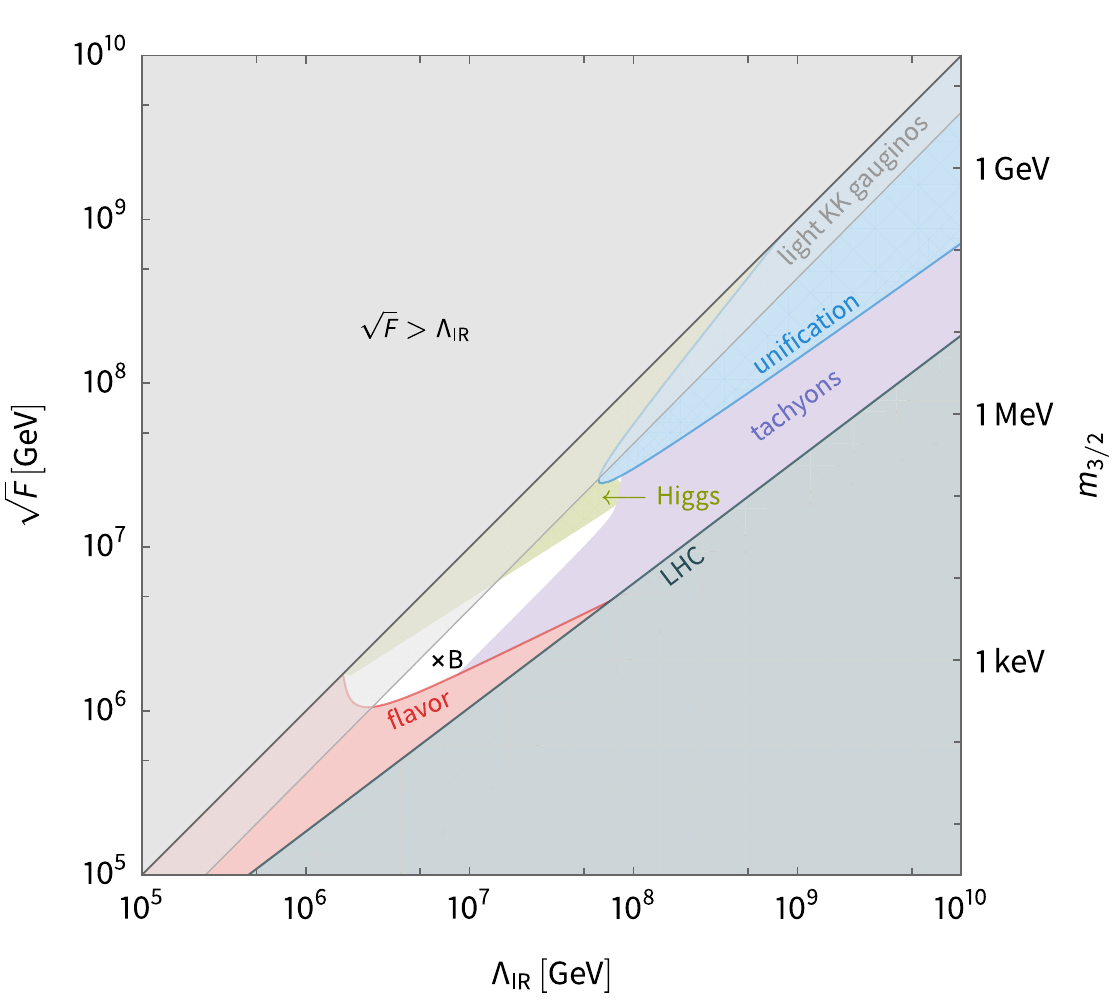}
  \caption{Plot of the constraints on the parameter space of our model in the ($\Lambda_{\text{IR}}, \sqrt{F}$) plane in the nonsinglet spurion case for $Y^{(5)} k = 1$.}
  \label{fig:parameterspace-xx}
\end{figure}

Together, the constraints discussed in Secs.~\ref{sec:parameterspace-pheno} and \ref{sec:parameterspace-tachyon} lead to restrictions on the $(\Lambda_{\text{IR}},\sqrt{F})$ parameter space, which is shown in Figs.~\ref{fig:parameterspace-x} and \ref{fig:parameterspace-xx} for the singlet spurion and nonsinglet spurion cases, respectively.

In the light gray region of each plot $\sqrt{F} > \Lambda_{\text{IR}}$, which is excluded as the dynamics of the spurion are restricted by $\Lambda_{\text{IR}}$ as a cutoff scale.
Along the edges of these regions, the shaded strip gives the area in which the next-to-lightest gaugino KK mode must be included in the low-energy spectrum.
The dark gray/blue regions give an estimate of the exclusions due to collider direct-detection limits.
The yellow region of Fig.~\ref{fig:parameterspace-x} shows our estimate of the BBN exclusions when the NLSP is the lightest neutralino or the lightest stau (also see Fig.~\ref{fig:parameterspace-darkmatter-x}). The corresponding limits in the nonsinglet spurion case (see Fig.~\ref{fig:parameterspace-darkmatter-xx}) are less strict than the collider constraint, and therefore not visible in Fig.~\ref{fig:parameterspace-xx}.
In blue are the regions in which $|\mu| > 100$~TeV, which is excluded in order to preserve gauge coupling unification.
In the red regions, the bulk hypermultiplet localization cannot be chosen such that all of the first- and second-generation sfermions have masses that are at least 100 TeV.
In green are the regions where we estimate that the stop masses are heavier than 100 TeV, and hence we expect the resulting Higgs boson mass to be too heavy to match the observed value.
In the purple region of Fig.~\ref{fig:parameterspace-xx}, one or more of the sfermions receives a tachyonic mass, either on the IR brane, or in the subsequent MSSM running.
The remaining white areas are the regions of interest, simultaneously satisfying all constraints. Within these regions, the flavor constraint, the observed Higgs boson mass, and radiative corrections impose additional restrictions on the hypermultiplet $c$ parameters.

The constraints favor two regions: either $\Lambda_{\text{IR}} \sim10^7$~GeV, with a keV-scale gravitino and a singlet spurion, or a GUT-scale value for $\Lambda_{\text{IR}}$, with an ${\sim}500$~GeV gravitino and a nonsinglet spurion. In Sec.~\ref{sec:numerical-spectrum}, we calculate detailed sparticle spectra for two benchmark scenarios (marked as A and B in Figs.~\ref{fig:parameterspace-x} and \ref{fig:parameterspace-xx}, respectively).

\subsection{Numerical results} \label{sec:numerical-spectrum}

Based on the constraints considered in Sec.~\ref{sec:parameterspace}, we select the regions of parameter space given in Table~\ref{tab:parameterpoints} as our benchmark scenarios. With these parameters we determine the sparticle mass spectrum and Higgs boson mass predicted by the partially composite supersymmetric model. The IR brane scale, $\Lambda_{\text{IR}}$, and the scale of supersymmetry breaking, $\sqrt{F}$, set the overall soft mass scale, and are chosen to comply with all phenomenological constraints in Sec.~\ref{sec:parameterspace-pheno}. $\tan \beta$ is determined by the measured Higgs boson mass and the sign of $\mu$ is set in order to achieve the correct pattern of EWSB. 

\begin{table}[t]
  \centering
  \caption{Selected parameter space sampling regions.}
  \begin{minipage}{0.55\textwidth}
    \renewcommand{\thefootnote}{\alph{footnote}}
    \centering
    \begin{tabular}{c c c}
      \toprule
                                    & A                         & B \\
      \midrule
      $\Lambda_{\text{IR}}$        & $2 \times 10^{16}$~GeV    & $6.5 \times 10^6$~GeV \\
      $\sqrt{F}$                   & $4.75 \times 10^{10}$~GeV & $2 \times 10^6$~GeV \\
      $\tan \beta$\footnotemark[1] & ${\sim} 3$                & ${\sim} 5$ \\
      $\operatorname{sign} \mu$    & $-1$                      & $-1$ \\
      $Y^{(5)} k$                  & $1$                       & $1$ \\
      \midrule
      spurion                      & singlet                   & nonsinglet \\
      $M_1$\footnotemark[1]        & $52.9$~TeV                & $14.60$~TeV \\
      $M_2$\footnotemark[1]        & $50.7$~TeV                & $22.9$~TeV \\
      $M_3$\footnotemark[1]        & $49.85$~TeV               & $38.94$~TeV \\
      \midrule
      $m_{3/2}$                    & $535$~GeV                 & $1$~keV \\
      \bottomrule
    \end{tabular}
    \footnotetext[1]{At scale $\Lambda_{\text{IR}}$.}
  \end{minipage}
  \label{tab:parameterpoints}
\end{table}

\subsubsection{Renormalization procedure}\label{sec:numerical-spectrum-renormalization}

To obtain pole mass predictions for the superpartners, we use the spectrum calculator
\textsc{FlexibleSUSY}~\cite{Athron:2014yba,Athron:2017fvs},
which incorporates elements of
\textsc{sarah}~\cite{Staub:2009bi,Staub:2010jh,Staub:2012pb,Staub:2013tta}
and
\textsc{softsusy}~\cite{Allanach:2001kg,Allanach:2013kza},
to run selected points down from the input scale (IR brane) to lower energy. To solve the renormalization boundary value problem, \textsc{FlexibleSUSY} employs a nested iterative algorithm, using the three-loop MSSM $\beta$ functions (the renormalization procedure includes components and corrections from~\cite{Degrassi:2001yf,Brignole:2001jy,Dedes:2002dy,Brignole:2002bz,Dedes:2003km,Jack:2003sx,Jack:2004ch}) between boundary conditions imposed at the high scale, $\Lambda_{\text{IR}}$, and the SM at the electroweak scale. Electroweak symmetry breaking is determined by numerical minimization of the loop-corrected Higgs potential, with the value of $\tan \beta$ and the Higgsino mass parameter $\mu$ determined iteratively. Loop-corrected pole masses are calculated from the full self-energies for each particle.

The renormalization procedure is made more complicated by the fact that the soft mass spectrum at the IR-brane scale depends on the values of the supersymmetric parameters (the gauge and Yukawa couplings and the Higgsino mass parameter) at that scale. As discussed in Sec.~\ref{sec:fermionhierarchy-5d}, the value of the 4D Yukawa couplings at the IR-brane scale must be known in order to choose the localizations of the bulk hypermultiplets to explain the fermion mass hierarchy; additionally, the radiative corrections included for the soft masses at the IR-brane scale explicitly incorporate gauge and Yukawa couplings as well as the Higgsino mass parameter. The IR-brane values of these parameters in turn depend (weakly) on the resulting sparticle pole masses. In order to consistently determine the sparticle mass spectrum, we must therefore apply the renormalization procedure iteratively.

\begin{figure}[t]
  \centering
  \includegraphics{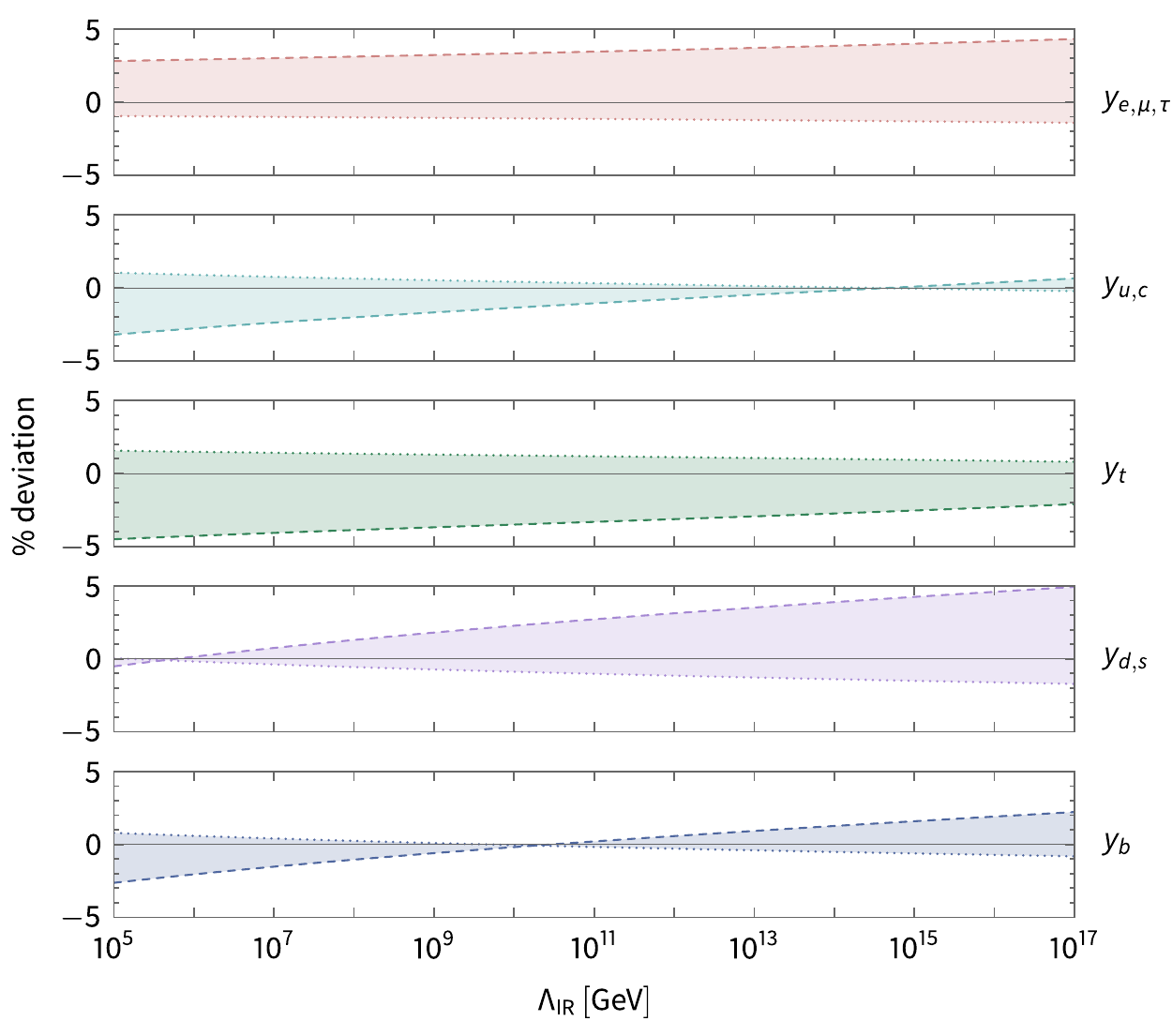}
  \caption{Plot of the percent deviation of the Yukawa couplings in the $\xoverline[1]{\text{DR}}$ scheme for $m_{\text{SUSY}} = 5$~TeV (dotted) and $m_{\text{SUSY}} = 100$~TeV (dashed) from the value for $m_{\text{SUSY}} = 10$~TeV, with $\tan \beta = 5$.}
  \label{fig:yukawa-uncertainty}
\end{figure}

To do this, we first obtain an initial estimate of the high-scale theory using \textsc{FlexibleSUSY} to extract the Yukawa and gauge couplings from low-energy experimental data and run them in the SM (including components and corrections from Refs.~\cite{Chetyrkin:1999qi,Melnikov:2000qh,Chetyrkin:2000yt,Degrassi:2012ry,Martin:2014cxa,Martin:2015eia,Bednyakov:2013eba,Buttazzo:2013uya,Vega:2015fna,Chetyrkin:2016ruf,Bednyakov:2015ooa}) up to a common supersymmetry scale, $m_{\text{SUSY}}$, where they are matched at tree level to the MSSM $\xoverline[1]{\text{DR}}$ couplings. The evolution is then continued under the two-loop MSSM RGEs (extracted from \textsc{sarah}) up to $\Lambda_{\text{IR}}$. The IR-scale couplings calculated using this procedure are insensitive to the details of the sparticle spectrum, since the tree-level matching procedure neglects threshold corrections at the scale $m_{\text{SUSY}}$, where the MSSM is matched to the SM. We accordingly use this procedure to provide coupling estimates in cases where we wish to make a general calculation without reference to a particular spectrum. The uncertainty resulting from the tree-level matching approximation can be quantified by varying $m_{\text{SUSY}}$ and observing the variation in the renormalized couplings. This is illustrated in Fig.~\ref{fig:yukawa-uncertainty}, where we plot the percent deviation of the Yukawa couplings under this renormalization procedure, relative to $m_{\text{SUSY}} = 10$~TeV. By this measure, the uncertainty in this Yukawa coupling estimate is less than 6\% throughout our parameter space.

To obtain higher precision coupling values, a particular spectrum must be specified by choosing localizations for the doublet bulk hypermultiplets. Given these, the IR-brane couplings calculated as above can be combined with educated guesses for $\tan \beta$ and $\mu$ to obtain an initial estimate of the IR-brane scale soft mass spectrum. This spectrum is then renormalized and EWSB computed using the full \textsc{FlexibleSUSY} MSSM routine. Unlike our spectrum-agnostic procedure above, this renormalization and matching procedure includes threshold effects from the soft masses, and the running values of the resulting supersymmetric parameters are spectrum-dependent. The values of the parameters extracted at the IR-brane scale are then used to construct an updated input spectrum and the procedure is repeated until the values of the input parameters converge.

\subsubsection{Higgs mass calculation}\label{sec:numerical-spectrum-higgsmass}

Due to the high scale of supersymmetry in our model, to calculate the neutral scalar Higgs pole mass we use
\textsc{hssusy}~\cite{Bagnaschi:2017xid,Athron:2014yba,Athron:2017fvs,Staub:2009bi,Staub:2010jh,Staub:2012pb,Staub:2013tta,Allanach:2001kg,Allanach:2013kza,Chetyrkin:1999qi,Melnikov:2000qh,Chetyrkin:2000yt,Degrassi:2012ry,Martin:2014cxa,Martin:2015eia,Bednyakov:2013eba,Buttazzo:2013uya,Vega:2015fna,Chetyrkin:2016ruf,Bednyakov:2015ooa}
and
\textsc{FlexibleEFTHiggs}~\cite{Athron:2016fuq,Athron:2014yba,Athron:2017fvs,Staub:2009bi,Staub:2010jh,Staub:2012pb,Staub:2013tta,Allanach:2001kg,Allanach:2013kza,Chetyrkin:1999qi,Melnikov:2000qh,Chetyrkin:2000yt,Degrassi:2012ry,Martin:2014cxa,Degrassi:2001yf,Brignole:2001jy,Dedes:2002dy,Brignole:2002bz,Dedes:2003km,Martin:2015eia,Bednyakov:2013eba,Buttazzo:2013uya,Vega:2015fna,Chetyrkin:2016ruf,Bednyakov:2015ooa}.
Both numerical methods are based on an effective field theory approach in which all non-SM particles are integrated out at a common threshold (namely, $m_{\text{SUSY}} = \sqrt{\smash[b]{m_{\tilde{t}_1} m_{\tilde{t}_2}}}$), at which point the theory is matched to the SM and run down to the electroweak scale, where the SM couplings are matched to experimental data and the Higgs pole mass is extracted. Theoretical uncertainty in this type of procedure arises in three areas:

\begin{enumerate}[label=\textbf{\arabic*.},labelindent=0pt,labelwidth=1em,leftmargin=!]

  \item \textbf{SM uncertainty:} \emph{uncertainty due to neglected higher-order corrections at the electroweak scale.}

  \noindent
  We consider two sources. The first source is missing corrections in the extraction of the SM running parameters from experimental data at the electroweak scale, which induce uncertainty that can be estimated as the effect of higher loop corrections on the Higgs pole mass. Here, we estimate the uncertainty by comparing the Higgs pole mass when 3-loop QCD corrections to the top Yukawa coupling are included the to the result for 2-loop top Yukawa coupling corrections. The second source is missing corrections in the calculation of the Higgs pole mass itself. Since such missing corrections lead to residual renormalization-scale dependence in the pole mass, we estimate this uncertainty by varying the scale at which the pole mass is calculated over the range $(\frac{1}{2} m_t, 2 m_t)$ and taking the difference between the maximum and minimum. The total SM uncertainty is the linear sum of these two contributions.

  \item \textbf{SUSY uncertainty:} \emph{uncertainty due to neglected threshold corrections in the matching of the SM to the MSSM.}

  \noindent
  In the EFT approach, the Higgs pole mass is primarily sensitive to new physics via threshold corrections from supersymmetric particles to the Higgs quartic coupling, at the scale where the Standard Model is matched to the MSSM. The matching in \textsc{hssusy} and \textsc{FlexibleEFTHiggs} is complete up through the 2-loop level, and includes some 3-loop corrections. The uncertainty due to the neglected higher-order corrections to the Higgs quartic coupling can be estimated by the residual matching-scale dependence in the Higgs pole mass. To calculate this, we vary the scale at which the SM is matched to the MSSM over the interval $(\frac{1}{2} m_{\text{SUSY}}, 2 m_{\text{SUSY}})$ and take the difference between the maximum and minimum value of the Higgs pole mass.

  \item \textbf{EFT uncertainty:} \emph{uncertainty due to neglected higher-dimensional operators in the SM EFT below the matching scale.}

  \noindent
  Both \textsc{hssusy} and \textsc{FlexibleEFTHiggs} only include terms up to order $\mathcal{O}(v/m_{\text{SUSY}})$ in the SM EFT. In a pure EFT approach, the uncertainty due to the missing terms of order $\mathcal{O}(v^2/m_{\text{SUSY}}^2)$ and higher can be estimated as a shift in the SUSY threshold corrections to the Higgs quartic coupling. Accordingly, in \textsc{hssusy}, we calculate this uncertainty as the shift in the Higgs pole mass induced by multiplying all 1-loop threshold corrections to the Higgs quartic coupling at the scale $m_{\text{SUSY}} $ by $(1 + v^2/m_{\text{SUSY}}^2)$. In \textsc{FlexibleEFTHiggs}, conversely, this uncertainty is not present, as the calculation departs from the pure EFT approach at low energy by switching to a diagrammatic calculation, which correctly resums leading and subleading logarithms to all orders.

\end{enumerate}
The total uncertainty is taken to be the linear sum of the SM, SUSY, and EFT uncertainties. As our Higgs mass estimate, we take the average of the \textsc{hssusy} and \textsc{FlexibleEFTHiggs} results, with uncertainty given by the union of the two calculated ranges.

\subsubsection{Superpartner mass spectrum}\label{sec:numerical-spectrum-polemasses}

To explore the parameter space of the benchmark scenarios given in Table~\ref{tab:parameterpoints}, we randomly sample over the estimated ranges of doublet $c$ parameters ($c_{{L_i},{Q_i}}$) that are consistent with all phenomenological constraints. The allowed ranges are principally determined by the FCNC constraints on the first- and second-generation sfermions and by the Higgs mass constraints on the third-generation squarks. EWSB and the large $D$-term radiative corrections discussed in Sec.~\ref{sec:parameterspace-tachyon} impose further limits on the $c$ parameters that can only be determined a posteriori in the numerical renormalization. These constraints can further limit the $c$-parameter ranges and introduce correlations among the $c$ parameters of successful spectra. Once the doublet $c$ parameters are specified, the singlet $c$ parameters are fixed according to \eqref{eq:yukawa-4d} to generate the SM fermion mass spectrum. In order to avoid introducing any new hierarchies with this mechanism, we additionally require that all the $c$ parameters are order-one numbers. In practice, we (generously) require $\pm c \lesssim 10$, providing effective upper and lower limits on all sfermion masses that hold in the absence of stronger constraints. 

\begin{figure}[t]
  \centering
  \includegraphics{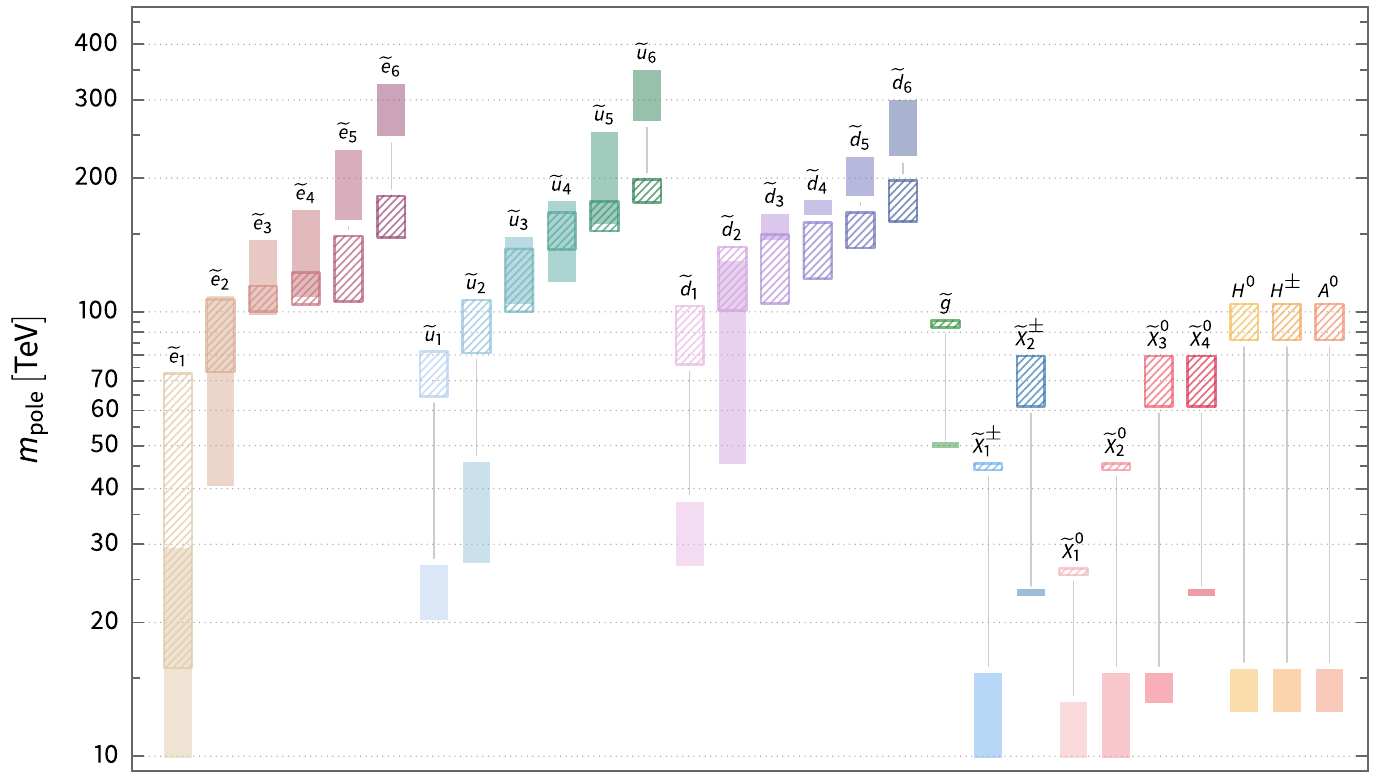}
  \caption{Predicted superpartner pole mass spectra for benchmark scenarios A (hatched) and B (solid) given in Table~\ref{tab:parameterpoints}.}
  \label{fig:spectrum-schematic-numeric}
\end{figure}

\begin{figure}[t]
  \centering
  \includegraphics{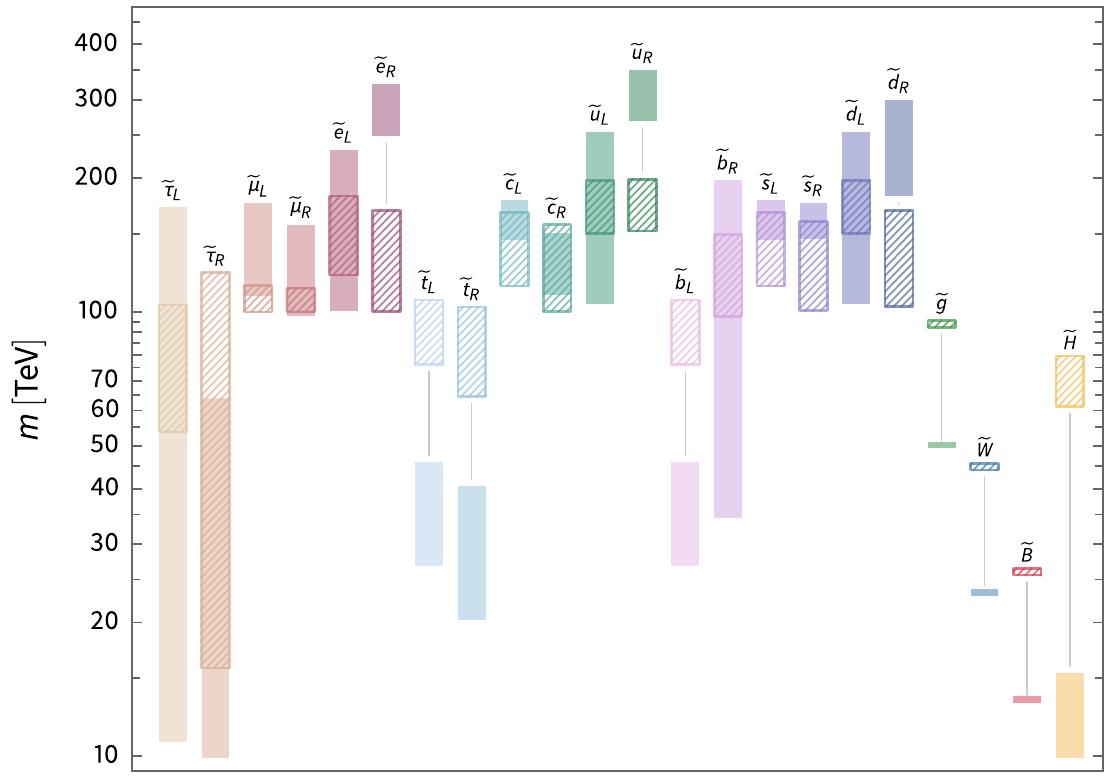}
  \caption{Predicted sfermion mass spectra in the gauge-eigenstate basis for benchmark scenarios A (hatched) and B (solid) given in Table~\ref{tab:parameterpoints}.}
  \label{fig:spectrum-schematic-numeric-interaction}
\end{figure}

In Fig.~\ref{fig:spectrum-schematic-numeric} we present the resulting superpartner pole mass spectra obeying all phenomenological constraints and consistent with the measured value of the Higgs boson mass (see Sec.~\ref{sec:numerical-spectrum-higgsmass} for details of the Higgs mass calculation). The corresponding ranges for the sfermion, gaugino, and Higgsino masses in the gauge-eigenstate basis are shown in Fig.~\ref{fig:spectrum-schematic-numeric-interaction}. In general, the spread in the masses is a result of the freedom in the bulk hypermultiplet localizations ($c$ parameters) remaining after the application of all constraints, combined with the uncertainty in the numerical calculations.

In both cases, the allowed mass ranges for the third-generation sfermions are relatively unconstrained on phenomenological grounds, and their limits are principally determined by the restriction of $c$-parameters to order-one numbers. In particular, we note that for the stops (below 100 TeV, these can be identified unequivocally with $\widetilde{u}_{1,2}$), the general mass scale ($\sqrt{\smash[b]{m_{\tilde{t}_1} m_{\tilde{t}_2}}}$) is broadly consistent with the observed Higgs mass throughout the allowed $c$-parameter ranges.

\begin{figure}[t]
  \centering
  \includegraphics{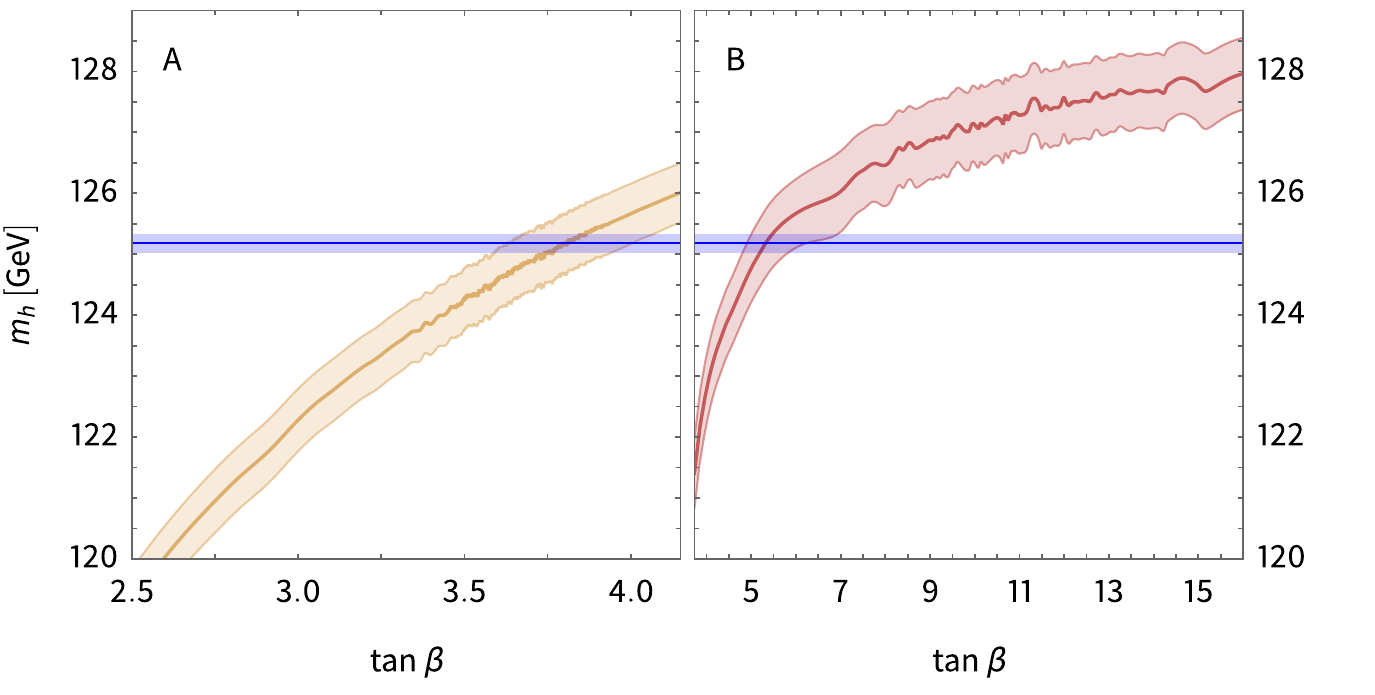}
  \caption{Predicted Higgs boson mass and uncertainty for benchmark scenarios A (left) and B (right) as functions of $\tan \beta$ at the scale $m_{\text{SUSY}} = \sqrt{\smash[b]{m_{\tilde{t}_1} m_{\tilde{t}_2}}}$. The horizontal blue line and surrounding region shows the observed Higgs mass value and its uncertainty.}
  \label{fig:higgs-mass}
\end{figure}

The observed Higgs boson mass provides stronger constraints indirectly, since the particular structure of EWSB in our model makes it sensitive to the scalar $D$-term corrections arising either from the extra dimension or in the MSSM running (see Sec.~\ref{sec:parameterspace-tachyon}). We show the Higgs boson mass estimates for both scenarios in Fig.~\ref{fig:higgs-mass} as smoothed functions of $\tan \beta$. For both benchmark scenarios, consistency with the observed Higgs boson mass restricts the allowed range of $D$-term corrections, such that the necessary value of $\tan \beta$ results from EWSB. This, first of all, introduces correlations among the heavy sfermion masses, such that the necessary corrections are obtained. 
The primary constraint arises on the $c$ parameters of the first-generation sfermions (the heaviest sfermions) $c_{L_1}$ and $c_{Q_1}$, which must be correlated such that $\Delta_{\mathcal{S}}, \mathcal{S} \sim m^2_{\tilde{u}_L,\tilde{d}_L} - 2 m^2_{\tilde{u}_R} + m^2_{\tilde{d}_R} - m^2_{\tilde{e}_L} + m^2_{\tilde{e}_R}$ are of the correct scale. 

\begin{figure}[t]
  \centering
  \includegraphics{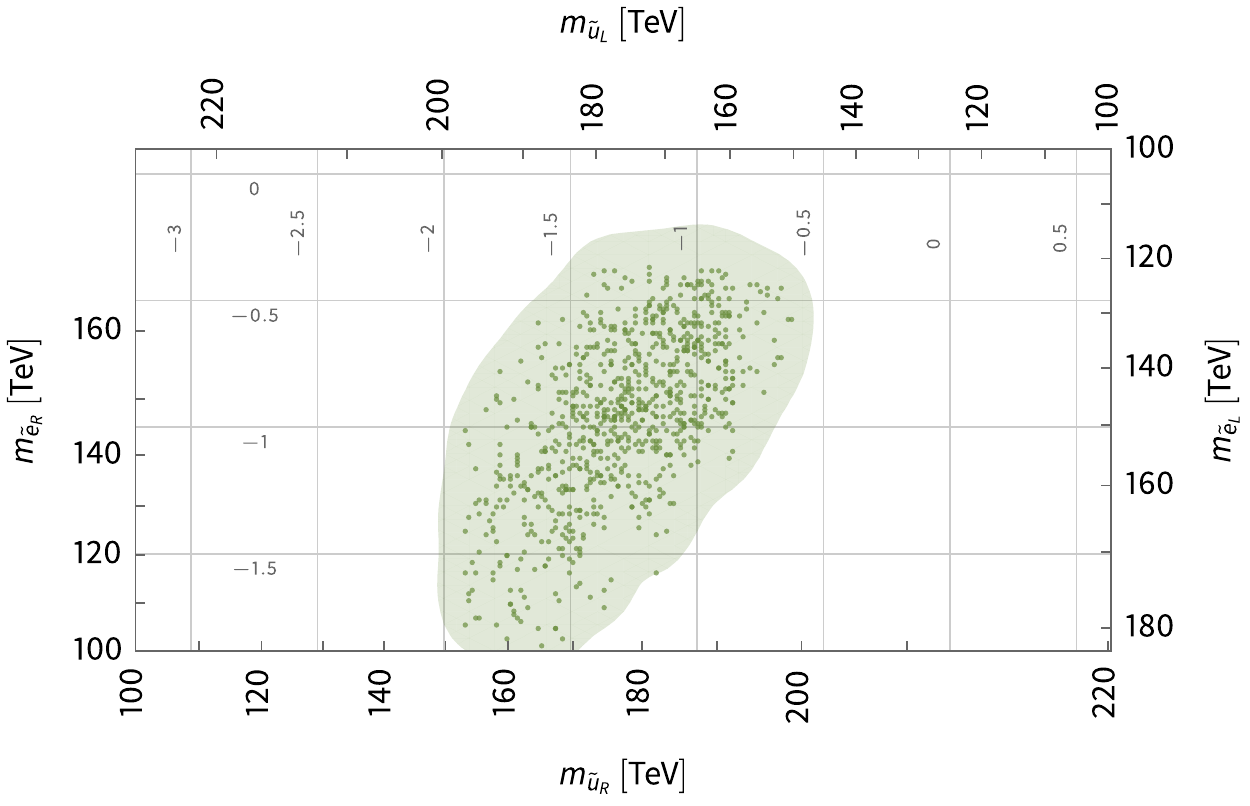}
  \caption{Correlation between first-generation slepton and up-squark masses for benchmark scenario A (given in Table~\ref{tab:parameterpoints}). The green region gives a smoothed estimate of the region preferred by the experimentally measured Higgs boson. The horizontal (vertical) gray lines give the range of $c_{L_1}$ ($c_{Q_1}$).}
  \label{fig:c-parameter-correlation}
\end{figure}

We show this correlation for scenario A in Fig.~\ref{fig:c-parameter-correlation}. For scenario A, the explanation of the SM fermion mass hierarchy typically requires that either $\widetilde{u}_L$ or $\widetilde{u}_R$ is the heaviest sfermion. Thus, in order to obtain $\Delta_{\mathcal{S}}, \mathcal{S} > 0$, as preferred by the Higgs boson mass, $m^2_{\tilde{e}_R}$ must be heavy enough to compensate for the negative contribution of $m^2_{\tilde{u}_R}$, if $m_{\tilde{u}_R} > m_{\tilde{u}_L}$. When $m_{\tilde{u}_L} > m_{\tilde{u}_R}$, then $m_{\tilde{e}_L} > m_{\tilde{e}_R}$ is allowed. Note that the Higgs mass measurement also constrains the allowed spread of the first-generation sfermion masses more than the condition imposed to suppress FCNCs, which merely restricts the masses to be above 100 TeV. Similarly, for scenario B, the Higgs boson mass prefers $\Delta_{\mathcal{S}}, \mathcal{S} < 0$, which is accomplished when $\widetilde{u}_R$ is the heaviest sfermion in the theory. $\Delta_{\mathcal{S}}, \mathcal{S} > 0$ is also allowed, but in this case,\ $m_{\tilde{u}_R}$ must be tuned against $m_{\tilde{e}_R}$. For this point we note that the Higgs boson mass also limits the mass ranges of the other first- and second-generation sfermions more strictly than necessary to suppress FCNCs.

For the third-generation sfermions, the $D$-term corrections necessary to obtain the observed Higgs boson mass may necessitate additional constraints in order to avoid tachyonic masses. For both scenarios, this effect imposes the lower limit on $m_{\tilde{\tau}_R}$, and for scenario B it also imposes the lower limit on $m_{\tilde{\tau}_L}$. These are the only constraints on the third-generation sfermion masses that are stronger than those set by the order-one limit on the $c$ parameters. For both scenarios, the right-handed stau may thus be the lightest sfermion, and $\widetilde{\tau}_1$ may accordingly be the NLSP. 

When $\widetilde{\tau}_1$ is heavy, the NLSP for both scenarios is $\widetilde{\chi}^{0}_{1}$.
For scenario A, $\widetilde{\chi}^{0}_{1}$ is Bino-like and $\widetilde{\chi}^{\pm}_{2}$, $\widetilde{\chi}^0_{2}$ are Wino-like, while the heavy charginos, $\widetilde{\chi}^{\pm}_{2}$, and the heavy neutralinos, $\widetilde{\chi}^{0}_{3,4}$, are Higgsino-like. This is reversed for scenario B, where $\widetilde{\chi}^{0}_{1,2,3}$ and $\widetilde{\chi}^{\pm}_{1}$ are Bino-like or Higgsino-like and $\widetilde{\chi}^{\pm}_{2}$,$\widetilde{\chi}^{0}_{4}$ are Wino-like. The spread in the masses of the Higgsino-like states is due to the spread of the Higgsino mass parameter $\mu$, which is fixed by EWSB. In both cases, $\mu$ correlates precisely with the soft mass parameter $m_{H_u}^2$, which is predominantly determined by Yukawa radiative corrections from $m^2_{\tilde{Q}_3}$ and $m^2_{\tilde{u}_3}$. Thus, the spread of $\mu$ is ultimately tied to freedom in $c_{Q_3}$, which as we discussed above, is limited only by the constraint that it is an order-one number. The masses of the heavy Higgs, $H^0$, $H^{\pm}$, and $A^0$ also scale with $\mu$, but about 10\% of their spread is due to the running of the soft masses $m_{H_u}^2$ and $m_{H_d}^2$. The spread in the gluino and the Bino-like and Wino-like states is due primarily to the uncertainty in the gauge and Yukawa couplings. In Figs.~\ref{fig:spectrum-schematic-numeric} and \ref{fig:spectrum-schematic-numeric-interaction} that spread is exaggerated for clarity. 

In both scenarios, the hierarchical structure of the mass spectrum is clear. The largest hierarchy occurs for $\widetilde{\tau}_1$, where we find ratios up to $m_{\tilde{u}_6,\tilde{d}_6}/ m_{\tilde{\tau}_1} \sim 13$ in the singlet spurion case and up to $m_{\tilde{u}_6} / m_{\tilde{\tau}_1} \sim 35$ in the nonsinglet spurion case. The hierarchy for the stops is relatively more modest: we find ratios up to $m_{\tilde{u}_6,\tilde{d}_6} / m_{\tilde{t}_1} \sim 3$ in the singlet spurion case and up to $m_{\tilde{u}_6} / m_{\tilde{t}_1} \sim 18$ in the nonsinglet spurion case. The size of these mass splittings, which cannot be generated by MSSM running alone, is a direct consequence of the hierarchy in the sfermion IR-brane soft mass boundary conditions, and hence is ultimately a signature of the SM fermion mass spectrum, mediated by the radiative corrections of the extra dimension and the MSSM.


\section{Conclusion}

We have presented a minimal supersymmetric model that uses partial compositeness to relate the SM fermion mass hierarchy to the sfermion mass hierarchy. This occurs by assuming that the SM gauge fields, Higgs sector, and the third-generation matter are (mostly) elementary, while the first two generations of matter are composite due to some unknown strong dynamics that confines at a scale $\Lambda_{\textrm{IR}}$. Hierarchies are then generated when elementary superfields linearly mix with supersymmetric operators that have large anomalous dimensions. Since the Higgs fields are elementary, the more composite the fermion, the lighter the corresponding fermion mass. The strong dynamics is also assumed to dynamically break supersymmetry, such that the composite sparticle states directly feel the supersymmetry breaking. The predominantly elementary states, such as the third-generation sfermions, Higgsinos, and gauginos, are therefore split from the much heavier first- and second-generation  composite sfermions. Thus, the partially composite supersymmetric model generically predicts that light (heavy) SM fermions, have heavy (light) sfermion superpartners. Moreover, since the gravity multiplet mixes with the stress-energy tensor (via an irrelevant term), the gravitino is much lighter than the gauginos. It therefore becomes the LSP that can play the role of dark matter.

To obtain quantitative predictions and model the unknown strong dynamics responsible for the composite states and large anomalous dimensions, we use the AdS/CFT correspondence to study a 5D version of our 4D model [where the strong dynamics is specifically due to a large-$N$ gauge theory (CFT)]. In a slice of AdS$_5$, the Higgs sector is confined to the UV brane, while the remaining MSSM superfields are located in the bulk. Supersymmetry breaking occurs on the IR brane. The MSSM fields are identified with the zero modes of the corresponding 5D fields. The zero-mode profile depends on a bulk mass (dimensionless) parameter $c$ that can be arbitrarily varied to localize the zero-mode superfield anywhere in the bulk. The fermion and sfermion mass hierarchy is now dictated by the 5D fermion geography. Since the Higgs fields are confined to the UV brane, the third-generation SM fermions are UV-localized, while the first- and second-generation SM fermions are IR-localized. This naturally leads to an \emph{inverted} sfermion mass hierarchy, where the first- and second-generation sfermions are heavy, while those of the third generation are light.

At tree level, the sfermion hierarchy may be exponentially large due to the suppressed coupling between the UV-localized fields and the supersymmetry-breaking sector. The mass scale of the third-generation sfermions is therefore set by radiative corrections from the heavy states, which transmit the breaking of supersymmetry at loop order and become the dominant soft mass contribution. At one loop in 5D, these corrections arise from bulk gauginos and scalars. Since the Higgs fields are localized on the UV brane, both the Higgs-sector soft masses and the soft trilinear scalar couplings ($a$-terms) are zero at tree level, but they, too, receive radiative corrections from the bulk. 

The overall scales in the 5D model can be fixed by imposing a number of phenomenological constraints: (i) the LSP gravitino is assumed to be the dark matter with a mass $\gtrsim 1$ keV; (ii) electroweak symmetry is broken, consistent with a 125 GeV Higgs boson; (iii) the first- and second-generation sfermions are at least as heavy as 100 TeV to ameliorate the supersymmetric flavor problem; (iv) the gaugino and Higgsino masses are constrained, so as to preserve gauge coupling unification as in the usual MSSM [assuming any underlying dynamics preserves SU(5)]; and (v) only the MSSM fields are present in the theory below the scale of compactification. The SM fermion mass spectrum is used to constrain the bulk fermion mass parameters $c_i$. The 5D model then predicts the sfermion masses at the IR-brane scale, $\Lambda_{\text{IR}}$, which are run down to lower energies using renormalization group equations. Since the boundary conditions for the sfermion masses are nonuniversal and flavor-dependent, tachyonic constraints that avoid charge- and color-breaking minima must be imposed to further restrict the parameter space.

The numerical results of our benchmark scenarios, given in Table \ref{tab:parameterpoints}, predict a hierarchical sfermion mass spectrum. The third-generation sfermions have masses in the approximate range 10--100 TeV (20--100 TeV for the stops), while the first- and second-generation sfermions have masses in the range 100--350 TeV. We do not obtain a unique prediction because we assume that there is no relation between the $c_{L,R}$ parameters of the left- and right-handed fermions. Nevertheless, the numerical results reveal some interesting features. Most obvious is the hierarchical nature of the spectrum. Typical MSSM running cannot produce a mass spectrum with widely separated sparticle masses, and thus, with minimal particle content, the origin of the mass hierarchy must necessarily reside in the high-scale boundary conditions. Such conditions are a generic feature of our model and result in a distinctive split spectrum. The nonuniversality of the sfermion boundary conditions is also visible at a finer level, as it is responsible for the presence of sizeable $D$-term radiative corrections to the scalar masses. Although the sign and magnitude of these corrections are highly constrained on tachyonic grounds, they can be favorable for EWSB and can offset negative contributions to the scalars that arise at two loops. Due to the structure of EWSB in our model (imposed by radiative corrections from the bulk), the predicted Higgs boson mass is also sensitive to $D$-term corrections, and the experimentally measured mass value can therefore indirectly constrain the heaviest mass scales in the theory. In fact, since the measured Higgs mass is broadly consistent with stop masses in the 10--100 TeV range (as predicted in both benchmark scenarios) it is primarily through this effect that it constrains our benchmark spectra.

Our model is not too different from the usual MSSM, where a hidden sector with strong dynamics is typically invoked to dynamically break supersymmetry (e.g., via gaugino condensation). The supersymmetry breaking is then mediated via gravity (or alternatively, gauge interactions) to the visible sector with universal boundary conditions for the sfermion masses. The difference in our model is that the first- and second-generations of matter are composites of the strong dynamics at some high scale $\Lambda_{\text{IR}}$. The composite states also directly feel the supersymmetry breaking (e.g., perhaps via a nonzero $F$-term of the underlying constituents), thereby giving rise to strongly flavor-dependent sfermion mass boundary conditions. Furthermore, assuming that the strong dynamics is SU(5) invariant (similar to what is imposed on the messenger sector in gauge-mediated models), gauge coupling unification is still preserved at the GUT scale ${\sim} 10^{16}$~GeV. 

In light of the Higgs boson discovery and its implications for the supersymmetric spectrum, our model thus provides a more predictive, splitlike supersymmetry scenario by explicitly relating the SM fermion mass hierarchy to the sfermion mass spectrum. It would be interesting to construct models of the nontrivial dynamics (perhaps going beyond large-$N$ theories) that may constrain the anomalous dimensions even further, and therefore lead to exact predictions for the sparticle spectrum. Nonetheless, the partially composite supersymmetric model provides the raision d'\^{e}tre for the inverted sfermion hierarchy with a gravitino LSP. The NLSP is typically a Bino, Higgsino, or right-handed stau which decays to the gravitino and could eventually be probed at a future 100 TeV collider. Alternatively, the heavy first- and second-generation sfermions could be indirectly probed via rare-decay experiments, such as the flavor-violating Mu2e experiment~\cite{Bartoszek:2014mya}, or experiments attempting to measure the electric dipole moment of the electron~\cite{Andreev:2018ayy}. Of course, with heavy superpartners, our model is tuned, and the question of why the overall scale of the sparticle spectrum is much heavier than the TeV scale remains a mystery. Perhaps this is just evidence of the multiverse, as speculated in split-supersymmetric models, or a supersymmetric relaxion mechanism is at play, or, instead, the tuning could be related to the strong dynamics of the supersymmetry-breaking sector. In any case, we have attempted to provide further rationale for why low-energy supersymmetry may be lurking at a scale of 10--1000 TeV.


\section*{Acknowledgments}

We thank Jason Evans, Ben Harling, and Alex Pomarol for useful discussions.
This work was supported by the U.S. Department of Energy Grant No. DE-SC0011842 at the University of Minnesota.



\numberwithin{equation}{section}

\newpage
\appendix


\section{Partial Compositeness}\label{app:holographic-mixing}

In this appendix we present details of the partial compositeness (equivalent to holographic mixing in 5D) for the chiral, vector, and gravity supermultiplets.

\subsection{Chiral supermultiplet}

\subsubsection{Complex scalar}

At the IR scale, the supersymmetric Lagrangian \eqref{eq:scalarLag} for the chiral supermultiplet has the component form
\begin{equation}
  \begin{split}
    \mathcal{L}_{\text{scalar}} 
      &= - \partial_{\mu} \phi^{\dagger} \partial^{\mu} \phi 
         + F^{\dagger} F
         + \varepsilon_\Phi \Lambda_{\text{IR}} \left(  \phi F^{c(1)} + F \phi^{c(1)} + \text{H.c.} \right) \\[1ex]
      &    \qquad
         - \partial_{\mu} \phi^{c (1) \dagger} \partial^{\mu} \phi^{c(1)} 
         + F^{c(1)\dagger} F^{c(1)}
         - \partial_{\mu} \phi^{(1)\dagger} \partial^{\mu} \phi^{(1)} 
         + F^{(1) \dagger} F^{(1)} \\[1ex]
      &    \qquad
         + m_{\Phi}^{(1)} \left( \phi^{(1)} F^{c(1)} + F^{(1)} \phi^{c(1)} + \text{H.c.} \right) \, ,
  \end{split}
\end{equation}
where $\varepsilon_\Phi$ is a dimensionless constant and $m_{\Phi}^{(1)}=g_\Phi^{(1)} \Lambda_{\text{IR}}$. Eliminating the auxiliary fields gives rise to the Lagrangian
\begin{equation}
  \begin{split}
    \mathcal{L}_{\text{scalar}} 
      &= - \partial_{\mu} \phi^{\dagger} \partial^{\mu} \phi 
         - \varepsilon_\Phi^2  \Lambda_{\text{IR}}^2 \phi^{\dagger} \phi 
         - \varepsilon_\Phi g_\Phi^{(1)} \Lambda_{\text{IR}}^2 \left( \phi^{\dagger} \phi^{(1)} + \text{H.c.} \right) \\[1ex]
      &    \qquad
         - \partial_{\mu} \phi^{(1)\dagger} \partial^{\mu} \phi^{(1)} 
         - m_{\Phi}^{(1)2} \phi^{(1)\dagger} \phi^{(1)} \\[1ex]
      &    \qquad
         - \partial_{\mu} \phi^{c(1)\dagger} \partial^{\mu} \phi^{c(1)} 
         - \left( g_\Phi^{(1)2} + \varepsilon_\Phi^2 \right) \Lambda_{\text{IR}}^2 \phi^{c(1)\dagger} \phi^{c(1)} \, ,
  \end{split}
\end{equation}
where, in the basis $( \phi, \phi^{(1)},\phi^{c(1)} )$, the mass matrix is given by
\begin{equation}
  m_{\phi}^2 
    = \left( 
        \begin{matrix} 
          \varepsilon_\Phi^2  & \varepsilon_\Phi g_\Phi^{(1)} & 0 \\ 
          \varepsilon_\Phi g_\Phi^{(1)}  & g_\Phi^{(1)2} & 0 \\ 
          0 & 0 & \varepsilon_\Phi^2 + g_\Phi^{(1)2} 
        \end{matrix} 
      \right)
      \Lambda_{\text{IR}}^2 \, .
\end{equation}
Note that there is a mass mixing between the elementary state $\phi$ and the composite state,
$\phi^{(1)}$. Nevertheless, when this matrix is diagonalized, there is a massless eigenstate which
can be written as:
\begin{equation}
  \label{eq:masslesseqn}
  |\phi_0 \rangle 
    \simeq \mathcal{N}_\Phi 
            \left\{ 
              | \phi \rangle 
            - \frac{\varepsilon_\Phi}{ {g_\Phi^{(1)}}} \, | \phi^{(1)}
            \rangle\right\} \, ,
\end{equation}
where $\mathcal{N}_\Phi$ is a normalization constant, while the massive eigenstates are given by
\begin{subequations}
  \label{eq:massiveeqn}
  \begin{align}
    |\phi_1 \rangle 
      &\simeq \mathcal{N}_\Phi 
              \left\{ 
                \frac{\varepsilon_\Phi}{ {g_\Phi^{(1)}}} \, | \phi \rangle 
              + | \phi^{(1)} \rangle 
              \right\} \, , \\[1ex]
    |\phi_2 \rangle &\simeq | \phi^{c(1)} \rangle \, .
  \end{align}
\end{subequations}
Thus, we see that the massless eigenstate is an admixture of the elementary and composite states and that the massive eigenstate is a complex scalar with mass squared $(\varepsilon_\Phi^2 + g_\Phi^{(1)2}) \Lambda_{\text{IR}}^2$. Note that the eigenstates are expressed in the mass-mixing basis, unlike the kinetic-mixing basis used in Ref.~\cite{Batell:2007jv}.
While both bases are equivalent at the level of mass eigenstates, supersymmetry breaking in the mass-mixing basis is shown in Sec.~\ref{sec:susybreaking-5d} to give consistent results with the 5D gravity model.

\subsubsection{Fermion}

Similarly, the fermion part of the supersymmetric Lagrangian \eqref{eq:scalarLag} at the IR scale is
given by:
\begin{equation}
  \begin{split}
    \mathcal{L}_{\text{fermion}} 
      &= i \psi^{\dagger} \bar{\sigma}^{\mu} \partial_{\mu} \psi
      + i \psi^{(1) \dagger} \bar{\sigma}^{\mu}\partial_{\mu} \psi^{(1)}
      + i \psi^{c(1)\dagger} \bar{\sigma}^{\mu} \partial_{\mu} \psi^{c(1)} \\[1ex]
      & \qquad 
      -\varepsilon_\Phi \Lambda_{\text{IR}}  \left(  \psi \psi^{c(1)} + \text{H.c.} \right) 
      - m_{\Phi}^{(1)} \left( \psi^{(1)} \psi^{c(1)} + \text{H.c.} \right) \, .
  \end{split}
\end{equation}
In the basis $( \psi, \psi^{c(1)}, \psi^{(1)} )$, this leads to the following fermion mass matrix:
\begin{equation}
  m_{\psi}
    = \frac{1}{2}
      \left( 
        \begin{matrix} 
          0 & \varepsilon_\Phi & 0 \\ 
          \varepsilon_\Phi & 0 & g_\Phi^{(1)} \\ 
          0 & g_\Phi^{(1)} & 0 
        \end{matrix} 
      \right)
      \Lambda_{\text{IR}} \, .
\end{equation}
The mass eigenstates correspond to a massless Weyl fermion and a massive Dirac state with mass $( \varepsilon_\Phi^2 + g_\Phi^{(1)2} \Lambda_{\text{IR}} )^{1/2}$. The massless eigenstate is given by
\begin{equation}
  \label{eq:zerofermioneigenstate}
  |\psi_0 \rangle 
    \simeq 
      \mathcal{N}_\Phi 
      \left\{ | \psi \rangle - \frac{\varepsilon_\Phi}{g_\Phi^{(1)}} \, | \psi^{(1)} \rangle \right\} \, ,
\end{equation}
while the massive eigenstates are
\begin{equation}
  |\psi_{1,2} \rangle 
    \simeq 
      \frac{\mathcal{N}_\Phi}{\sqrt{2}} 
      \left\{ \frac{\varepsilon_\Phi}{g_
        \Phi^{(1)}} | \psi \rangle 
      + | \psi^{(1)} \rangle 
      \pm \sqrt{1 + \frac{\varepsilon_\Phi^2}{g_\Phi^{(1)2}} } \, | \psi^{c(1)} \rangle
\right\}~.
\end{equation}
Thus, partial compositeness leads to a massless complex scalar and Weyl fermion which combine into a chiral supermultiplet.

\subsection{Vector supermultiplet}

The component fields of the vector supermultiplet can be identified by noting that the IR Lagrangian \eqref{eq:vectorsuperL} is invariant under the supergauge transformations 
\begin{subequations}
  \begin{align}
    V &\rightarrow V + i \left( \Omega^{\dagger} - \Omega \right) \, , \\[1ex]
    V^{(1)} + \frac{\Phi_V^{(1)} + \Phi_V^{(1)\dagger} }{\sqrt{2} g_V^{(1)} \Lambda_{\text{IR}}} 
      &\rightarrow 
         V^{(1)} 
       + \frac{\Phi_V^{(1)} + \Phi_V^{(1)\dagger} }{\sqrt{2} g_V^{(1)} \Lambda_{\text{IR}}} 
       - i \frac{{\varepsilon}_V}{g_V^{(1)}} \left( \Omega^{\dagger} - \Omega \right) \, ,
  \end{align}
\end{subequations}
where $\Omega$ is a chiral superfield gauge-transformation parameter. Choosing the Wess-Zumino gauge for $V$, the superfields then take the form
\begin{subequations}
  \label{eq:componentvecL}
  \begin{align}
    V &= \theta^{\dagger} \bar{\sigma}^{\mu} \theta A_{\mu} 
       + \theta^{\dagger} \theta^{\dagger} \theta \lambda 
       + \theta \theta \theta^{\dagger} \lambda^{\dagger} 
       + \frac{1}{2} \theta \theta \theta^{\dagger} \theta^{\dagger} D \, , \\[1ex]
    V^{(1)} &= \theta^{\dagger} \sigma^{\mu} \theta \widetilde{A}_{\mu}^{(1)} 
             + \theta^{\dagger} \theta^{\dagger} \theta \lambda^{(1)} 
             + \theta \theta \theta^{\dagger} \lambda^{(1)\dagger} 
             + \frac{1}{2} \theta \theta \theta^{\dagger} \theta^{\dagger} D^{(1)} \, , \\[1ex]
  \Phi_V^{(1)} &= \begin{aligned}[t]
                  &  \phi_V^{(1)} + \sqrt{2} \theta \chi^{(1)} 
                   + \theta \theta F_V^{(1)} \\[1ex]
                  &  \quad
                   + i \theta^{\dagger} \bar{\sigma}^{\mu} \theta \partial_{\mu} \phi_V^{(1)}
                   - \frac{i}{\sqrt{2}} \theta \theta \theta^{\dagger} \bar{\sigma}^{\mu} \partial_{\mu} \chi^{(1)} 
                   + \frac{1}{4} \theta \theta \theta^{\dagger} \theta^{\dagger} \Box\phi_V^{(1)} \, ,
                  \end{aligned} 
  \end{align}
\end{subequations}
where $\Box = \partial_{\mu} \partial^{\mu}$. Here, $\widetilde{A}_\mu^{(1)}$, $\phi_{V}^{(1)}$, $\lambda^{(1)}$, and $\chi^{(1)}$ are dynamical (composite) fields, while $F_V^{(1)}$ and $D^{(1)}$ are auxiliary (composite) fields. Next, we diagonalize the mass term for the component Lagrangians.

\subsubsection{Gauge field}\label{app:gaugefield}

Using \eqref{eq:vectorsuperL} and \eqref{eq:componentvecL}, the gauge field component Lagrangian becomes
\begin{equation}
  \mathcal{L}_{\text{gauge}} 
    = -\frac{1}{4} F^{\mu \nu} F_{\mu \nu} 
    - \frac{1}{4} \widetilde{F}^{(1) \mu \nu} \widetilde{F}_{\mu \nu}^{(1)} 
    - \frac{1}{2} \Lambda_{\text{IR}}^2 \left( \varepsilon_V A_{\mu} 
    + g_V^{(1)} \widetilde{A}_{\mu}^{(1)} + \partial_{\mu} \varphi^{(1)} \right)^2 \, ,
\label{eq:gaugeLag}
\end{equation}
which is invariant under the gauge transformations 
\begin{subequations}
  \begin{align}
    A_\mu &\rightarrow A_\mu + \partial_\mu \zeta \, , \\[1ex]
    A_{\mu}^{(1)} \equiv \widetilde{A}_\mu^{(1)} + \frac{\partial_{\mu} \varphi^{(1)}}{g_V^{(1)}} 
      &\rightarrow A_\mu^{(1)} - \frac{{\varepsilon}_V}{g_V^{(1)}} \partial_\mu \zeta \, ,
  \end{align}
\end{subequations}
where $\zeta$ is a gauge parameter and $\varphi^{(1)} \equiv \frac{i}{2 \Lambda_{\text{IR}} } ( \phi_V^{(1)} - \phi_V^{(1)\ast} )$. Note that in the limit of no mixing between the elementary and composite sectors ($\varepsilon_V=0)$, the lowest-lying massive composite state, $A_\mu^{(1)}$, has a mass $m_V^{(1)} = g_V^{(1)} \Lambda_{\text{IR}}$. The appearance of a mass term for $A_\mu$ is similar to what happens for vector-meson dominance in QCD~\cite{OConnell:1995nse}.

The mass eigenstates are obtained by diagonalizing the mass squared term in \eqref{eq:gaugeLag}, where in the basis $(A_\mu, A_\mu^{(1)})$, the mass matrix is given by
\begin{equation}
  m_{A}^2 
    = \left( 
        \begin{matrix} 
          \varepsilon_V^2  & \varepsilon_V g_V^{(1)} \\
          \varepsilon_V g_V^{(1)} & g_V^{(1)2} \\
        \end{matrix} 
      \right) 
      \Lambda_{\text{IR}}^2 \, .
\end{equation}
This gives rise to one massless and one massive eigenstate. For canonical kinetic terms, the massless gauge boson eigenstate becomes
\begin{equation}
  \label{eq:gaugeboson0mode}
  |A_{\mu 0} \rangle 
    \simeq 
      \mathcal{N}_V 
      \left\{ |A_{\mu} \rangle - \frac{{\varepsilon}_V}{g_V^{(1)}} | A_{\mu}^{(1)} \rangle \right\} \, ,
\end{equation}
while the massive gauge boson eigenstate is given by
\begin{equation}
  | A_{\mu 1} \rangle 
    \simeq 
      \mathcal{N}_V 
      \left\{ \frac{{\varepsilon}_V}{g_V^{(1)}} |A_{\mu}\rangle + |A_{\mu}^{(1)}\rangle \right\} \, ,
\end{equation}
where $\mathcal{N}_V$ is a normalization constant. The eigenstates are expressed in the mass-mixing basis instead of the kinetic-mixing basis used in Ref.~\cite{Batell:2007jv}, since supersymmetry is assumed to be broken in this basis. The massless eigenmode \eqref{eq:gaugeboson0mode} now transforms under a gauge transformation as $A_{\mu0} \rightarrow A_{\mu0} + ( 1 + \varepsilon_V^2 / g_V^{(1)2} ) \partial_\mu \zeta$, while, as expected,  $A_{\mu1}$ no longer transforms.

The massive eigenstate $A_{\mu 1}$ obtains a mass
\begin{equation}
  m_{V_1}^2 = \bigl( {\varepsilon}_V^2 + g_V^{(1)2} \bigr) \Lambda_{\text{IR}}^2
            \simeq \left[ g_s^2( \Lambda_{\text{IR}}) + g_V^{(1)2} \right] \Lambda_{\text{IR}}^2 \, ,
\end{equation}
where, in the second expression, we have used $g_s( \Lambda_{\text{IR}}) \simeq \frac{\tilde{\varepsilon}_V}{\sqrt{Z_V}} \frac{\sqrt{N}}{4\pi} = {\varepsilon}_V$, which follows from the large-$N$ corrections to the elementary-field gauge coupling $g_s$. The diagonal gauge-field Lagrangian for the two-state system is then given by
\begin{equation}
  \mathcal{L}_{\text{gauge}} 
    = - \frac{1}{4} F_0^{\mu \nu} F_{0\mu \nu} 
      - \frac{1}{4} F_1^{\mu \nu} F_{1\mu \nu}
      - \frac{1}{2} m_{V_1} ^2 A_{1\mu}^2 \, ,
\end{equation}
where the gauge coupling of the massless mode is obtained from
\begin{equation}
  g_s(\Lambda_{\text{IR}}) \psi^{\dagger} \bar{\sigma}^{\mu} A_{\mu} \psi 
    - g_V^{(1)} \psi^{(1)\dagger} \bar{\sigma}^{\mu} A_{\mu}^{(1)} \psi^{(1)}
    + g_V^{(1)} \psi^{c(1)\dagger} \bar{\sigma}^{\mu} A_{\mu}^{(1)} \psi^{c(1)} 
    = g \psi_0^{\dagger} \bar{\sigma}^{\mu} A_{\mu 0} \psi_0 + \dotsb
\end{equation}
Using \eqref{eq:zerofermioneigenstate} and \eqref{eq:gaugeboson0mode}, this leads to the expression
\begin{equation}
  \frac{1}{g^2} 
    = \left[ 
        \mathcal{N}_{\Phi}^2 \, 
        \mathcal{N}_V 
        \left(
          g_s( \Lambda_{\text{IR}}) 
        + g_V^{(1)} \frac{\varepsilon_{\Phi}^2}{g_{\Phi}^2} \frac{\varepsilon_V}{g_V^{(1)}} 
        \right)
       \right]^{-2}
    \simeq \frac{1}{g_s^2( \Lambda_{\text{IR}})}+ \frac{1}{g_V^{(1)2}} \, .
\label{eq:zerogc}
\end{equation}

Finally, note that by eliminating $D$ and $D^{(1)}$ in the scalar field part of the Lagrangian, one can check that the real part of the composite scalar field $\phi_{V}^{(1)}$ also obtains a mass, identical to that of the gauge field $A_{\mu 1}$.

\subsubsection{Gaugino}

The gaugino part of the vector supermultiplet Lagrangian \eqref{eq:vectorsuperL} is given by
\begin{equation}
  \label{eq:gauginoLag}
  \begin{split}
    \mathcal{L}_{\text{gaugino}} 
      &= i \lambda^{\dagger} \bar{\sigma}^{\mu} \partial_{\mu} \lambda
       + i \lambda^{(1)\dagger}\bar{\sigma}^{\mu} \partial_{\mu} \lambda^{(1)} 
       + i \chi^{(1)\dagger} \bar{\sigma}^{\mu} \partial_{\mu} \chi^{(1)} \\[1ex]
      &  \qquad
       - \Lambda_{\text{IR}} 
         \left( {\varepsilon}_V \lambda \chi^{(1)} + g_V^{(1)} \lambda^{(1)} \chi^{(1)} + \text{H.c.} \right) \, .
  \end{split}
\end{equation}
In the limit $\varepsilon_V = 0$, the massive composite state is a Dirac fermion with mass $m_V^{(1)} = g_V^{(1)} \Lambda_{\text{IR}}$. Using the basis $( \lambda, \chi^{(1)}, \lambda^{(1)} )$ and the orthogonal matrix
\begin{equation}
  O_{1/2} 
    = \frac{1}{\sqrt{g_V^{(1)2} + {\varepsilon}_V^2}} 
      \left( 
        \begin{matrix} g_V^{(1)} & 0 & -\varepsilon_V \\ 
        \frac{\varepsilon_V}{\sqrt{2}} & - \sqrt{\frac{g_V^{(1)2} + \varepsilon_V^2}{2}} & \frac{g_V^{(1)}}{\sqrt{2}} \\ 
        \frac{\varepsilon_V}{\sqrt{2}} & \sqrt{\frac{g_V^{(1)2} + \varepsilon_V^2}{2}} & \frac{g_V^{(1)}}{\sqrt{2}} 
        \end{matrix} 
      \right) \, ,
\end{equation}
the mass terms in \eqref{eq:gauginoLag} can be diagonalized via
\begin{equation}
  \label{eq:diaggaugino}
  O_{1/2} 
  \left( 
    \begin{matrix} 
      0 & \frac{\varepsilon_V}{2} & 0 \\ 
      \frac{\varepsilon_V}{2} & 0 & \frac{g_V^{(1)}}{2} \\ 
      0 & \frac{g_V^{(1)}}{2} & 0 
    \end{matrix} 
  \right) 
  O_{1/2}^T 
    = \left( 
        \begin{matrix} 
          0 & 0 & 0 \\ 
          0 & - \frac{\sqrt{ {\varepsilon}_V^2 + g_V^{(1)2}}}{2} & 0 \\ 
          0 & 0 & \frac{\sqrt{ {\varepsilon}_V^2 + g_V^{(1)2}}}{2} 
        \end{matrix} 
      \right) \, ,
\end{equation}
which gives rise to a massless Weyl fermion and a massive Dirac fermion. The massless gaugino eigenstate is given by
\begin{equation}
  \label{eq:masslessgaugino}
  |\lambda_0 \rangle 
    \simeq 
      \mathcal{N}_V 
      \left\{ |\lambda \rangle - \frac{{\varepsilon}_V}{g_V^{(1)}} \, | \lambda^{(1)} \rangle \right\} \, .
\end{equation}
Thus, the massless gaugino is an admixture of the elementary gaugino $\lambda$ and the composite gaugino $\lambda^{(1)}$.

The massive Dirac fermion state has the decomposition
\begin{align}
  | \lambda_{1,2} \rangle 
    &\simeq 
       \frac{\mathcal{N}_V}{\sqrt{2}} 
       \left\{ 
         \frac{{\varepsilon}_V}{g_V^{(1)}} \, | \lambda \rangle 
         + | \lambda^{(1)} \rangle \pm \sqrt{ 1 + \frac{{\varepsilon}_V^2}{g_V^{(1)2}} } \, | \chi^{(1)} \rangle 
         \right\} \, , \nonumber \\[1ex]
    &\simeq 
       \frac{\mathcal{N}_V}{\sqrt{2}} 
       \left\{
         \frac{1}{g_V^{(1)}\sqrt{2\zeta_V \log \bigl( \frac{\Lambda_{\text{UV}}}{\Lambda_{\text{IR}}} \bigr)} } \, 
         | \lambda \rangle 
       + | \lambda^{(1)} \rangle \pm | \chi^{(1)} \rangle 
       \right\} \, ,
\end{align}
assuming $\varepsilon_V \ll 1$ and dropping terms of $\mathcal{O}(\varepsilon_V^2)$ and higher in the second expression, with mass eigenvalues
\begin{equation}
  \label{eq:Diracgauginomass}
  m_{V_{1,2}} \simeq \pm \sqrt{g_s^2 \left( \Lambda_{\text{IR}} \right)+ g_V^{(1)2} } \, \Lambda_{\text{IR}} \, .
\end{equation}
This mass agrees with that obtained for the gauge field $A_{\mu 1}$ and scalar field $\phi_{V}^{(1)}$, as expected by supersymmetry.

\subsection{Gravity supermultiplet}

To identify the component fields of the gravity supermultiplet, we note that the IR Lagrangian \eqref{eq:IRgravityLag} is invariant under the supergauge transformations
\begin{subequations}
  \begin{align}
    H_\mu &\rightarrow H_\mu + \Delta_\mu \, , \\[1ex]
    H^{(1)}_\mu &\rightarrow H^{(1)}_\mu - \frac{{\varepsilon}_H}{g_H^{(1)}} \Delta_\mu \, ,
  \end{align}
\end{subequations}
where $\Delta_{\mu}$ is a real superfield gauge-transformation parameter. Choosing an analog of the Wess-Zumino gauge for $H_{\mu}$, the superfields take the form
\begin{subequations}
  \begin{align}
    H_{\mu} 
      &= - \frac{1}{\sqrt{2}} \theta^{\dagger} \bar{\sigma}^{\nu} \theta\,h_{\mu \nu} 
         - i \theta \theta \theta^{\dagger} \lambda_{\mu}^{\dagger} 
         + i \theta^{\dagger} \theta^{\dagger} \theta \lambda_{\mu} 
         + \frac{1}{2} \theta \theta \theta^{\dagger} \theta^{\dagger} D_{\mu} + \dotsb \\[1ex]
    H_{\mu}^{(1)} 
      &= C_{\mu}^{(1)} 
        - i \theta \omega_{\mu}^{(1)} 
        + i \theta^{\dagger} \omega_{\mu}^{(1)\dagger} 
        - \theta^{\dagger} \bar{\sigma}^{\nu} \theta\, V_{\mu \nu}^{(1)} 
        - i \theta \theta \theta^{\dagger} \left( \lambda_{\mu}^{(1)\dagger} 
        - \frac{i}{2} \bar{\sigma}^{\nu} \partial_{\nu} \omega_{\mu}^{(1)} \right) \nonumber \\[1ex]
      &   \qquad
        + i \theta^{\dagger} \theta^{\dagger} \theta 
            \left( \lambda_{\mu}^{(1)} - \frac{i}{2} \sigma^{\nu} \partial_{\nu} \omega_{\mu}^{(1)\dagger} \right) 
        + \frac{1}{2} \theta \theta \theta^{\dagger} \theta^{\dagger} 
          \left( D_{\mu}^{(1)} + \frac{1}{2} \Box C_{\mu}^{(1)} \right) 
        + \dotsb
  \end{align}
\end{subequations}
where we have neglected terms with auxiliary fields. The gravity component fields are then defined
to be 
\begin{subequations}
  \begin{align}
    h_{\mu \nu}^{(1)} 
      &\equiv \frac{V_{\mu \nu}^{(1)} + V_{\nu \mu}^{(1)}}{\sqrt{2}} \, , \\[1ex]
    C_{\mu}^{(1)} 
      &\equiv \frac{1}{\sqrt{3}\Lambda_{\text{IR}}} \frac{1}{g_H^{(1)}} h_{\mu}^{(1)} \, , \\[1ex]
    \frac{1}{2} \psi_{\mu}^{(1)} 
      &\equiv \lambda_{\mu}^{(1)} + \frac{1}{3} \sigma_{\mu} \bar{\sigma}^{\rho} \lambda_{\rho}^{(1)} \, \\[1ex]
    b^{(1)\sigma} 
      &\equiv D^{(1)\sigma} + \frac{1}{2} \epsilon^{\nu \mu \kappa \sigma} \partial_{\kappa} V_{\nu \mu}^{(1)} \, , \\[1ex]
    \omega_{\mu}^{(1)} 
      &\equiv \frac{1}{\Lambda_{\text{IR}}} \frac{1}{2 g_H^{(1)}} \chi_\mu^{(1)} \, .
  \end{align}
\end{subequations}
Note that $h_{\mu \nu}^{(1)}$, $\psi_{\mu}^{(1)}$, $\chi_{\mu}^{(1)}$, and $h_{\mu}^{(1)}$ are dynamical (composite) fields, while $b_{\mu}^{(1)}$ is an auxiliary (composite) field.

\subsubsection{Graviton}

Using \eqref{eq:IRgravityLag} the graviton field component Lagrangian becomes
\begin{equation}
  \label{eq:gravitonLag}
  \mathcal{L}_{\text{graviton}} 
    = \frac{1}{\sqrt{2}} h_{\mu \nu} E^{\mu \nu} 
    + \frac{1}{\sqrt{2}} h_{\mu \nu}^{(1)} E^{(1) \mu \nu}
    - \frac{1}{2} \Lambda_{\text{IR}}^2 \left( {\varepsilon}_H h_{\mu \nu} 
    + g_H^{(1)}h_{\mu \nu}^{(1)} \right)^2 \, ,
\end{equation}
where $E_{\mu \nu} \equiv \frac{1}{\sqrt{2}} ( \partial_{\mu} \partial_{\nu} h_{\lambda}^{\lambda} + \Box h_{\mu \nu} - \partial_{\mu} \partial^{\lambda} h_{\lambda \nu} - \partial_{\nu} \partial^{\lambda} h_{\lambda \mu} - \eta_{\mu \nu} \Box h_{\lambda}^{\lambda} + \eta_{\mu \nu} \partial^{\lambda} \partial^{\rho} h_{\lambda \rho} )$.
The Lagrangian is invariant under the gauge transformation
\begin{subequations}
  \begin{align}
    h_{\mu \nu} 
      &\rightarrow h_{\mu \nu} - \frac{1}{2} \left( \partial_\mu \zeta_{\nu} + \partial_{\nu} \zeta_{\mu} \right) \, , \\[1ex]
    h_{\mu \nu}^{(1)} 
      &\rightarrow 
         h_{\mu \nu}^{(1)} 
       + \frac{1}{2} \frac{ {\varepsilon}_H}{g_H^{(1)}} \left( \partial_\mu \zeta_{\nu} 
       + \partial_{\nu} \zeta_{\mu} \right) \, ,
  \end{align}
\end{subequations}
where $\zeta_{\mu}$ is a gauge parameter. Note that in the limit of no mixing between the elementary and composite sectors ($\varepsilon_H = 0)$, the lowest-lying massive composite state, $h_{\mu \nu}^{(1)}$, has a mass $m_H^{(1)} = g_H^{(1)} \Lambda_{\text{IR}}$.

The mass eigenstates are obtained by diagonalizing the mass term in \eqref{eq:gravitonLag}. The massless graviton eigenstate is then
\begin{equation}
  \label{eq:graviton0mode}
  |h_{\mu \nu 0} \rangle 
    \simeq 
      \mathcal{N}_H 
      \left\{ |h_{\mu \nu} \rangle - \frac{ {\varepsilon}_H}{g_H^{(1)}} \, | h_{\mu \nu}^{(1)} \rangle \right\} \, ,
\end{equation}
and the massive graviton eigenstate is given by
\begin{equation}
  | h_{\mu \nu1} \rangle 
    \simeq 
      \mathcal{N}_H 
      \left\{ \frac{ {\varepsilon}_H}{g_H^{(1)}} \, |h_{\mu \nu}\rangle +| h_{\mu \nu}^{(1)}\rangle \right\} \, ,
\end{equation}
where $\mathcal{N}_H$ is a normalization constant. The massless eigenmode $h_{\mu \nu 0}$ now transforms under a gauge transformation as $h_{\mu \nu 0} \rightarrow h_{\mu \nu 0} - \frac{1}{2} ( 1 + \varepsilon_H^2 / g_H^{(1)2} ) ( \partial_\mu \zeta_{\nu} + \partial_{\nu} \zeta_{\mu} )$, while, as expected, $h_{\mu \nu 1}$ no longer transforms. The massive eigenstate $h_{\mu \nu 1}$ obtains a mass $m_{H_1}^2 \simeq ( {\varepsilon}_H^2 + g_H^{(1)2} ) \Lambda_{\text{IR}}^2$. The diagonal gravity Lagrangian for the two-state system is then given by
\begin{equation}
  \mathcal{L}_{\text{graviton}}
    = \frac{1}{\sqrt{2}} h_0^{\mu \nu} E_{\mu \nu 0} 
    + \frac{1}{\sqrt{2}} h_1^{\mu \nu} E_{\mu \nu 1}
    - \frac{1}{2} m_{H_1}^2 h_{\mu \nu 1}^2 \, .
\end{equation}
By eliminating $b_{\mu}$ and $b_{\mu}^{(1)}$ in the vector-field part of the Lagrangian, one can check that the composite vector field $h_{\mu}^{(1)}$ also obtains a mass identical to that of the graviton field $h_{\mu \nu 1}$. For simplicity, we have not included the scalar components, but they obtain a similar mass by supersymmetry.

\subsubsection{Gravitino}

The gravitino part of the gravity supermultiplet Lagrangian \eqref{eq:IRgravityLag} at the IR scale is given by
\begin{equation}
  \begin{split}
    \mathcal{L}_{\text{gravitino}} 
      &= - \frac{1}{2} \epsilon^{\mu \nu \rho \kappa} \psi_{\mu} \sigma_{\nu} \partial_{\rho} \psi_{\kappa}^{\dagger} 
         - \frac{1}{2} 
           \epsilon^{\mu \nu \rho \kappa} \psi_{\mu}^{(1)} \sigma_{\nu} \partial_{\rho} \psi_{\kappa}^{(1) \dagger} 
         - \frac{1}{2} 
           \epsilon^{\mu \nu \rho \kappa} \chi_{\mu}^{(1)} \sigma_{\nu} \partial_{\rho} \chi_{\kappa}^{(1) \dagger} \\[1ex]
      &    \qquad
         - \frac{1}{4} 
           \Lambda_{\text{IR}} 
           \left( {\varepsilon}_H \psi_{\mu} \left[ \sigma^{\mu}, \bar{\sigma}^{\nu} \right] 
           \chi_{\nu}^{(1)} 
         + g_H^{(1)} \psi_{\mu}^{(1)} \left[ \sigma^{\mu}, \bar{\sigma}^{\nu} \right] \chi_{\nu}^{(1)} + \text{H.c.} \right) \, ,
  \end{split}
\label{eq:gravitinoLag}
\end{equation}
where $\psi_{\mu}^{(1)}$ and $\chi_{\mu}^{(1)}$ are both contained in the tensor supermultiplet $H_{\mu}^{(1)}$. In the limit $\varepsilon_H = 0$ the massive composite state is a Dirac fermion with mass $m_H^{(1)}$.

Using the basis $( \psi_\mu, \chi_\mu^{(1)},\psi_\mu^{(1)} )$, the mass term in \eqref{eq:gravitinoLag} can be diagonalized via the transformation
\begin{equation}
  \label{eq:diaggravitino}
  O_{3/2} 
  \left( 
    \begin{matrix} 
      0 & \frac{{\varepsilon}_H}{2} & 0 \\ 
      \frac{{\varepsilon}_H}{2} & 0 & \frac{g_H^{(1)}}{2} \\ 
      0 & \frac{g_H^{(1)}}{2} & 0 
    \end{matrix} 
  \right) 
  O_{3/2}^T 
    = \left( 
        \begin{matrix} 
          0 & 0 & 0 \\ 
          0 & - \frac{\sqrt{ {\varepsilon}_H^2 + g_H^{(1)2}}}{2} & 0 \\ 
          0 & 0 & \frac{\sqrt{ {\varepsilon}_H^2 + g_H^{(1)2}}}{2} 
        \end{matrix} 
      \right) \, ,
\end{equation}
with the orthogonal matrix
\begin{equation}
  O_{3/2} 
  = \frac{1}{\sqrt{g_H^{(1)2} + {\varepsilon}_H^2}} 
    \left( 
      \begin{matrix} 
        g_H^{(1)} & 0 & -\varepsilon_H \\ 
        \frac{\varepsilon_H}{\sqrt{2}} & - \sqrt{\frac{g_H^{(1)2} + \varepsilon_H^2}{2}} & \frac{g_H^{(1)}}{\sqrt{2}} \\ 
        \frac{\varepsilon_H}{\sqrt{2}} & \sqrt{\frac{g_H^{(1)2} + \varepsilon_H^2}{2}} & \frac{g_H^{(1)}}{\sqrt{2}} 
      \end{matrix} 
    \right) \, ,
\end{equation}
which gives rise to a massless Weyl fermion and a massive Dirac fermion. The massless gravitino eigenstate then becomes
\begin{equation}
  \label{eq:masslessgravitino}
  |\psi_{\mu 0} \rangle 
    \simeq 
      \mathcal{N}_H 
      \left\{ |\psi_{\mu} \rangle - \frac{ {\varepsilon}_H}{g_H^{(1)}} | \psi_{\mu}^{(1)} \rangle \right\} \, ,
\end{equation}
Thus, the massless gravitino is an admixture of the elementary gravitino $\psi_{\mu}$ and the composite gravitino $\psi_{\mu}^{(1)}$.

The massive Dirac fermion state is given by
\begin{align}
  | \psi_{\mu 1,2} \rangle 
    &\simeq 
       \frac{\mathcal{N}_H}{\sqrt{2}} 
       \left\{ 
         \frac{ {\varepsilon}_H}{g_H^{(1)}} \, | \psi_{\mu} \rangle 
       + | \psi_{\mu}^{(1)} \rangle 
       \pm \sqrt{1+ \frac{{\varepsilon}_H^2}{g_H^{(1)2}}} \, | \chi_{\mu}^{(1)} \rangle 
       \right\} \, , \nonumber \\[1ex]
    &\simeq 
       \frac{\mathcal{N}_H}{\sqrt{2}} 
       \left\{ 
         \frac{1}{g_H^{(1)} \sqrt{\smash[b]{\zeta_H}} } \frac{\Lambda_{\text{IR}}}{\Lambda_{\text{UV}}} \, | \psi_\mu \rangle 
        + | \psi_\mu^{(1)} \rangle \pm | \chi_\mu^{(1)} \rangle 
       \right\} \, ,
\end{align}
where $\varepsilon_H \ll 1$ and terms of $\mathcal{O}(\varepsilon_H^2)$ and higher have been dropped in the second expression. The mass eigenvalues are
\begin{equation}
  m_{H_{1,2}} = \pm \sqrt{{\varepsilon}_H^2 + g_H^{(1)2}} \Lambda_{\text{IR}} \, ,
\end{equation}
which agrees with that obtained for the graviton field $h_{\mu \nu}^{(1)}$ and the vector field $h_{\mu}^{(1)}$, as expected by supersymmetry.


\section{Bulk Zero-mode Profiles in a Slice of AdS}\label{app:zero-modes}

The quadratic part of the 5D bulk action of a hypermultiplet containing complex scalar fields
$\phi$ and $\phi^c$ and Dirac fermion $\Psi$ living in a slice of AdS$_5$ is given by~\cite{Gherghetta:2000qt}
\begin{align}
  S_5 = \int d^5 x \, \sqrt{-g} \,
        \Bigl[
          - |\partial_M \phi|^2 
          - m_\phi^2 |\phi|^2 
          - |\partial_M \phi^c|^2 
&         - m_{\phi^c}^2 |\phi^c|^2 \nonumber \\[1ex]
&         + i \bar{\Psi} \Gamma^M D_M \Psi 
          - i m_{\Psi} \bar{\Psi} \Psi \,
        \Bigr] \, ,
\end{align}
where $g = \det (g_{MN})$ is the determinant of the AdS metric \eqref{eq:ads5-metric} and the curved space covariant derivative $D_M = \partial_M + \omega_M$ includes the spin connection $\omega_M$.
The bulk masses are given by
\begin{subequations}
  \begin{align}
   m_\Psi &= c \sigma^{\prime} \, , \\
   m_{\phi,\phi^c}^2 &= a k^2 + b \sigma^{\prime\prime} \, ,
  \end{align}
\end{subequations}
where $\sigma = k |y|$ and $a, b, c$ are dimensionless parameters. Performing a KK decomposition, the zero-mode profiles (with respect to a flat metric) are~\cite{Grossman:1999ra, Gherghetta:2000qt}
\begin{align}
  \tilde{f}_{\Psi_{L,R}}^{(0)}(y)
    &=  e^{- \frac{3}{2} k |y|} f_{\Psi_{L,R}}^{(0)}(y)
     = \sqrt{\frac{ (\frac{1}{2} \mp c ) k}{e^{2(\frac{1}{2} \mp c)\pi k R} - 1}} e^{(\frac{1}{2} \mp c ) k |y|} \, , \label{eq:zeromode-fermion} \\[1ex]
  \tilde{f}_{\phi}^{(0)}(y)
    &=  e^{- k |y|} f_{\phi}^{(0)}(y)
     = \sqrt{\frac{(b - 1)k}{e^{2 (b - 1) \pi k R} - 1}} e^{(b - 1) k |y|} \, , \label{eq:zeromode-scalar}
\end{align}
where the upper (lower) sign is used for the $L$ $(R)$ component and $a = b(b - 4)$ must be satisfied for a massless scalar mode. This is automatic for a hypermultiplet, where supersymmetry requires that $b = \frac{3}{2} \mp c$, such that $\tilde{f}_{\Psi_{L,R}}^{(0)}(y) = \tilde{f}_{\phi_{L,R}}^{(0)}(y) \propto e^{(\frac{1}{2} \mp c) k |y|}$ as expected for the zero-mode fermions and scalar fields in a hypermultiplet.

In a vector supermultiplet, the profile for the zero mode of the gauge boson is
\begin{equation} \label{eq:zeromode-gauge}
  \tilde{f}_A^{(0)}(y) = f_A^{(0)}(y) = \frac{1}{\sqrt{2 \pi R}} \text{ ,}
\end{equation}
while that for the gaugino corresponds to $f_{\Psi_L}^{(0)}(y)$ in \eqref{eq:zeromode-fermion} with $c = \frac{1}{2}$, and therefore matches
\eqref{eq:zeromode-gauge}. Similarly for the graviton supermultiplet the zero-mode graviton profile is given by \eqref{eq:zeromode-scalar} with $b = 0$, and for the gravitino the profile is $f_{\Psi_L}^{(0)}(y)$ in \eqref{eq:zeromode-fermion} with $c = \frac{3}{2}$, which again matches by supersymmetry.

When supersymmetry is broken as in Sec.~\ref{sec:susybreaking-5d} by the IR-brane operators \eqref{eq:susybreaking-action-sfermion} and \eqref{eq:susybreaking-action-gaugino}, the sfermion and gaugino zero modes acquire soft masses $m_{\phi}$ and $M_\lambda$. The boundary mass terms on the IR brane induce a backreaction on the KK profiles of these fields. In this case, the zero-mode profiles are determined in the same manner as the profiles of massive KK states:
\begin{align}
  \tilde{f}_{\phi}^{(0)}(y)
    &= N_{\phi} \, e^{k |y|}
       \bigg[ \,
         J_{2 - b} \Big( \frac{ m_{\phi}}{k} e^{k |y|} \Big)
       - \frac{J_{1 - b}(\frac{m_{\phi}}{k})}{Y_{1 - b}(\frac{m_{\phi}}{k})}
         Y_{2 - b} \Big( \frac{ m_{\phi}}{k} e^{k |y|} \Big) \,
       \bigg] \, , \label{eq:zeromode-sfermion-generalized} \\[1ex]
  \tilde{f}_{\lambda}^{(0)}(y)
    &= N_{\lambda} \, e^{k |y|}
       \bigg[ \,
         J_1 \Big( \frac{ M_{\lambda}}{k} e^{k |y|} \Big)
       - \frac{J_0(\frac{M_{\lambda}}{k})}{Y_0(\frac{M_{\lambda}}{k})}
         Y_1 \Big( \frac{ M_{\lambda}}{k} e^{k |y|} \Big) \,
       \bigg] \, , \label{eq:zeromode-gaugino-generalized}
\end{align}
where $N_{\phi, \lambda}$ are determined by the normalization conditions~\cite{Casagrande:2008hr}
\begin{align}
  \int_{-\pi R}^{\pi R} dy \, \left( \tilde{f}^{(0)}_{\phi}(y) \right)^2
    &= 1 \, , \label{eq:kk-orthonormality-sfermion} \\[1ex]
  \int_{-\pi R}^{\pi R} dy \, \left( \tilde{f}^{(0)}_{\lambda}(y) \right)^2
    &= 1 + g^2 (2 \pi k R)
           \frac{1}{2}
           \frac{F}{\Lambda_{\text{IR}} M_{\lambda}}
           \frac{1}{k}
           \left( \tilde{f}^{(0)}_{\lambda}(\pi R) \right)^2 \, . \label{eq:kk-orthonormality-gaugino}
\end{align}

When supersymmetry is broken, the super-Higgs effect gives rise to the gravitino coupling \eqref{eq:susybreaking-action-gravitino} on the UV brane, and the gravitino acquires a mass $m_{3/2}$ \eqref{eq:KKgrav}. This boundary mass term induces a backreaction on the gravitino KK profiles, such that the zero-mode profile is given
\begin{equation}
  \tilde{f}_{3/2}^{(0)}(y)
    = N_{3/2} \, e^{k |y|}
      \bigg[ \,
        J_2 \Big( \frac{ m_{3/2}}{k} e^{k |y|} \Big)
      - \frac{J_1(\frac{m_{3/2}}{k})}{Y_1(\frac{m_{3/2}}{k})}
        Y_2 \Big( \frac{ m_{3/2}}{k} e^{k |y|} \Big) \,
      \bigg] \, , \label{eq:zeromode-gravitino-generalized}
\end{equation}
where
\begin{equation}
  \int_{-\pi R}^{\pi R} dy \, \left( \tilde{f}^{(0)}_{3/2}(y) \right)^2
    = 1 + \frac{1}{2}
          \frac{F}{\sqrt{3} M_P m_{3/2}}
          \frac{1}{k}
          \left( \tilde{f}^{(0)}_{3/2}(0) \right)^2 \, , \label{eq:kk-orthonormality-gravitino}
\end{equation}
determines the normalization $N_{3/2}$.


\section{Radiative Corrections in a Slice of AdS}\label{app:radiativecorrections}

In this appendix we calculate the radiative corrections to the scalar masses as well as soft mass parameters in the Higgs sector arising from the 5D bulk in our model.

\subsection{Bulk scalar soft masses}\label{app:radiativecorrections-scalar}

Although the sfermions receive soft masses at tree level from their couplings to the supersymmetry-breaking sector on the IR brane, as discussed in Sec.~\ref{sec:susybreaking-5d}, quantum corrections from the bulk can become significant for UV-localized fields. At one loop, these corrections arise from the gauge sector, Yukawa couplings, and $D$-term interactions. Here, we consider the corrections to a generic bulk scalar soft mass squared in a supersymmetric theory.

\subsubsection{Gauge-sector corrections}\label{app:radiativecorrections-scalar-gauge}

In the bulk, scalars couple to gauge bosons and gauginos, generating soft mass corrections at one loop that take the form
\begin{equation}
  \label{eq:scalar-correction-gauge}
  (\Delta m^2_{\phi_i})_g = 4 g^2 C(R_{\phi_i}) \, \Pi^{\phi_i}_g \, ,
\end{equation}
where $i$ indexes the scalar field, $g$ is the (4D) gauge coupling at the IR-brane scale, $C(R)$ is the quadratic Casimir [in the SU(5) normalization] of the representation $R$, and
\begin{equation}
  \label{eq:scalar-pi-gauge}
  \Pi^{\phi_i}_g
    = \frac{2 \pi k R}{k}
      \int \frac{d^4 p}{(2\pi)^4}
      \int_{-\pi R}^{\pi R} dy
      \left( \tilde{f}^{(0)}_{\phi_i}(y) \right)^2 \left( G_A(p, y, y) - e^{-3k|y|}  G_{\lambda}(p, y, y) \right) \, ,
\end{equation}
in the limit in which we neglect the external momentum. If we parametrize the amount of supersymmetry breaking on the IR brane as $F/\Lambda_{\text{IR}} = \xi^2 k e^{-\pi k R}$, the gauge boson, $G_A$, and gaugino, $G_{\lambda}$, bulk propagators take the forms
\begin{subequations}
  \begin{align}
    G_A(p_E, y, y)
      &= \frac{1}{2} z^2 k \,
         \frac{S^0_1(x_{\text{UV}}, x) \, S^1_0(x, x_{\text{IR}})}{T^0_0(x_{\text{UV}}, x_{\text{IR}})} \, , \\[1ex]
    G_{\lambda}(p_E, y, y)
      &= \frac{1}{2} z^5 k^4 \,
         S^0_1(x_{\text{UV}}, x)
         \frac{S^1_0(x, x_{\text{IR}}) - i g^2 \pi k R \, \xi^2 \, T_1(x, x_{\text{IR}})}
              {T^0_0(x_{\text{UV}}, x_{\text{IR}}) - i g^2 \pi k R \, \xi^2 \, S^0_1(x_{\text{UV}}, x_{\text{IR}})} \, ,
  \end{align}
\end{subequations}
where $p_E = - i p$ is the Euclidean momentum (the general expressions for Euclidean 5D mixed position-momentum propagators, normalized to the interval $0 \le y \le \pi R$ are given in Ref.~\cite{Gherghetta:2000kr}), $z = e^{k |y|} / k$ is a conformal coordinate along the fifth dimension such that $z_{\text{UV}} = 1/k$ and $z_{\text{IR}} = e^{\pi k R}/k$, and the auxiliary functions
\begin{subequations}
  \begin{align}
    S^{\alpha_1}_{\alpha_2}(x_1, x_2) &= I_{\alpha_1}(x_1) \, K_{\alpha_2}(x_2) + K_{\alpha_1}(x_1) \, I_{\alpha_2}(x_2) \, , \\
    T^{\alpha_1}_{\alpha_2}(x_1, x_2) &= I_{\alpha_1}(x_1) \, K_{\alpha_2}(x_2) - K_{\alpha_1}(x_1) \, I_{\alpha_2}(x_2) \, ,
  \end{align}
\end{subequations}
are combinations of the modified Bessel functions $I, K$. The natural argument of the Bessel functions is the dimensionless variable $x = p_E z$. In the following, we suppress repeated indices: i.e. $T_{\alpha}(x_1, x_2) \equiv T^{\alpha}_{\alpha}(x_1, x_2)$.

That the loop contribution to the scalar soft masses squared is purely a supersymmetry-breaking effect can be seen in the difference
\begin{align}
  &G_A(p_E, y, y) - e^{-3k|y|} G_{\lambda}(p_E, y, y) \nonumber \\
  &      \qquad
     = - \frac{1}{2}
         \frac{1}{p_E}
         \frac{z^2}{z_{\text{UV}} z_{\text{IR}}}
         \frac{S^0_1(x_{\text{UV}}, x)}{T_0(x_{\text{UV}}, x_{\text{IR}})}
         \frac{i g^2 \pi k R \, \xi^2 \, S^0_1(x_{\text{UV}}, x)}
              {T_0(x_{\text{UV}}, x_{\text{IR}}) - i g^2 \pi k R \, \xi^2 \, S^0_1(x_{\text{UV}}, x_{\text{IR}})} \, ,
\end{align}
which vanishes (by cancellation) when supersymmetry is unbroken $(\xi=0)$. When the gaugino acquires a mass, this cancellation is shifted, and the scalar receives a correction that is quadratically divergent.

This divergence arises since the addition of boundary masses as in Sec.~\ref{sec:susybreaking-5d} deforms the superpartner KK wavefunction profiles, which results in a difference between the effective couplings of the gauge boson and of the gauginos, leading to a parametrically hard breaking of supersymmetry on the IR brane. The extra dimension protects the UV brane and the bulk from the hard breaking, but on the IR brane, where there is no finite distance separating the scalar mode from the source of supersymmetry breaking, quantum corrections to the scalar masses squared are not finite, and sensitive to the cutoff scale. We note that this divergent behavior is a peculiar feature of warped spaces and does not occur in the flat case, where this type of supersymmetry breaking is globally realized (and hence does not lead to a local distortion of field profiles). Thus, the flat-space breaking is soft, and quantum corrections are finite and independent of the cutoff scale~\cite{Antoniadis:1997zg,Antoniadis:1998sd,Delgado:1998qr,Gherghetta:2000kr}.

In the 4D dual theory, this divergence is a result of the breakdown of the perturbativity of the loop expansion as the gauge coupling becomes strong. Since we lose control over the corrections near the compactification scale, we wish to extract a well-defined, finite portion of the correction associated with the long-range physics, which includes the breaking of supersymmetry. 
The correspondence between the renormalization scale in the 4D dual theory and position in the fifth dimension suggests an appropriate regularization procedure in which we scale the effective IR brane seen by the propagators in the loop with position in the extra dimension~\cite{Randall:2001gb}. The equivalent procedure in the 5D perspective is to isolate the portion of the loop correction due to the compactification of the theory, absorbing the remaining infinite part into a counterterm~\cite{Georgi:2000ks,Cheng:2002iz,Puchwein:2003jq,Freitas:2017afm}. Since the presence of the IR brane in AdS explicitly breaks 5D Lorentz symmetry, this finite correction is nonlocal, associated with a winding around the compact dimension.
This purely curvature-dependent contribution to \eqref{eq:scalar-pi-gauge} can be extracted by employing a cutoff $\Lambda = k e^{-\pi k R}$ on the 4D momentum integral. We have checked that both of these renormalization methods are numerically equivalent, and therefore may be used interchangeably. For calculational convenience, we employ the simple cutoff scheme.

\begin{figure}[t]
  \centering
  \includegraphics{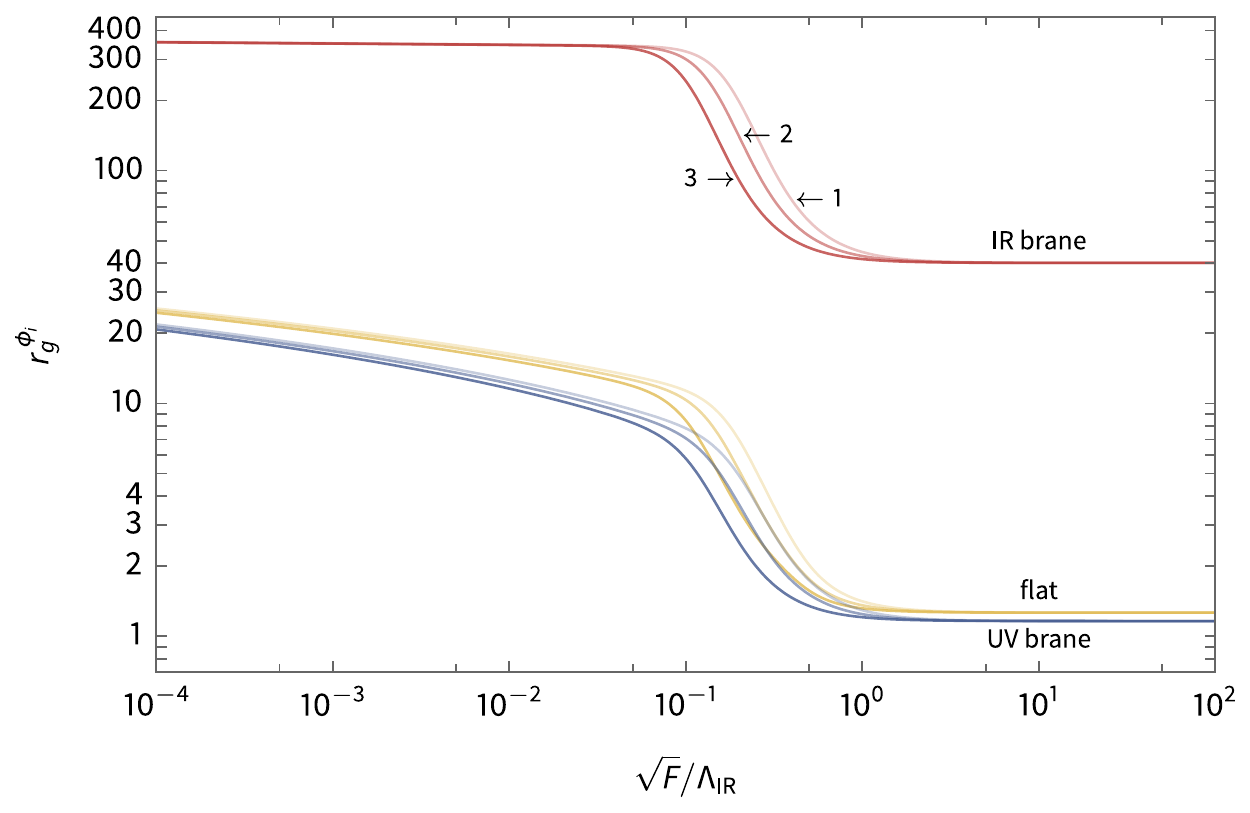}
  \caption{Plot of the coefficient $r^{\phi_i}_g$, which parametrizes the one-loop gauge corrections to bulk scalar soft masses squared, for the U(1), SU(2), and SU(3) gauge groups as a function of the relative supersymmetry breaking on the IR brane, $\sqrt{F}/\Lambda_{\text{IR}}$, for $\Lambda_{\text{IR}} = 10^7$~GeV.}
  \label{fig:scalar-gauge-loopcoefficient}
\end{figure}

After regularization, the resulting finite part of the correction can be parametrized in terms of the gaugino mass as
\begin{equation}
  \label{eq:scalar-correction-gauge-r}
  (\Delta m^2_{\phi_i})_{g} = \frac{r^{\phi_i}_g}{8\pi^2} 4 g^2 C(R_{\phi}) \, M_{\lambda}^2 \, ,
\end{equation}
where
\begin{equation}
  \label{eq:scalar-r-gauge}
  r^{\phi_i}_g = 8 \pi^2 \frac{\operatorname{Re} i \Pi^{\phi_i}_g}{M_{\lambda}^2}
\end{equation}
is a positive parameter that depends on the amount of supersymmetry breaking as well as the localization of the bulk hypermultiplet containing the scalar field (and on the IR-brane scale).
We plot $r^{\phi_i}_g$ in Fig.~\ref{fig:scalar-gauge-loopcoefficient} as a function of $\sqrt{F}/\Lambda_{\text{IR}}$ at $\Lambda_{\text{IR}} = 10^7$~GeV for three cases of the hypermultiplet localization: the two limits in which the scalar is confined to the UV and IR branes ($\pm c_i \rightarrow \infty$ and $\pm c_i \rightarrow -\infty$) as well as the case $\frac{1}{2}$ (flat) in between. For each localization, we give the U(1) (light), SU(2) (medium), and SU(3) (dark) contributions. The effect of the supersymmetry breaking saturates for $\sqrt{F}/\Lambda_{\text{IR}} \gg 1$ and $r^{\phi_i}_g$ approaches a constant value that is independent of the gauge group.

\subsubsection{Yukawa corrections}\label{app:radiativecorrections-scalar-yukawa}

The bulk scalar soft masses squared also receive contributions from their Yukawa couplings on the UV brane. At one loop, these corrections involve a boundary Higgs field and a scalar or fermion field from a different bulk hypermultiplet, taking the form:
\begin{equation}
  \label{eq:scalar-correction-yukawa}
  (\Delta m^2_{\phi_{L,R}})_y = y^2 \Pi^{\phi_{L,R}}_y \, ,
\end{equation}
where
\begin{equation}
  \label{eq:scalar-pi-yukawa}
  \Pi^{\phi_{L,R}}_y
    =
      \bigg( \frac{\tilde{f}^{(0)}_{\phi_{L,R}}(0)}{\tilde{f}^{(0)}_{\psi_L}(0) \, \tilde{f}^{(0)}_{\psi_R}(0)} \bigg)^2
      \int \frac{d^4 p}{(2\pi)^4} \,
      \left( G_{\phi_{R,L}}^{\text{UV}}(p) - G_{\psi_{R,L}}^{\text{UV}}(p) \right) \, .
\end{equation}
Here, the bulk scalar, $G_{\phi}^{\text{UV}}$, and fermion, $G_{\psi}^{\text{UV}}$, propagators are evaluated on the UV brane and take the forms
\begin{subequations}
  \begin{align}
    G_{\phi_{L,R}}^{\text{UV}}(p_E)
      &= \frac{1}{2} \frac{1}{p_E}
         \frac{ 2 p_E z_{\text{IR}} \, S^{\alpha}_{\beta}(x_{\text{UV}}, x_{\text{IR}}) - \xi^4 \, T_{\alpha}(x_{\text{UV}}, x_{\text{IR}})}
              { 2 p_E z_{\text{IR}} \, T_{\beta}(x_{\text{UV}}, x_{\text{IR}}) - \xi^4 \, S^{\beta}_{\alpha}(x_{\text{UV}}, x_{\text{IR}})} \, ,
         \label{eq:UVpropagator-scalar} \\[1ex]
    G_{\psi_{L,R}}^{\text{UV}}(p_E)
      &= \frac{1}{2} \frac{1}{p_E} \frac{S^{\alpha}_{\beta}(x_{\text{UV}}, x_{\text{IR}})}{T_{\beta}(x_{\text{UV}}, x_{\text{IR}})} \, ,
         \label{eq:UVpropagator-fermion}
  \end{align}
\end{subequations}
where $\alpha = | c \pm \frac{1}{2} |$,
\begin{equation}
  \beta = \frac{(c \pm \tfrac{1}{2})(c \mp \tfrac{1}{2})}{| c \pm \tfrac{1}{2} |} \, ,
\end{equation}
and $c$ is the fermion bulk mass parameter that specifies the localization of fields in the hypermultiplet.

The supersymmetry-breaking contribution arises from the difference between the scalar and fermion loops:
\begin{align}
   G_{\phi_{L,R}}^{\text{UV}}(p_E) - G_{\psi_{L,R}}^{\text{UV}}(p_E)
     = \frac{1}{2}
       \frac{1}{p_E^3}
  &    \frac{1}{z_{\text{UV}} z_{\text{IR}}}
       \frac{1}{T_{\beta}(x_{\text{UV}}, x_{\text{IR}})} \nonumber \\[1ex]
  &    \times
       \frac{\xi^4}
            { 2 x_{\text{IR}} \, T_{\beta}(x_{\text{UV}}, x_{\text{IR}}) -  \xi^4 \, S^{\beta}_{\alpha}(x_{\text{UV}}, x_{\text{IR}})} \, .
       \label{eq:UVpropagator-mod-scalar}
\end{align}

\begin{figure}[t]
  \centering
  \includegraphics{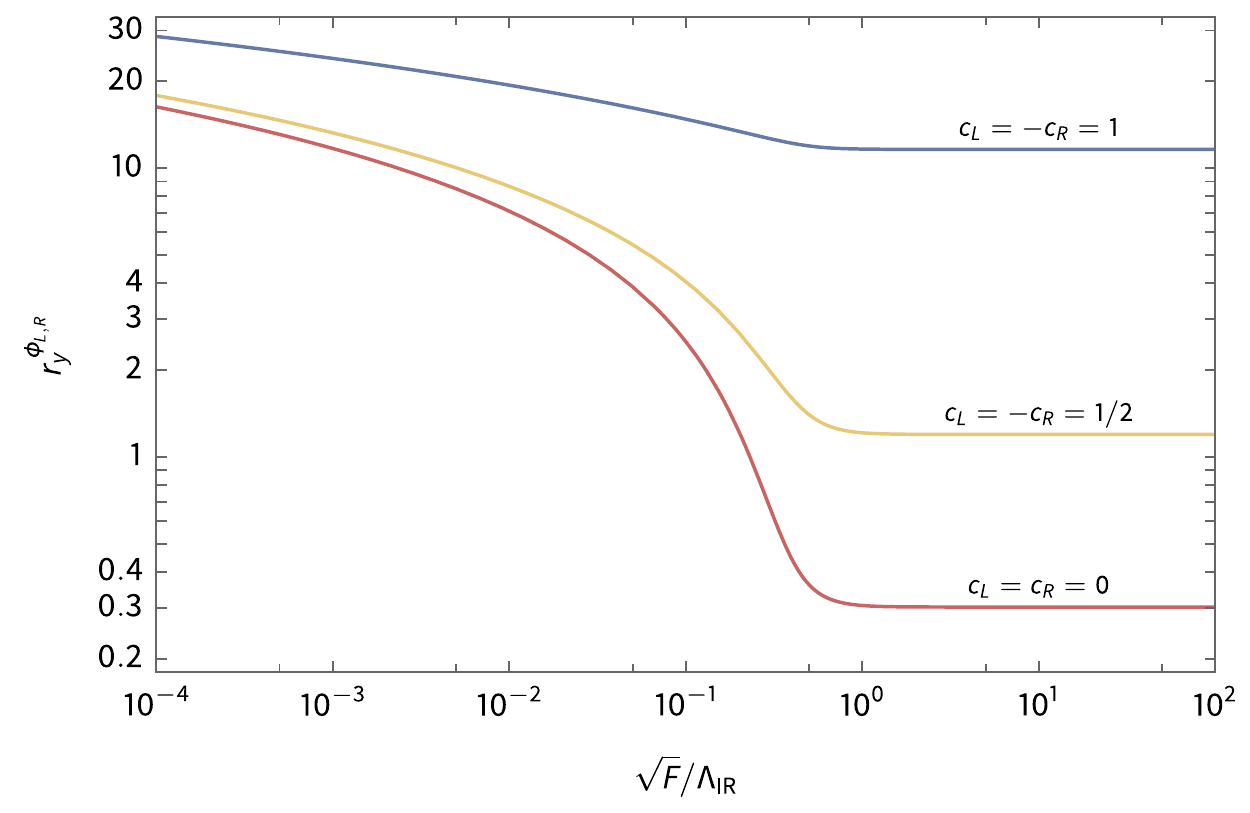}
  \caption{Plot of the coefficient $r^{\phi_{L,R}}_y$, which parametrizes the one-loop Yukawa corrections to bulk scalar soft masses squared as a function of the relative supersymmetry breaking on the IR brane, $\sqrt{F}/\Lambda_{\text{IR}}$, for $\Lambda_{\text{IR}} = 10^7$~GeV.}
  \label{fig:scalar-yukawa-loopcoefficient}
\end{figure}

When supersymmetry is broken, the correction is finite and negative and can be parametrized in terms of the soft scalar masses as
\begin{equation}
  \label{eq:scalar-correction-yukawa-r}
  (\Delta m^2_{\phi_{L,R}})_y = - \frac{r^{\phi_{L,R}}_y}{8 \pi^2} y^2 m^2_{\phi_{R,L}} \, ,
\end{equation}
where
\begin{equation}
  \label{eq:scalar-r-yukawa}
  r^{\phi_{L,R}}_y
    = - 8 \pi^2 \frac{\operatorname{Re} i \Pi^{\phi_{L,R}}_y }{m_{\phi_{R,L}}^2}
\end{equation}
is positive and depends on the amount of supersymmetry breaking and the localizations of the bulk hypermultiplets.
In Fig.~\ref{fig:scalar-yukawa-loopcoefficient} we plot $r^{\phi_{L,R}}_y$ as a function of $\sqrt{F}/\Lambda_{\text{IR}}$ for three choices of bulk hypermultiplet localizations: $c_L = -c_R = 0, \frac{1}{2}, 1$. The behavior of $r^{\phi_{L,R}}_y$ is similar to that of $r^{\phi_i}_g$, saturating to a constant for $\sqrt{F}/\Lambda_{\text{IR}} \gg 1$.

\subsubsection{\textit{D}-term corrections}\label{app:radiativecorrections-scalar-dterm}

In models such as ours, where the pattern of supersymmetry breaking is nonuniversal in flavor-space, Fayet-Iliopoulos U(1)$_Y$ $D$-term corrections to scalar soft masses squared due to supersymmetry breaking arise. At one loop, the contributing diagrams are of tadpole form, involving a bulk scalar field and a vector supermultiplet auxiliary field:
\begin{equation}
  \label{eq:scalar-correction-dterm}
  (\Delta m_{{\phi}_i}^2)_D = \frac{3}{5} g_1^2 Y({\phi}_i) \sum_j Y({\phi}_j) \, (\Pi^{\phi_i}_D)_{\phi_j} \, ,
\end{equation}
where
\begin{align}
  (\Pi^{\phi_i}_D)_{\phi_j}
    = \frac{2\pi k R}{k}
  &   \int \frac{d^4 p}{(2\pi)^4}
      \int_{-\pi R}^{\pi R} dy \, \left( \tilde{f}^{(0)}_{\phi_i}(y) \right)^2 \nonumber \\[1ex]
  &   \times
      \int_{-\pi R}^{\pi R} dy^{\prime} \,
      G_{D}(0, y, y^{\prime}) \, e^{-2k|y^{\prime}|}
      \left( G_{\phi_j}(p, y^{\prime}, y^{\prime}) - G_{\phi^c_j}(p, y^{\prime}, y^{\prime}) \right) \, ,
      \label{eq:scalar-pi-dterm-j}
\end{align}
$Y$ is the hypercharge, and the sum is over all bulk scalars. Here $\phi^c_i$ is the $\mathcal{N} = 2$ supersymmetric scalar partner of $\psi_i$, a member of the same hypermultiplet, but odd under the orbifold symmetry. The bulk scalar, $G_{\phi}$ and $G_{\phi^c}$, and auxiliary field, $G_D$, propagators take the forms
\begin{subequations}
  \begin{align}
    G_{\phi_{L,R}}(p_E, y, y)
      &= \frac{1}{2} z^4 k^3 \,
         S^{\beta}_{\alpha}(x_{\text{UV}}, x)
         \frac{ 2 p_E z_{\text{IR}} \, S^{\alpha}_{\beta}(x, x_{\text{IR}}) - \xi^4 \, T_{\alpha}(x, x_{\text{IR}})}
              { 2 p_E z_{\text{IR}} \, T_{\beta}(x_{\text{UV}}, x_{\text{IR}}) - \xi^4 \, S^{\beta}_{\alpha}(x_{\text{UV}}, x_{\text{IR}})} \, ,
         \label{eq:bulkpropagator-scalar}  \\[1ex]
    G_{\phi^c_{L,R}}(p_E, y, y)
      &= \frac{1}{2} z^4 k^3
         \frac{T_{\alpha}(x_{\text{UV}}, x) \, T_{\alpha}(x, x_{\text{IR}})}{T_{\alpha}(x_{\text{UV}}, x_{\text{IR}})} \, ,
         \label{eq:bulkpropagator-scalar-odd} \\[1ex]
    G_{D}(0, y, y^{\prime})
      &= \frac{k}{2 \pi k R}  \label{eq:eq:bulkpropagator-vectorauxiliary} \, ,
  \end{align}
\end{subequations}
where we have evaluated the auxiliary field propagator in the zero-momentum limit, as there is no momentum flow into the scalar loop. Since the auxiliary field has no $y$-dependence in this limit, we can use the orthonormality condition for the scalar zero-mode profile \eqref{eq:kk-orthonormality-sfermion} to do the first integral over the fifth dimension, leaving
\begin{equation}
  \label{eq:scalar-pi-dterm}
  (\Pi^{\phi_i}_D)_{\phi_j}
  \rightarrow
  \Pi^D_{\phi_j}
    = \int \frac{d^4 p}{(2\pi)^4}
      \int_{-\pi R}^{\pi R} dy \, e^{-2k|y|}
      \left( G_{\phi_j}(p, y, y) - G_{\phi^c_j}(p, y, y) \right) \, .
\end{equation}
This quantity is independent of the localization of the external scalar, indicating that the sum
\begin{equation}
  \sum_i Y({\phi}_i) \, \Pi^D_{\phi_i}
\end{equation}
gives a universal correction for all scalars (bulk and boundary) that are charged under the U(1) gauge symmetry.

\begin{figure}[t]
  \centering
  \includegraphics{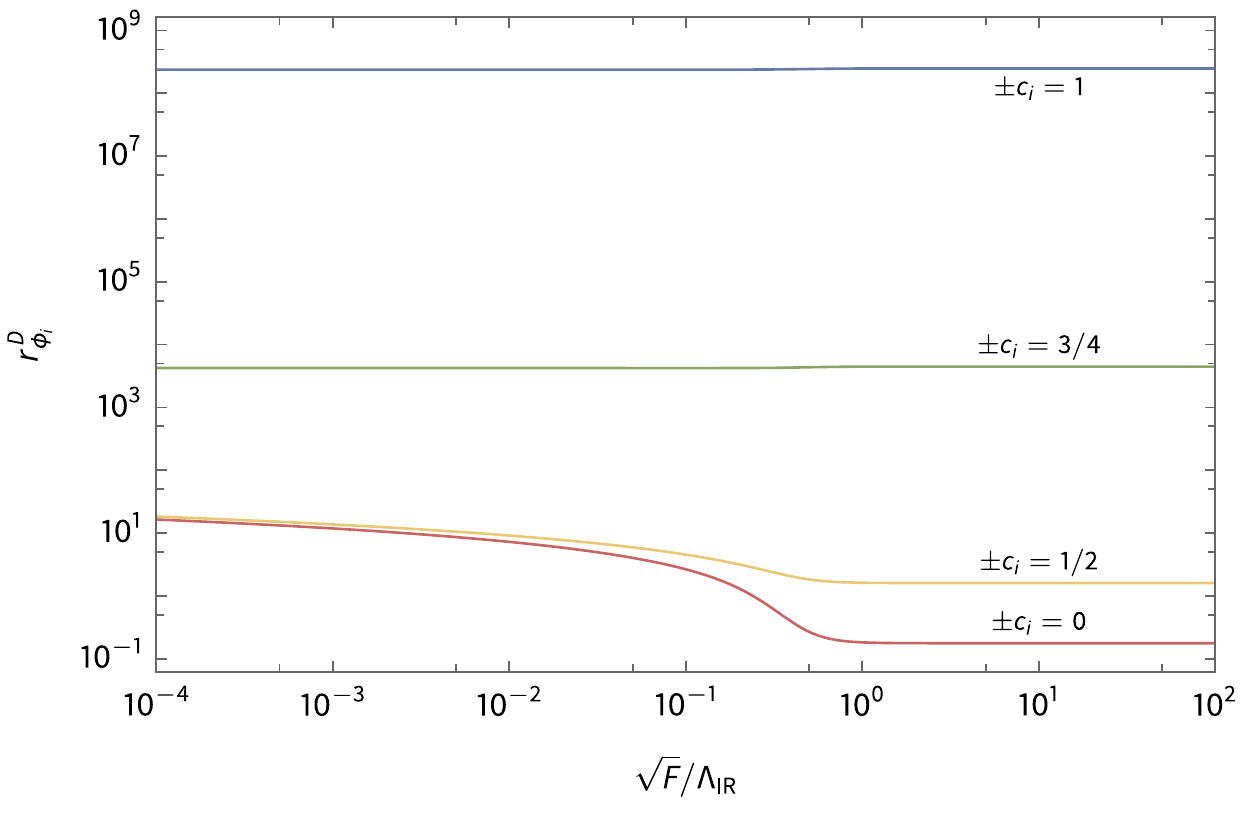}
  \caption{Plot of the coefficient $r^D_{\phi_i}$, which parametrizes the one-loop corrections to a bulk scalar soft masses squared due to a $D$-term coupling with the bulk scalar $\phi_i$ as a function of the relative supersymmetry breaking on the IR brane, $\sqrt{F}/\Lambda_{\text{IR}}$, for $\Lambda_{\text{IR}} = 10^7$~GeV.}
  \label{fig:scalar-dterm-loopcoefficient}
\end{figure}

As with \eqref{eq:scalar-correction-gauge}, the loop integral here is divergent. However, the leading divergences in this case arise even when supersymmetry is unbroken.\footnote{Both quadratic and linear divergences arise in this manner; the former on the branes, and the latter in both the bulk and on the branes~\cite{Hirayama:2003kk}. The quadratic divergences depend on the hypercharge, and hence vanish in the trace over the scalars, provided that the sum of the scalar hypercharges is zero (as is true for the MSSM field content). The linear divergences can be absorbed in a renormalization of the hypermultiplet bulk mass when regularized with a position-dependent cutoff.} The $D$-term contribution to the soft scalar mass squared arising purely as a supersymmetry-breaking effect can be extracted by considering the difference between loop integrals in broken and unbroken supersymmetry: schematically,
\begin{equation}
  ( \Pi^D_{\phi_i} )_{\xi} = \Pi^D_{\phi_i} - \bigl[ \Pi^D_{\phi_i} \bigr]_{\xi = 0} \, .
\end{equation}
This is equivalent to evaluating the loop integral using the difference in propagators:
\begin{align}
  &G_{\phi_{L,R}}(p_E, y, y) - \big[ G_{\phi_{L,R}}(p_E, y, y) \big]_{\xi = 0} \nonumber \\[1ex]
  &  \qquad
   = \frac{1}{2}
     \frac{1}{p_E}
     \frac{z^4}{z_{\text{UV}}^3 z_{\text{IR}}}
     \frac{S^{\beta}_{\alpha}(x_{\text{UV}}, x)}{T_{\beta}(x_{\text{UV}}, x_{\text{IR}})}
     \frac{\xi^4 \, S^{\beta}_{\alpha}(x_{\text{UV}}, x)}
          { 2 p_E z_{\text{IR}} \, T_{\beta}(x_{\text{UV}}, x_{\text{IR}}) -  \xi^4 \, S^{\beta}_{\alpha}(x_{\text{UV}}, x_{\text{IR}})} \, .
     \label{eq:bulkpropagator-mod-scalar}
\end{align}
The supersymmetry-breaking contribution extracted in this manner is linearly divergent. After regularizing the integrals, the resulting finite part of the correction is negative and can be parametrized in terms of the scalar mass:
\begin{equation}
  \label{eq:scalar-correction-dterm-r}
  (\Delta m_{\phi_i}^2)_D
    = - \frac{1}{8 \pi^2} \frac{3}{5} g_1^2 Y(\phi_i) \sum_j Y(\phi_j) \, r^D_{\phi_j} m_{\phi_j}^2
    \equiv - \frac{1}{8 \pi^2} \frac{3}{5} g_1^2 Y(\phi_i) \, \Delta_{\mathcal{S}} \, ,
\end{equation}
where
\begin{equation}
  \label{eq:scalar-r-dterm}
  r^D_{\phi_i} = - 8 \pi^2 \frac{\operatorname{Re} i ( \Pi^D_{\phi_i} )_{\xi} }{m_{\phi_i}^2}
\end{equation}
is positive and depends on the amount of supersymmetry breaking as well as the localizations of the bulk scalars. In Fig.~\ref{fig:scalar-dterm-loopcoefficient} we plot $r^D_{\phi_i}$ as a function of $\sqrt{F}/\Lambda_{\text{IR}}$, considering four cases for the hypermultiplet localization of the internal scalar: $c_i = 0, \frac{1}{2}, \frac{3}{4}, 1$. The largeness of $r^D_{\phi_i}$ when the bulk hypermultiplet is highly UV-localized is due to the exponential smallness of the scalar tree level mass in that case, rather than the absolute magnitude of the correction.

\subsection{Higgs soft scalar masses}\label{app:radiativecorrections-higgs}

When the Higgs fields are confined to the UV brane, they have no direct couplings to the supersymmetry-breaking sector on the IR brane. The Higgs-sector soft mass terms are therefore zero at tree level, and are generated instead at higher loop order by radiative corrections involving bulk fields that transmit the supersymmetry breaking from the IR brane.
Here, we consider the one-loop corrections to the soft mass squared for a generic Higgs field completely localized on the UV brane. A similar one-loop analysis was performed in the 4D KK formalism for the Higgs sector in unbroken supersymmetry in Ref.~\cite{Bouchart:2011va}.

\subsubsection{Gauge-sector corrections}\label{app:radiativecorrections-higgs-gauge}

The contributions to the Higgs scalar soft masses squared from the gauge sector arise from loops of bulk gauge bosons and gauginos. At one loop, the corrections involve one bulk field and one boundary field and induce the soft mass correction~\cite{Mirabelli:1997aj, Antoniadis:1998sd, Delgado:1998qr, Gherghetta:2003he, Abel:2010vb}
\begin{equation}
  \label{eq:higgs-correction-gauge}
  (\Delta m^2_{H_i})_g = 4 g^2 C(R_{H_i}) \, \Pi^H_g \, ,
\end{equation}
where $i = u, d$ indexes the Higgs field and
\begin{equation}
  \label{eq:higgs-pi-gauge}
  \Pi^H_g = \frac{2 \pi k R}{k} \int \frac{d^4 p}{(2\pi)^4} \left( G_A^{\text{UV}}(p) - G_{\lambda}^{\text{UV}}(p) \right) \, .
\end{equation}
The gauge boson, $G^{\text{UV}}_A$, and gaugino, $G^{\text{UV}}_{\lambda}$, bulk propagators are evaluated on the UV brane where the Higgs fields are localized, taking the forms
\begin{subequations}
  \begin{align}
  \label{eq:UVpropagator-gaugeboson}
    G_A^{\text{UV}}(p_E)
      &= \frac{1}{2} \frac{1}{p_E} \frac{S^1_0(x_{\text{UV}}, x_{\text{IR}})}{T_0(x_{\text{UV}}, x_{\text{IR}})} \, ,\\[1ex]
    G_{\lambda}^{\text{UV}}(p)
      &= \frac{1}{2}
         \frac{1}{p_E}
         \frac{S^1_0(x_{\text{UV}}, x_{\text{IR}}) - i g^2 \pi k R \, \xi^2 \, T_1(x_{\text{UV}}, x_{\text{IR}})}
              {T_0(x_{\text{UV}}, x_{\text{IR}}) - i g^2 \pi k R \, \xi^2 \, S^0_1(x_{\text{UV}}, x_{\text{IR}})} \, .
    \label{eq:UVpropagator-gaugino}
  \end{align}
\end{subequations}
We note that this correction \eqref{eq:higgs-correction-gauge} is a special case of the general bulk scalar soft mass squared correction \eqref{eq:scalar-correction-gauge}, corresponding to the limit in which the scalar is confined to the UV brane.

\begin{figure}[t]
  \centering
  \includegraphics{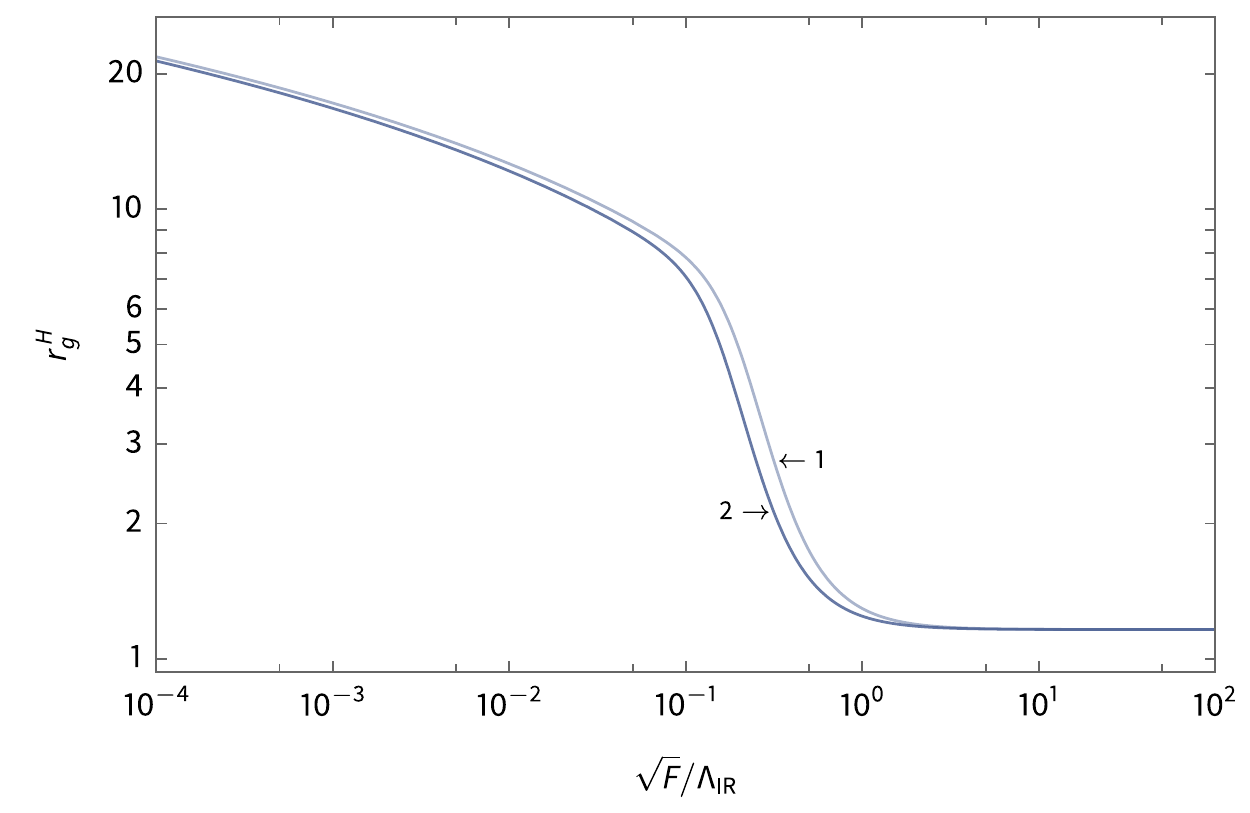}
  \caption{Plot of the coefficient $r^H_g$, which parametrizes the one-loop gauge corrections to the Higgs soft masses squared, for the U(1) and SU(2) gauge groups as a function of the relative supersymmetry breaking on the IR brane, $\sqrt{F}/\Lambda_{\text{IR}}$, for $\Lambda_{\text{IR}} = 10^7$~GeV.}
  \label{fig:higgs-gauge-loopcoefficient}
\end{figure}

The contribution of the loop integral from supersymmetry breaking is extracted in the propagator difference:
\begin{align}
  G_A^{\text{UV}}(p_E) - G_{\lambda}^{\text{UV}}(p_E)
    = - \frac{1}{2}
        \frac{1}{p_E^3}
  &     \frac{1}{z_{\text{UV}} z_{\text{IR}}}
        \frac{1}{T_0(x_{\text{UV}}, x_{\text{IR}})} \nonumber \\[1ex]
  &     \times
        \frac{i g^2 \pi k R \, \xi^2}
             {T_0(x_{\text{UV}}, x_{\text{IR}}) - i g^2 \pi k R \, \xi^2 \, S^0_1(x_{\text{UV}}, x_{\text{IR}})} \, .
      \label{eq:UVpropagator-mod-gauge}
\end{align}
The loop integral \eqref{eq:higgs-pi-gauge} is finite, unlike the bulk scalar case \eqref{eq:scalar-pi-dterm}, due to the finite separation between the Higgs fields on UV brane and the supersymmetry-breaking sector on the IR brane. The resulting contribution to the Higgs soft masses squared can be parametrized in terms of the gaugino mass as:
\begin{equation}
  \label{eq:higgs-correction-gauge-r}
  (\Delta m^2_{H_i})_g = \frac{r^H_g}{8\pi^2} 4 g^2 C(R_{H_i}) \, M_{\lambda}^2 \, ,
\end{equation}
where
\begin{equation}
  \label{eq:higgs-r-gauge}
  r^H_g = 8 \pi^2 \frac{\operatorname{Re} i \Pi^H_g}{M_{\lambda}^2}
\end{equation}
depends on the amount of supersymmetry breaking. We plot $r^H_g$ in Fig.~\ref{fig:higgs-gauge-loopcoefficient} for the U(1) and SU(2) gauge groups as a function of $\sqrt{F}/\Lambda_{\text{IR}}$. This behavior matches the UV-brane limit of the bulk scalar corrections in Fig.~\ref{fig:scalar-gauge-loopcoefficient} and reproduces that found in Ref.~\cite{Abel:2010vb} up to an order-one shift due to a difference in the definition of the supersymmetry-breaking gaugino IR-brane operator and the UV cutoff of the 4D momentum integration.

\subsubsection{Yukawa corrections}\label{app:radiativecorrections-higgs-yukawa}

Similarly, the Higgs soft masses squared receive contributions via their Yukawa interactions with bulk fermions and scalars. At one loop, the corrections each involve two bulk fields and take the form
\begin{equation}
  \label{eq:higgs-correction-yukawa}
  (\Delta m^2_{H_i})_y = y^2 \Pi^H_y \, ,
\end{equation}
where
\begin{align}
  \Pi^H_y
    = {} & \bigg( \frac{1}{ \tilde{f}^{(0)}_{\psi_L}(0) \, \tilde{f}^{(0)}_{\psi_R}(0) } \bigg)^2 \nonumber \\[1ex]
         & \times
           \int \frac{d^4 p}{(2\pi)^4} \,
           \left(
             G_{{\phi}_L}^{\text{UV}}(p) \, G_{F_R}^{\text{UV}}(p)
           + G_{{\phi}_R}^{\text{UV}}(p) \, G_{F_L}^{\text{UV}}(p)
           - 2 p^2 G_{\psi_L}^{\text{UV}}(p) \, G_{\psi_R}^{\text{UV}}(p)
           \right) \, .
           \label{eq:higgs-pi-yukawa}
\end{align}
The UV-brane bulk scalar and fermion propagators are given in \eqref{eq:UVpropagator-scalar} and \eqref{eq:UVpropagator-fermion}. The bulk hypermultiplet auxiliary field propagator $G_{F}^{\text{UV}}$ (in the style of Ref.~\cite{ArkaniHamed:2001mi}) is likewise evaluated on the UV brane and takes the form
\begin{equation}
  G_{F_{L,R}}^{\text{UV}}(p_E)
    = - p_E^2 \big[ G_{{\phi}_{L,R}}^{\text{UV}}(p_E) \big]_{\xi = 0}
    = - p_E^2 \, G_{\psi_{L,R}}^{\text{UV}}(p_E) \, .
\end{equation}

\begin{figure}[t]
  \centering
  \includegraphics{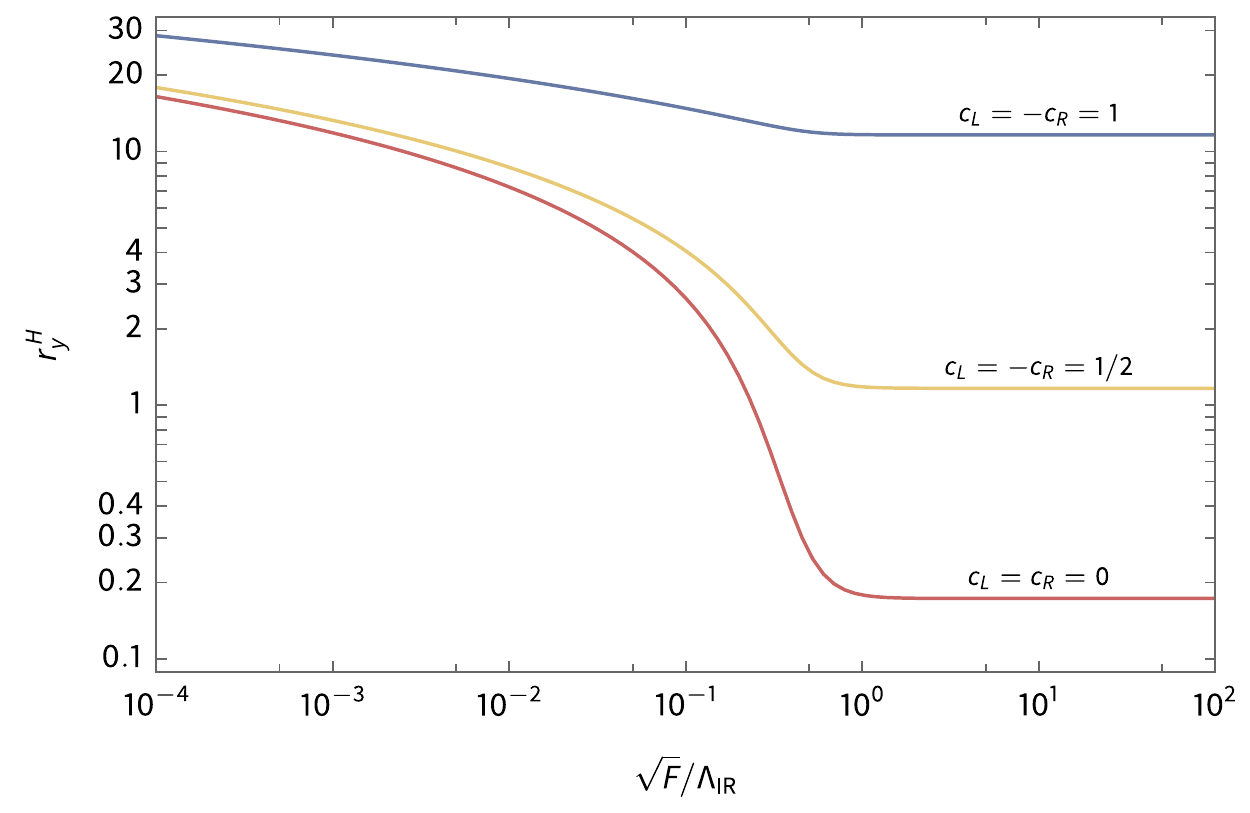}
  \caption{Plot of the coefficient $r^H_y$, which parametrizes the one-loop Yukawa corrections to the Higgs soft masses squared, as a function of the relative supersymmetry breaking on the IR brane, $\sqrt{F}/\Lambda_{\text{IR}}$, for $\Lambda_{\text{IR}} = 10^7$~GeV.}
  \label{fig:higgs-yukawa-loopcoefficient}
\end{figure}

As with the bulk scalar Yukawa corrections, the contribution to the loop integral from supersymmetry breaking is extracted in the propagator differences \eqref{eq:UVpropagator-mod-scalar}. The resulting contribution to the Higgs soft masses squared is finite and negative and can be parametrized in terms of the scalar masses squared as
\begin{equation}
  \label{eq:higgs-correction-yukawa-r}
  (\Delta m^2_{H_i})_y = - \frac{r^H_{y}}{8\pi^2} y^2 ( m_{\phi_L}^2 + m_{\phi_R}^2 ) \, ,
\end{equation}
where
\begin{equation}
  \label{eq:higgs-r-yukawa}
  r^H_y = - 8 \pi^2 \frac{\operatorname{Re} i \Pi^H_y}{( m_{\phi_L}^2 + m_{\phi_R}^2 )}
\end{equation}
is positive and depends on the amount of supersymmetry breaking as well as the localization of the bulk fields.
In Fig.~\ref{fig:higgs-yukawa-loopcoefficient} we plot $r^H_y$ as a function of $\sqrt{F}/\Lambda_{\text{IR}}$ for three choices of the hypermultiplet localization: $c_L = - c_R = 0, \frac{1}{2}, 1$.

\subsubsection{\textit{D}-term corrections}\label{app:radiativecorrections-higgs-dterm}

Since, as discussed above, the $D$-term corrections to the scalar soft masses squared are independent of the localization of the scalar,
the Higgs-sector corrections on the boundary take the same form as the corrections in the bulk:
\begin{equation}
  \label{eq:higgs-correction-dterm}
  (\Delta m_{H_i}^2)_{D} = \frac{3}{5} g_1^2 \, Y(H_i) \sum_j Y(\phi_j) \, \Pi^D_{\phi_j} \, ,
\end{equation}
where $\Pi^D_{\phi_j}$ is defined in \eqref{eq:scalar-pi-dterm}. We regulate as discussed in Sec.~\ref{app:radiativecorrections-scalar-dterm}, resulting in a finite negative contribution to the Higgs soft masses squared, parametrized in terms of the scalar mass as
\begin{equation}
  \label{eq:higgs-correction-dterm-r}
  (\Delta m^2_{H_i})_D
    = - \frac{1}{8\pi^2} \frac{3}{5} g_1^2 \, Y(H_i) \sum_j Y(\phi_j) \, r^D_{\phi_j} m_{\phi_j}^2\, ,
\end{equation}
where $r^D_{\phi_j}$ is defined in \eqref{eq:scalar-r-dterm}.

\subsection{Higgs soft \textit{b}-term}\label{sec:radiativecorrection-bterm}

\begin{figure}[t]
  \centering
  \includegraphics{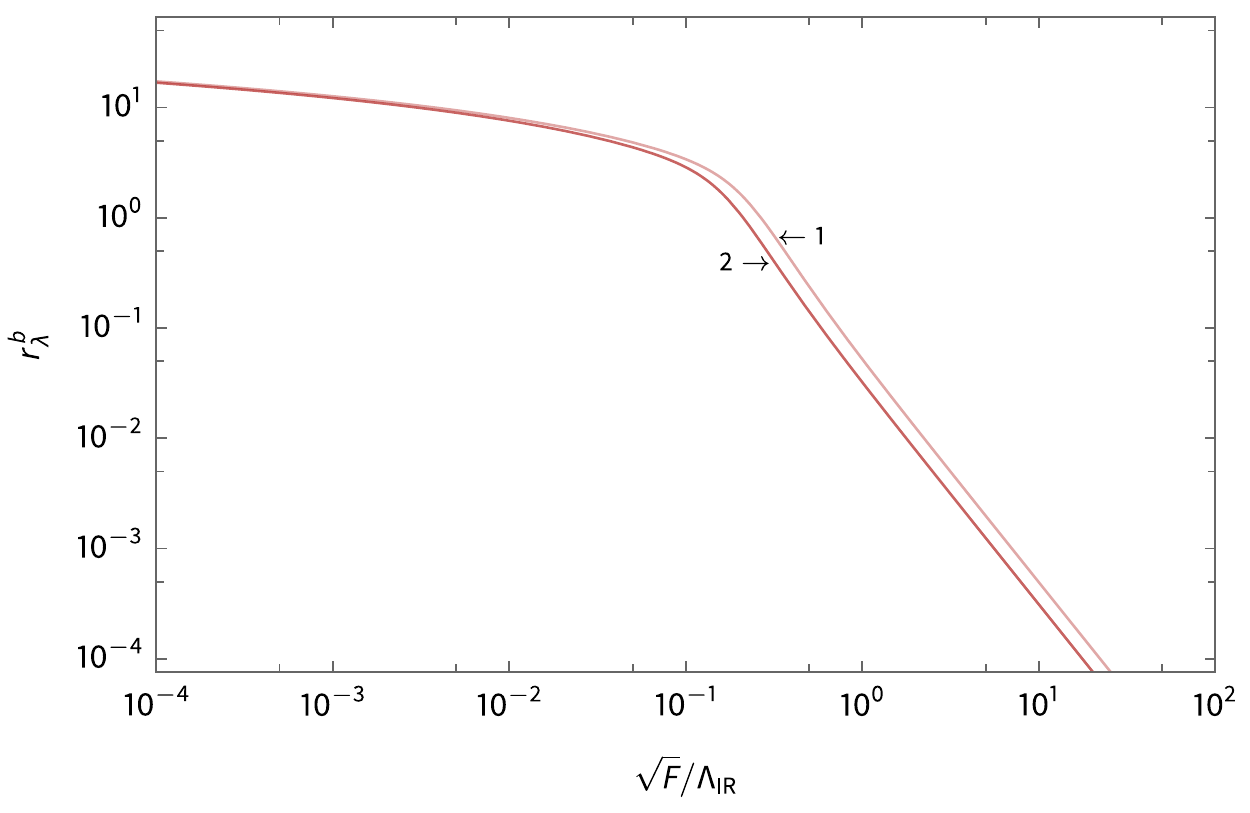}
  \caption{Plot of the coefficient $r^b_{\lambda}$, which parametrizes the one-loop gaugino corrections to the Higgs soft $b$-term, as a function of the relative supersymmetry breaking on the IR brane, $\sqrt{F}/\Lambda_{\text{IR}}$, for $\Lambda_{\text{IR}} = 10^7$~GeV.}
  \label{fig:softb-gaugino-loopcoefficient}
\end{figure}

The Higgs soft $b$-term, like the Higgs soft masses squared, is zero at tree level, and so is generated at loop order once supersymmetry is broken. In this case, the only corrections which contribute at one loop arise in the gauge sector, involving one bulk gaugino field and one boundary Higgsino field
\begin{equation}
  \label{eq:softb-correction-gaugino}
  (\Delta b)_{\lambda} = 4 g^2 T^a(R_{H_u}) \, T^a(R_{H_d}) \, \Pi^b_{\lambda} \, ,
\end{equation}
where $T^a(R)$ is the generator of the gauge group in representation $R$, and
\begin{equation}
  \label{eq:softb-pi-gaugino}
  \Pi^b_{\lambda}
    = - \frac{2 \pi k R}{k}
        \int \frac{d^4 p}{(2\pi)^4}
        \frac{\mu}{p^2}
        G_{M_\lambda}^{\text{UV}}(p) \, .
\end{equation}
The Majorana mass-mixing component of the gaugino propagator, $G_{M_\lambda}^{\text{UV}}(p)$, evaluated on the UV brane, is given by~\cite{Nomura:2003qb}
\begin{equation}
  \label{eq:UVpropagator-gaugino-majorana}
  G_{M_{\lambda}}^{\text{UV}}(p_E)
    = - \frac{1}{2}
        \frac{1}{p_E^2}
        \frac{z_{\text{UV}}}{z_{\text{IR}}}
        \frac{i}{S^0_1(x_{\text{UV}}, x_{\text{IR}})}
        \frac{1}{T_0(x_{\text{UV}}, x_{\text{IR}}) - i g^2 \pi k R \, \xi^2 \, S^0_1(x_{\text{UV}}, x_{\text{IR}})} \, .
\end{equation}

Unlike the previous loop corrections, the $b$-term one-loop gaugino correction depends explicitly on supersymmetry breaking through the Majorana gaugino mass, and no complementary loop of superpartners is present. The resulting contribution is finite and negative, and we parametrize it in terms of the Higgsino mass $\mu$ and the gaugino mass as:
\begin{equation}
  \label{eq:softb-correction-gaugino-r}
  (\Delta b)_{\lambda} = - \frac{r^b_{\lambda}}{8 \pi^2} 4 g^2 T^a(R_{H_u}) \, T^a(R_{H_d}) \, \mu M_{\lambda} \, ,
\end{equation}
where
\begin{equation}
  \label{eq:softb-r-gaugino}
  r^b_{\lambda} = - 8 \pi^2 \frac{\operatorname{Re} i \Pi^b_{\lambda}}{\mu M_{\lambda}}
\end{equation}
is positive and depends on the amount of supersymmetry breaking and $\mu$. We plot $r^b_{\lambda}$ in Fig.~\ref{fig:softb-gaugino-loopcoefficient} as a function of $\sqrt{F}/\Lambda_{\text{IR}}$ for the U(1) (lighter) and SU(2) (darker) gauge groups.
In the limit $\sqrt{F}/\Lambda_{\text{IR}} \gg 1$, $r^b_{\lambda}$ tends to zero as $1/\xi$. This is a result of the fact that in the twisted limit the gaugino mass is pure Dirac and the Majorana mixing that generates the soft coupling disappears. When $\sqrt{F}/\Lambda_{\text{IR}} \ll 1$, the effect of the supersymmetry breaking saturates, and $r^b_{\lambda}$ approaches a constant value.

\begin{figure}[t]
  \centering
  \includegraphics{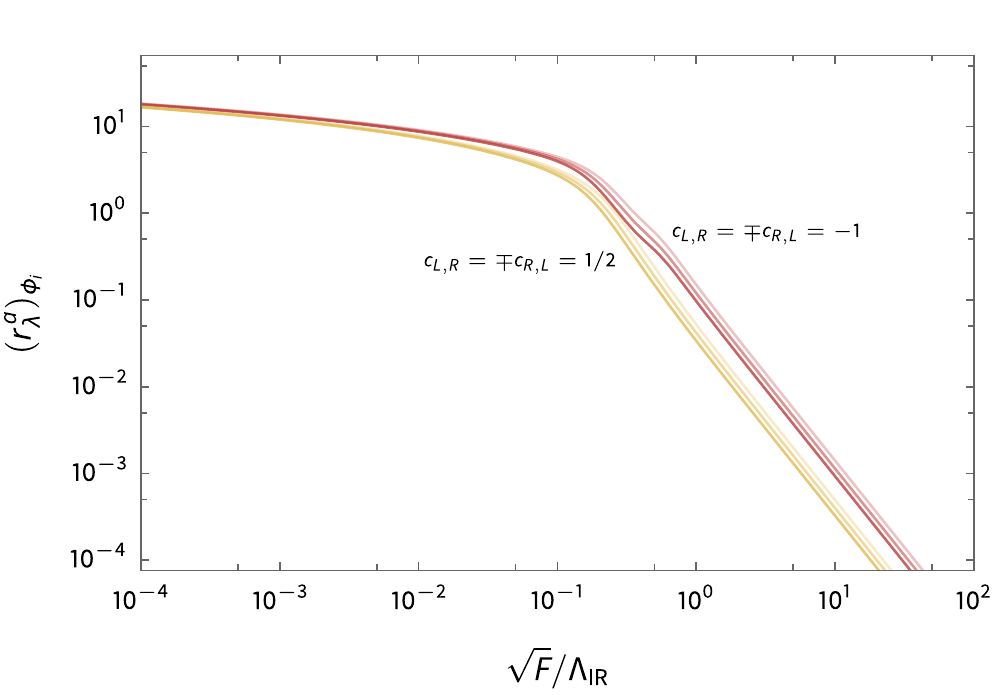}
  \caption{Plot of the coefficient $(r^a_{\lambda})_{\phi_{L,R}}$, which parametrizes the one-loop gaugino corrections to the trilinear soft scalar couplings, as a function of the relative supersymmetry breaking on the IR brane, $\sqrt{F}/\Lambda_{\text{IR}}$, for $\Lambda_{\text{IR}} = 10^7$~GeV.}
  \label{fig:softa-gaugino-loopcoefficient-1}
\end{figure}

\begin{figure}[t]
  \centering
  \includegraphics{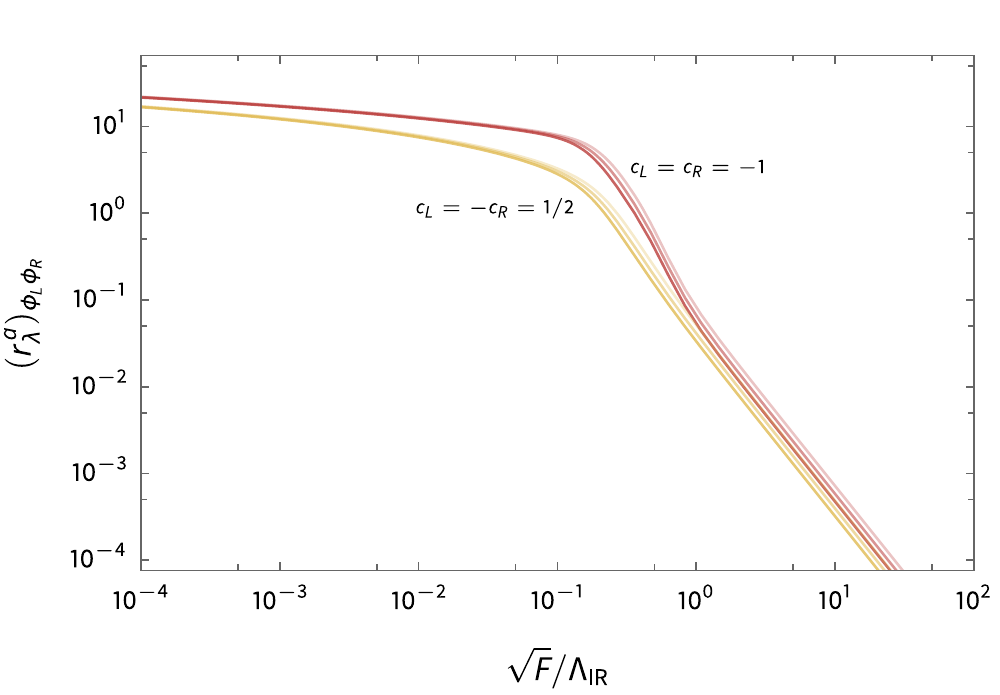}
  \caption{Plot of the coefficient $(r^a_{\lambda})_{\phi_L \phi_R}$, which parametrizes the one-loop gaugino corrections to the trilinear soft scalar couplings, as a function of the relative supersymmetry breaking on the IR brane, $\sqrt{F}/\Lambda_{\text{IR}}$, at $\Lambda_{\text{IR}} = 10^7$~GeV.}
  \label{fig:softa-gaugino-loopcoefficient-2}
\end{figure}

\subsection{Trilinear soft scalar couplings (\textit{a}-terms)}\label{app:radiativecorrections-aterms}

The soft $a$-term interactions, like the Higgs-sector soft terms, are zero at tree level but are generated at loop order once supersymmetry is broken. At one loop, the only nonvanishing corrections arise from loops of bulk gauginos, bulk fermions, and Higgsinos:
\begin{align}
  (\Delta a)_{\lambda}
    = 4 y g^2
      \Big[ \,
  &     T^a(R_H) \, T^a(R_{\phi_L}) \, (\Pi^a_{\lambda})_{\phi_L} \nonumber \\
  &     \quad
      + T^a(R_H) \, T^a(R_{\phi_R}) \, (\Pi^a_{\lambda})_{\phi_R}
      + T^a(R_{\phi_L}) \, T^a(R_{\phi_R}) \, (\Pi^a_{\lambda})_{\phi_L \phi_R} \,
      \Big] \, ,
      \label{eq:softa-correction-gaugino}
\end{align}
where
\begin{subequations}
  \begin{align}
    (\Pi^a_{\lambda})_{\phi_{L,R}}
      &= \begin{aligned}[t]
           - \frac{2 \pi k R}{k}
             \int \frac{d^4 p}{(2\pi)^4}
         &   \int_{-\pi R}^{\pi R} dy \,
             \frac{\tilde{f}^{(0)}_{\phi_{L,R}}(y)}{\tilde{f}^{(0)}_{\psi_{L,R}}(0)}
             \frac{\tilde{f}^{(0)}_{\phi_{R,L}}(0)}{\tilde{f}^{(0)}_{\psi_{R,L}}(0)} \,
             e^{-3k|y|} \\[1ex]
         &   \times
             G_{\psi_{L,R}}(p, 0, y) \, G_{M_{\lambda}}(p, y, 0)~,
         \end{aligned}
         \label{eq:softa-pi-gaugino-1} \\[1ex]
    (\Pi^a_{\lambda})_{\phi_L \phi_R}
      &= \begin{aligned}[t]
           - \frac{2 \pi k R}{k}
             \int \frac{d^4 p}{(2\pi)^4}
             \int_{-\pi R}^{\pi R} dy
         &   \int_{-\pi R}^{\pi R} dy^{\prime} \,
             p^2 \,
             \frac{\tilde{f}^{(0)}_{\phi_L}(y)}{\tilde{f}^{(0)}_{\psi_L}(0)}
             \frac{\tilde{f}^{(0)}_{\phi_R}(y^{\prime})}{\tilde{f}^{(0)}_{\psi_R}(0)} \,
             e^{-3k|y|} \,
             e^{-3k|y^{\prime}|} \\[1ex]
         &   \times
             G_{\psi_L}(p, 0, y) \, G_{M_{\lambda}}(p, y, y^{\prime}) \, G_{\psi_R}(p, y^{\prime}, 0)~,
         \end{aligned}
         \label{eq:softa-pi-gaugino-2}
  \end{align}
\end{subequations}
are the contributing loop integrals. The bulk gaugino Majorana mass-mixing propagator and the bulk fermion propagator take the forms
\begin{subequations}
  \begin{align}
    G_{M_{\lambda}}(p_E, y, y^{\prime})
      &= \begin{aligned}[t]
            - \frac{1}{2}
              \frac{(z z^{\prime})^{5/2}}{z_{\text{UV}}^4 z_{\text{IR}}}
           &  \frac{i}{S^0_1(x_{\text{UV}}, x_{\text{IR}})} \\[1ex]
           &  \times
              \frac{i S^0_1(x_{\text{UV}}, x_{>}) \, S^0_1(x_{\text{UV}}, x_{<})}{T_0(x_{\text{UV}}, x_{\text{IR}}) - i g^2 \pi k R \, \xi^2 \, S^0_1(x_{\text{UV}}, x_{\text{IR}})} \, ,
          \end{aligned}
          \label{eq:bulkpropagator-gaugino-majorana} \\[1ex]
    G_{\psi_{L,R}}(p_E, 0, y)
      &= \frac{1}{2}
         \frac{1}{p_E} (z k)^{5/2}
         \frac{S^{\alpha}_{\beta}(x, x_{\text{IR}})}{T_{\beta}(x_{\text{UV}}, x_{\text{IR}})} \, ,
  \end{align}
\end{subequations}
where the values $z_{>(<)}$ represent the greater (lesser) of $z$ and $z^{\prime}$.

As with the soft $b$-term, this gaugino correction alone depends explicitly on supersymmetry breaking, without the presence of the corresponding loop of superpartners. The loop integrals \eqref{eq:softa-pi-gaugino-1} and \eqref{eq:softa-pi-gaugino-2} are finite, despite their extension into the bulk. The resulting correction is negative and can be parametrized in terms of the gaugino mass as
\begin{align}
  \label{eq:softa-correction-gaugino-r}
  (\Delta a)_{\lambda}
    = - \frac{1}{8\pi^2} 4 y g^2
        \Big[ \,
          (r^a_{\lambda})_{\phi_L} T^a(R_H) \, T^a(R_{\phi_L})
  &     + (r^a_{\lambda})_{\phi_R} T^a(R_H) \, T^a(R_{\phi_R}) \nonumber \\[1ex]
  &     + (r^a_{\lambda})_{\phi_L \phi_R} T^a(R_{\phi_L}) \, T^a(R_{\phi_R}) \,
        \Big]
        M_{\lambda} \, ,
\end{align}
where
\begin{subequations}
  \begin{align}
    (r^a_{\lambda})_{\phi_{L,R}}
      &= - 8 \pi^2 \frac{\operatorname{Re} i (\Pi^a_{\lambda})_{\phi_{L,R}}}{M_{\lambda}} \, , \label{eq:softa-r-gaugino-1} \\[1ex]
    (r^a_{\lambda})_{\phi_L \phi_R}
      &= - 8 \pi^2 \frac{\operatorname{Re} i (\Pi^a_{\lambda})_{\phi_L \phi_R}}{M_{\lambda}}   \label{eq:softa-r-gaugino-2}
  \end{align}
\end{subequations}
are positive and depend on the amount of supersymmetry breaking and the localizations of the bulk hypermultiplets. We plot $(r^a_{\lambda})_{\phi_{L,R}}$ in Fig.~\ref{fig:softa-gaugino-loopcoefficient-1} and $(r^a_{\lambda})_{\phi_L \phi_R}$ in Fig.~\ref{fig:softa-gaugino-loopcoefficient-2} as functions of $\sqrt{F}/\Lambda_{\text{IR}}$ for two choices of hypermultiplet localizations: $c_L = - c_R = - 1$ and $c_L = - c_R \ge \frac{1}{2}$ (the correction saturates for UV-localized fields). The U(1) (light), SU(2) (medium), and SU(3) (dark) contributions are given in each case. The coefficients $(r^a_{\lambda})_{\phi_{L,R}}$ and $(r^a_{\lambda})_{\phi_L \phi_R}$ exhibit behavior similar to $r^b_{\lambda}$, vanishing as $1/\xi$ in the limit $\sqrt{F}/\Lambda_{\text{IR}} \gg 1$.


\section{Analytical Estimate of Scalar Mass Running in the MSSM}\label{app:tachyon-analytics}

\begin{figure}[t]
  \centering
  \includegraphics{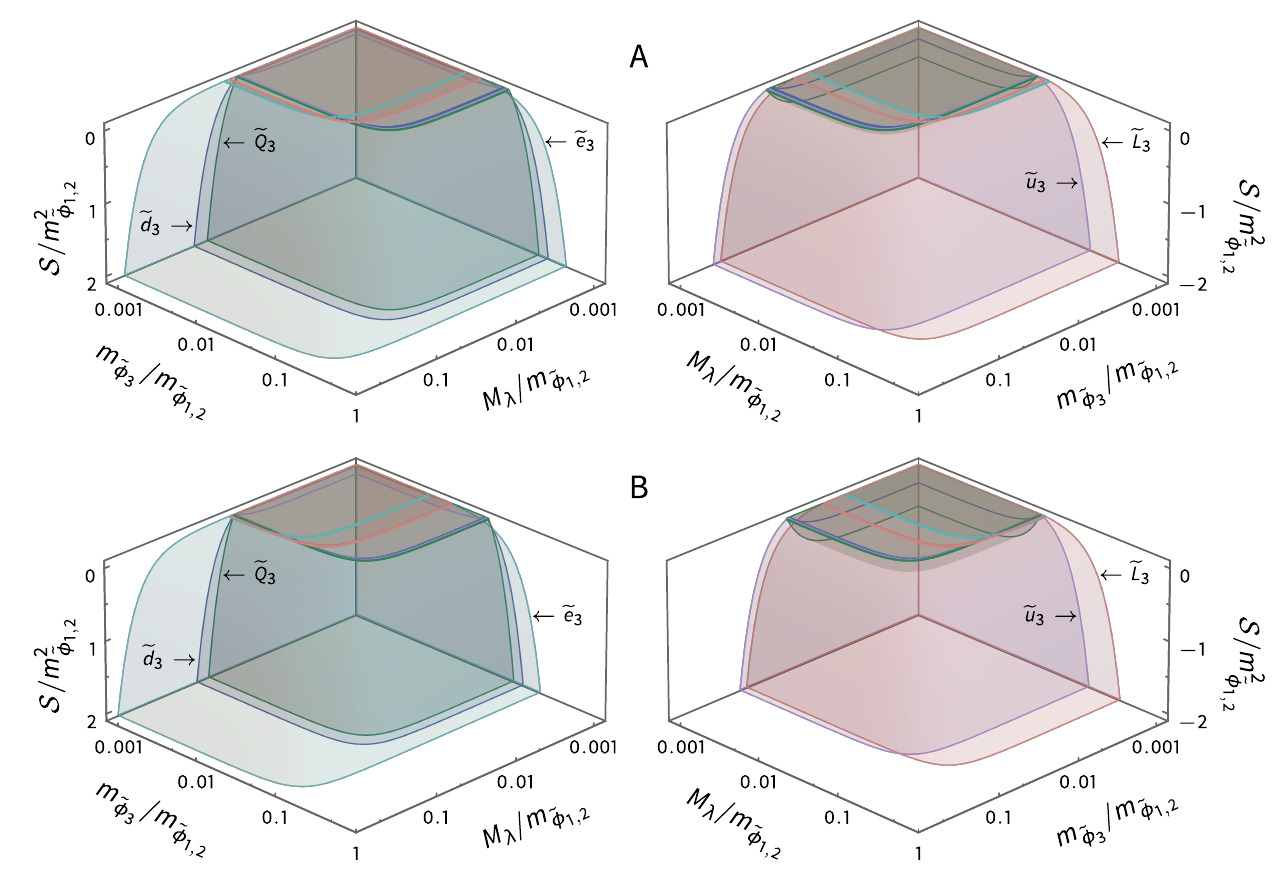}
  \caption{Estimate of the constraint on the sfermion masses in the MSSM for positive (left) and negative (right) values of the trace $\mathcal{S}$, where $\Lambda_{\text{IR}} = 2 \times 10^{16}$~GeV (upper row) and $\Lambda_{\text{IR}} = 6.5 \times 10^{6}$~GeV (lower row), corresponding to the scenarios A and B given in Table~\ref{tab:parameterpoints}. The shaded regions are excluded. We take $m_{\tilde{\phi}_{1,2}} = 100$~TeV and $m_{\text{SUSY}} = 10$~TeV.}
  \label{fig:tachyon-bounds-3d}
\end{figure}

\begin{figure}[t]
  \centering
  \includegraphics{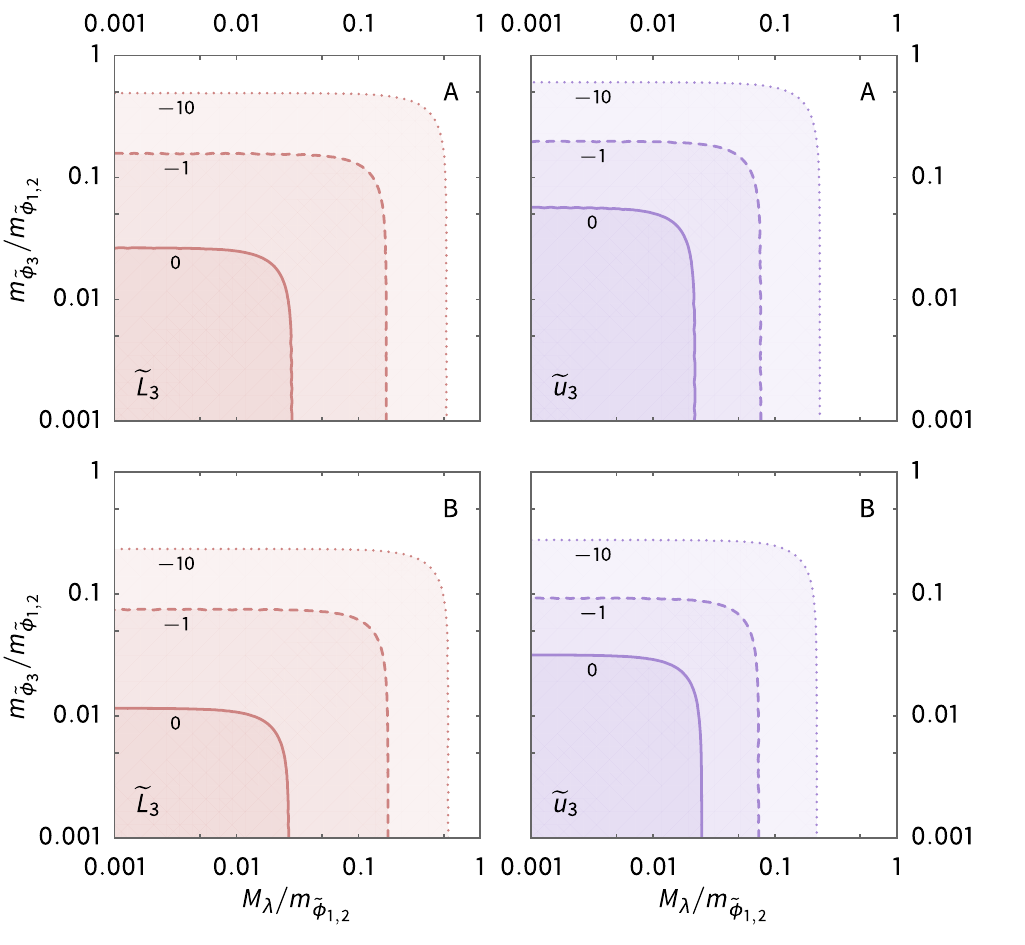}
  \caption{Contours of the estimated constraint on the sfermion masses in the MSSM for negative values of the ratio $\mathcal{S} / m^2_{\tilde{\phi}_{1,2}}$, where $\Lambda_{\text{IR}} = 2 \times 10^{16}$~GeV (upper row) and $\Lambda_{\text{IR}} = 6.5 \times 10^{6}$~GeV (lower row), corresponding to the scenarios A and B given in Table~\ref{tab:parameterpoints}. The shaded regions are excluded in each case. We take $m_{\tilde{\phi}_{1,2}} = 100$~TeV and $m_{\text{SUSY}} = 10$~TeV.}
  \label{fig:tachyon-bounds-2d-negative}
\end{figure}

\begin{figure}[t]
  \centering
  \includegraphics{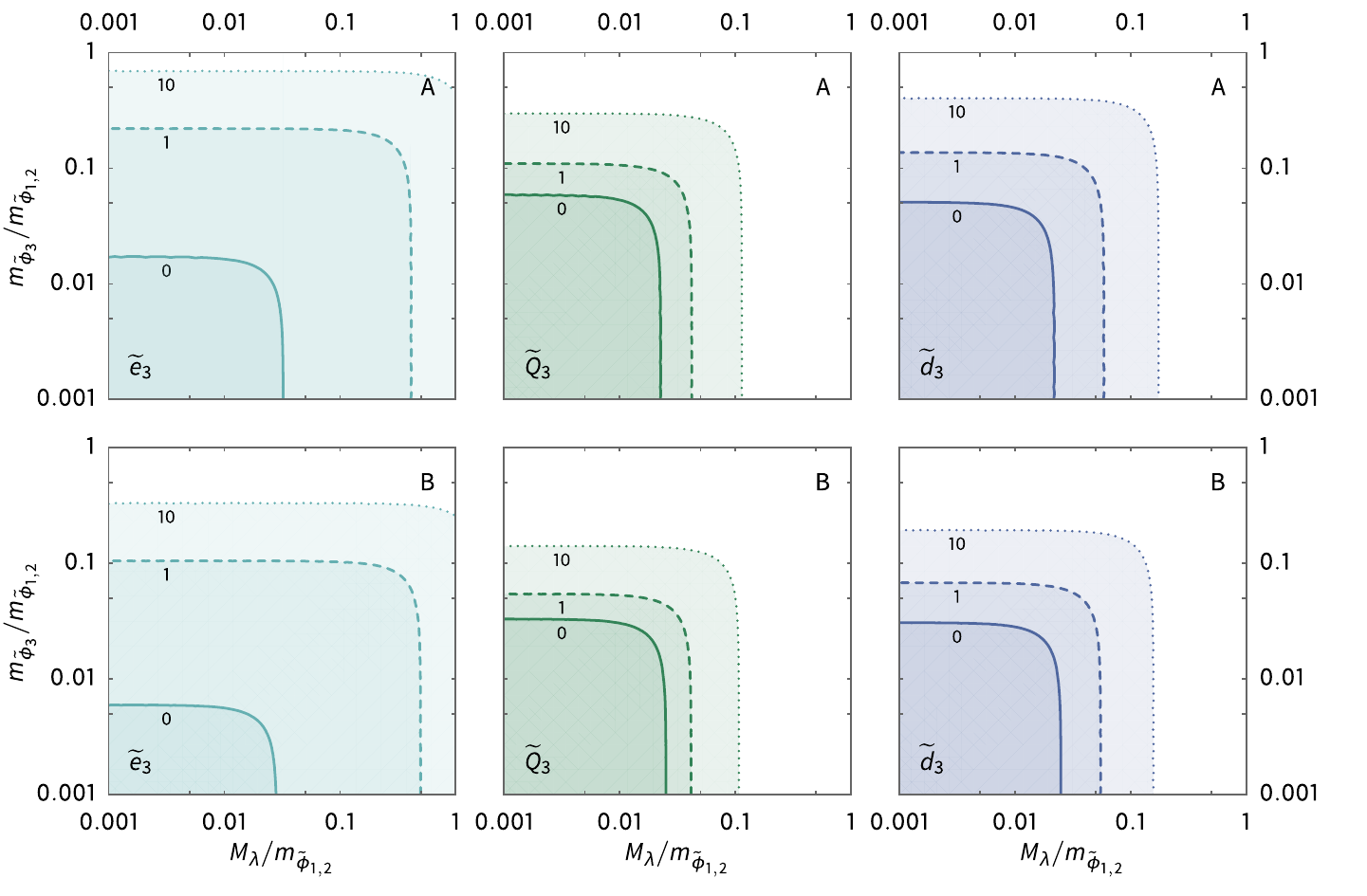}
  \caption{Contours of the estimated constraint on the sfermion masses in the MSSM for positive values of the ratio $\mathcal{S} / m^2_{\tilde{\phi}_{1,2}}$, where $\Lambda_{\text{IR}} = 2 \times 10^{16}$~GeV (upper row) and $\Lambda_{\text{IR}} = 6.5 \times 10^{6}$~GeV (lower row), corresponding to the scenarios A and B given in Table~\ref{tab:parameterpoints}. The shaded regions are excluded in each case. We take $m_{\tilde{\phi}_{1,2}} = 100$~TeV and $m_{\text{SUSY}} = 10$~TeV.}
  \label{fig:tachyon-bounds-2d-positive}
\end{figure}

The limits induced in the MSSM on scalar soft mass parameters by the terms \eqref{eq:scalar-correction-dterm-mssm} and \eqref{eq:scalar-correction-twoloop-mssm} can be estimated analytically. Following Ref.~\cite{ArkaniHamed:1997ab}, we write the RGE for the soft mass squared of a third-generation sfermion $\widetilde{\phi}_3$ as
\begin{align}
  \frac{d}{dt} m_{\tilde{\phi}_3}^2
    \simeq  - \frac{8}{16 \pi^2} \sum_a g_a^2 C^a(R_{\tilde{\phi}_3}) \, M_{\lambda_a}^2 
           &+ \frac{1}{16 \pi^2} \frac{6}{5} g_1^2 Y(\widetilde{\phi}_3) \, \mathcal{S} \nonumber \\[1ex]
           &+ \frac{4}{(16 \pi^2)^2} \sum_a g_a^2 C^a(R_{\tilde{\phi}_3}) \, m_{\tilde{\phi}_{1,2}}^2 \, ,
              \label{eq:scalar-rge-mssm}
\end{align}
where $C^a(R)$ is the quadratic Casimir [in the SU(5) normalization] of the representation $R$, $Y$ is the hypercharge, and $t$ is the logarithmic scale parameter. The parameters $M_{\lambda_a}$ are the gaugino masses and $m_{\tilde{\phi}_{1,2}}$ is the characteristic soft mass scale for the first- and second-generation sfermions. The $D$-term contribution, parametrized by $\mathcal{S}$, can be positive or negative and typically is of order the scale of $m_{\tilde{\phi}_{1,2}}^2$. Starting the running from a high scale (such as $\Lambda_{\text{IR}}$), we decouple the $D$-term and the two loop contribution at the mass scale of the heavy scalars $m_{\tilde{\phi}_{1,2}}$ (which we take to be constant) and the gaugino contribution at the lower scale $m_{\text{SUSY}}$. 

The scalar RGE (\ref{eq:scalar-rge-mssm}) can be solved analytically. We take the running gaugino masses to unify with a value $M_{\lambda_a}(m_{\text{GUT}}) = M_{\lambda}$ at the scale $m_{\text{GUT}} = 2 \times 10^{16}$~GeV. Enforcing $m_{\tilde{\phi}_3}(m_{\text{SUSY}}) > 0$ as a tachyon condition,\footnote{This is aggressive, as the $\xoverline[1]{\text{DR}}$ soft masses may take negative values, while the corresponding pole masses remain positive. See Ref.~\cite{Pierce:1996zz} for further details.} we obtain lower limits on the ratios $m_{\tilde{\phi}_3}(\Lambda_{\text{IR}}) / m_{\tilde{\phi}_{1,2}}$, $M_{\lambda} / m_{\tilde{\phi}_{1,2}}$, and $\mathcal{S}/ m_{\tilde{\phi}_{1,2}}^2$. These limits are shown in Figs.~\ref{fig:tachyon-bounds-3d}, \ref{fig:tachyon-bounds-2d-negative}, and \ref{fig:tachyon-bounds-2d-positive} for each third-generation sfermion. 

On the qualitative level, the limits are collectively weakest when $\mathcal{S} = 0$. The presence of a positive $D$-term contribution ameliorates the limits for the sfermions with negative hypercharge ($\widetilde{L}_3$ and $\widetilde{u}_3$), but worsens those of the positive hypercharge fields ($\widetilde{e}_3$, $\widetilde{Q}_3$, and $\widetilde{d}_3$), and vice versa when the $D$-term contribution is negative. Quantitatively, we expect these tachyon bounds to be accurate up to about an order of magnitude.


\newpage
\printbibliography[heading=bibintoc]


\end{document}